\documentclass{aa} 

\usepackage{graphicx}
\usepackage{xcolor}
\usepackage{natbib}
\bibpunct{(}{)}{;}{a}{}{,} 
\usepackage{txfonts}
\usepackage{hyperref}
\usepackage{upgreek}
\usepackage{footmisc}
\usepackage{CJK}

\newcommand\Gaia{\textit{Gaia }}
\newcommand\GnT{GALAH+TGAS}
\newcommand\GnTtotal{23096}
\newcommand\GnTused{7066}

\newcommand\importance[1]{#1} 
\newcommand\challenge[1]{#1} 
\newcommand\agetrend[1]{When we look at the trend with ages, we find #1} 
\newcommand\conclusion[1]{#1} 

\begin{document} 

\begin{CJK*}{UTF8}{gbsn}

   \title{The GALAH survey: An abundance, age, and kinematic inventory of the solar neighbourhood made with TGAS}

   \subtitle{}

   \author{
	{S.~Buder}\inst{1}\thanks{email: \href{mailto:buder@mpia.de}{buder@mpia.de}}\fnmsep\thanks{Fellow of the International Max Planck Research School for Astronomy \& Cosmic Physics at the University of Heidelberg},
	{K.~Lind}\inst{1,2},
	{M.~K.~Ness}\inst{3,4},
	{M.~Asplund}\inst{5,6},
	{L.~Duong}\inst{5},
	{J.~Lin}\inst{5},
	{J.~Kos}\inst{7,8},
	{L.~Casagrande}\inst{5,6},
	{A.~R.~Casey}\inst{9,10},
	{J.~Bland-Hawthorn}\inst{7,6}\fnmsep\thanks{Miller Professor, Miller Institute, University of California Berkeley, CA 94720, USA},
	{G.~M.~De~Silva}\inst{7,11},
	{V.~{D'Orazi}}\inst{12},
	{K.~C.~Freeman}\inst{5},
	{S.~L.~Martell}\inst{13,6},
	{K.~J.~Schlesinger}\inst{5},
	{S.~Sharma}\inst{7,6},
	{J.~D.~Simpson}\inst{13},
	{D.~B.~Zucker}\inst{11},
	{T.~Zwitter}\inst{8},	
	{K.~\v{C}otar}\inst{8},
	{A.~Dotter}\inst{14},
	{M.~R.~Hayden}\inst{7,6},
	{E.~A.~Hyde}\inst{15},
	{P.~R.~Kafle}\inst{16},
	{G.~F.~Lewis}\inst{7},
	{D.~M.~Nataf}\inst{17},
	{T.~Nordlander}\inst{5,6},
	{W.~Reid}\inst{15,11},
	{H.-W.~Rix}\inst{1},
	{\'{A}. Sk\'{u}lad\'{o}ttir}\inst{1},
	{D.~Stello}\inst{13,18,7,6},
	{Y.-S.~Ting ({丁\!源\!森})}\inst{19,20,21},
	{G.~Traven}\inst{8},
	{R.~F.~G.~Wyse}\inst{17},
	\and
	the~GALAH~collaboration
	}

   \titlerunning{The GALAH survey: An abundance, age, and kinematic inventory of the solar neighbourhood made with TGAS}
   \authorrunning{S. Buder et al.}

   \institute{ 
	Max Planck Institute for Astronomy (MPIA), Koenigstuhl 17, D-69117 Heidelberg
	\and 
	Department of Physics and Astronomy, Uppsala University, Box 516, SE-751 20 Uppsala, Sweden
	\and 
	Department of Astronomy, Columbia University, Pupin Physics Laboratories, New York, NY 10027, USA
	\and 
	Center for Computational Astrophysics, Flatiron Institute, 162 Fifth Avenue, New York, NY 10010, USA
	\and 
	Research School of Astronomy \& Astrophysics, Mount Stromlo Observatory, Australian National University, ACT 2611, Australia
	\and 
	Center of Excellence for Astrophysics in Three Dimensions (ASTRO-3D), Australia
	\and 
	Sydney Institute for Astronomy (SIfA), School of Physics, A28, The University of Sydney, NSW, 2006, Australia
	\and 
	Faculty of Mathematics and Physics, University of Ljubljana, Jadranska 19, 1000 Ljubljana, Slovenia
	\and 
	School of Physics and Astronomy, Monash University, Australia
	\and 
	Faculty of Information Technology, Monash University, Australia
	\and 
	Department of Physics and Astronomy, Macquarie University, Sydney, NSW 2109, Australia
	\and 
	Istituto Nazionale di Astrofisica, Osservatorio Astronomico di Padova, vicolo dell'Osservatorio 5, 35122, Padova, Italy
	\and 
	School of Physics, University of New South Wales, Sydney, NSW 2052, Australia
	\and 
	Harvard-Smithsonian Center for Astrophysics, Cambridge, MA 02138, USA
	\and 
	Western Sydney University, Locked Bag 1797, Penrith South, NSW 2751, Australia
	\and 
	ICRAR, The University of Western Australia, 35 Stirling Highway, Crawley, WA 6009, Australia
	\and 
	Department of Physics and Astronomy, The Johns Hopkins University, Baltimore, MD 21218, USA
	\and 
	Stellar Astrophysics Centre, Department of Physics and Astronomy, Aarhus University, DK-8000, Aarhus C, Denmark
	\and 
	Institute for Advanced Study, Princeton, NJ 08540, USA
	\and 
	Department of Astrophysical Sciences, Princeton University, Princeton, NJ 08544, USA
	\and 
	Observatories of the Carnegie Institution of Washington, 813 Santa Barbara Street, Pasadena, CA 91101, USA
    }

   \date{Received 12 04, 2018; accepted DD MM, YYYY}
 
 \abstract
{The overlap between the spectroscopic Galactic Archaeology with HERMES (GALAH) survey and \Gaia provides a high-dimensional chemodynamical space of unprecedented size. 
We present a first analysis of a subset of this overlap, of \GnTused~dwarf, turn-off, and sub-giant stars. These stars have spectra from the GALAH survey and high parallax precision from the \Gaia DR1 Tycho-\Gaia Astrometric Solution. We investigate correlations between chemical compositions, ages, and kinematics for this sample.
Stellar parameters and elemental abundances are derived from the GALAH spectra with the spectral synthesis code \textsc{Spectroscopy Made Easy}. We determine kinematics and dynamics, including action angles, from the \Gaia astrometry and GALAH radial velocities. Stellar masses and ages are determined with Bayesian isochrone matching, using our derived stellar parameters and absolute magnitudes.
We report measurements of Li, C, O, Na, Mg, Al, Si, K, Ca, Sc, Ti, V, Cr, Mn, Co, Ni, Cu, Zn, Y, as well as Ba and we note that we employ non-LTE calculations for Li, O, Al, and Fe. 
We show that the use of astrometric and photometric data improves the accuracy of the derived spectroscopic parameters, especially $\log g$.
Focusing our investigation on the correlations between stellar age, iron abundance [Fe/H], and mean alpha-enhancement [$\upalpha$/Fe] of the magnitude-selected sample, we recover the result that stars of the high-$\alpha$ sequence are typically older than stars in the low-$\upalpha$ sequence, the latter spanning iron abundances of $-0.7 < \mathrm{[Fe/H]} <  +0.5$. 
While these two sequences become indistinguishable in [$\upalpha$/Fe] vs. [Fe/H] at the metal-rich regime, we find that age can be used to separate stars from the extended high-$\upalpha$ and the low-$\upalpha$ sequence even in this regime. 
When dissecting the sample by stellar age, we find that the old stars ($> 8\,\mathrm{Gyr}$) have lower angular momenta $L_z$ than the Sun, which implies that they are on eccentric orbits and originate from the inner disk. Contrary to some previous smaller scale studies we find a continuous evolution in the high-$\upalpha$-sequence up to super-solar [Fe/H] rather than a gap, which has been interpreted as a separate "high-$\upalpha$ metal-rich" population.
Stars in our sample that are younger than $10\,\mathrm{Gyr}$, are mainly found on the low $\upalpha$-sequence and show a gradient in $L_z$ from low [Fe/H] ($L_z > L_{z,\odot}$) towards higher [Fe/H] ($L_z < L_{z,\odot}$), which implies that the stars at the ends of this sequence are likely not originating from the close solar vicinity.
}
   \keywords{Surveys -- Galaxy: solar neighborhood, evolution -- stars: fundamental parameters, abundances, kinematics and dynamics \vspace{-1.0cm}}

   \maketitle
%
$\,$
\newpage
\section{Introduction}

The first \Gaia data release \citep{Gaia2016}, which includes the Tycho-\Gaia Astrometric Solution (TGAS), is a milestone of modern astronomy and has delivered positions, proper motions, and parallaxes for more than two million stars \citep{Michalik2015, Lindegren2016}. This has marked the beginning of the \Gaia era, which will shed new light on our understanding of the formation and evolution of our Galaxy. The astrometric information delivered by \Gaia will be particularly important when used in combination with spectroscopic quantities obtained using large ground-based surveys \citep[e.g.][]{AllendePrieto2016, McMillan2018, Helmi2017, Kushniruk2017, PriceWhelan2017}. Taken together, these data will prove particularly powerful in testing our models of Galactic assembly. The ensemble of GALAH \citep{DeSilva2015} and \Gaia data have now begun to provide a high dimensionality mapping of the chemodynamical space of the nearby disk. In this study, we use the combination of GALAH and TGAS to describe the chemical, temporal, and kinematical distributions of nearby stars in the disk of the Milky Way. 

The paper is organised as follows: We first introduce GALAH and discuss the strengths of combining this survey with the astrometric information provided by the \Gaia satellite. In Sect.~\ref{observation}, we outline the observational strategy for GALAH. In Sect.~\ref{analysis} we explain our estimation of  stellar properties, including stellar parameters, chemical composition, and ages. The results of our analysis are presented in Sect.~\ref{results} where we show the abundance and age trends for our sample and discuss the observed distribution of disk stars as a function of chemical composition, age and kinematics. In Sect.~\ref{discussion}, we discuss the implications of our findings and make suggestions for further studies with GALAH (DR2) and \Gaia (DR2) in the concluding section of the paper.

\subsection{The GALAH survey}

The GALAH survey is a ground-based, high-resolution stellar spectroscopic survey. It is executed with the High Efficiency and Resolution Multi-Element Spectrograph (HERMES) fed by the Two Degree Field (2dF) f/3.3 top end at the Anglo-Australian Telescope \citep{Barden2010, Brzeski2011, Heijmans2012, Farrell2014, Sheinis2015}. The overall scientific motivation for GALAH is presented in \citet{DeSilva2015}. The survey's primary goal is the chemical tagging experiment, as proposed by \citet{FreemanBlandHawthorn2002} and described in detail by \citet{BlandHawthorn2010a}. Chemical tagging offers the promise of linking stars that were born together via their chemical composition. As proposed by recent simulations \citep{Ting2015, Ting2016}, this promise can be best explored with a high dimensionality in chemical space. Consequently, the spectrograph has been optimised to measure up to 30 different elements (more in very bright stars), covering a multitude of different nucleosynthesis channels, depending on the stellar type and evolution. 

The GALAH survey selection function is both simple (magnitude limited) and will ensure that almost all stars observed by the GALAH survey are also measured by the \Gaia satellite, see Sect.~\ref{observation} and \citet{DeSilva2015}.
Our target selection limit of $V \leq 14$ corresponds to \Gaia's peak performance with distance uncertainties for all GALAH stars expected to be better than 1\%.
Once GALAH is completed, this will provide both chemical and kinematical information for up to a million stars. These data will directly inform Galactic archaeology pursuits, enabling the empirical construction of the distribution function of stellar properties and populations (chemical composition, age, position, orbits). The sample analysed here, is a first step in this direction.

For the analysis of the whole GALAH survey data, a combination of classical spectrum synthesis with \textsc{Spectroscopy Made Easy} (SME) by \citet{Piskunov2017} and with a data-driven propagation via \textsc{The Cannon} \citep{Ness2015} is used. Prior to this study, data releases of GALAH \citep{Martell2017}, TESS-HERMES \citep{Sharma2018} and K2-HERMES \citep{Wittenmyer2018} were based only on spectroscopic input and provided the stellar properties obtained with the \textsc{The Cannon}. We stress that this work focuses on the first part of the usual GALAH analysis routine and is based only on a spectroscopic analysis with SME, but also includes photometric and astrometric input for the analysis. The stars used in this analysis will be a subset of the training set used for \textsc{The Cannon} in later data releases \citep[GALAH DR2;][]{Buder2018b}.

\subsection{The strengths of combining GALAH and \Gaia DR1 TGAS}

Combining \Gaia DR1 TGAS with high-resolution spectroscopy provides a variety of opportunities, ranging from an improved analysis of stellar parameters up to the expansion of the chemical space by the kinematical one.

Purely spectroscopic analyses may suffer from inaccuracies due to degeneracies in effective temperature ($T_\mathrm{eff}$), surface gravity ($\log g$), and chemical abundances. This is a consequence of simplified assumptions about stellar spectra and the subsequent construction of incomplete stellar models, for example assuming a 1D hydrostatic atmosphere and chemical compositions scaled with solar values and the metallicities. Previous high-resolution spectroscopic studies \citep[see e.g.][]{Bensby2014, Martell2017} have found inconsistencies between purely spectroscopic parameter estimates and those also based on photometric and asteroseismic information. Furthermore, many studies find that unphysical low surface gravities are estimated for G and K-type main sequence stars from spectroscopy alone \citep{Sousa2011, Adibekyan2012}. Cool dwarfs ($T_\mathrm{eff} < 4500\,\mathrm{K}$) are particularly challenging to study in the optical regime, because of the weakening of the singly ionised lines that are used to constrain the ionisation equilibrium, and due to the increasing influence of molecular blends as well as the failure of 1D LTE modelling; see e.g. \citet{Yong2004}. Adding further (non-spectroscopic) information may alleviate these problems.  Asteroseismic as well as interferometric and bolometric flux measurements for dwarf and turn-off stars are, however, still expensive and published values rare (especially for stars in the Southern hemisphere). \Gaia DR2 and later releases will provide astrometric information for all observed GALAH stars and numerous stars that have been observed by other spectroscopic surveys, e.g. APOGEE \citep{Majewski2017}, RAVE \citep{Kunder2017}, Gaia-ESO \citep{Gilmore2012, Randich2013}, and LAMOST \citep{Cui2012}. The estimation of bolometric luminosities using both astrometric and photometric information will therefore be feasible for a large sample of stars in the near future.

Astrometric data, in combination with photometric and spectroscopic information, can also improve extinction estimates and narrow down uncertainties in the estimation of stellar ages from theoretical isochrones. Parallaxes and apparent magnitudes can also be used to identify binary systems with main sequence stars, which are not resolved by spectroscopy. This is important, because the companion contributes light to the spectrum and can hence significantly contaminate the analysis results \citep{ElBadry2018,ElBadry2018b}.

Prior to the first \Gaia data release (DR1), the most notable observational chemodynamical studies were performed using the combination of the astrometric data from {\sc Hipparcos} \citep{ESA1997, vanLeeuwen2007} and Tycho-2 \citep{Hog2000} with additional observations by the Geneva-Copenhagen-Survey \citep{Nordstroem2004, Casagrande2011} and high-resolution follow-up observations \citep[e.g. ][]{Bensby2014}. Another approach, including post-correction of spectroscopic gravities has been adopted by \cite{DelgadoMena2017} for the HARPS-GTO sample. Large scale analyses have however been limited by the precision of astrometric measurements by {\sc Hipparcos} to within the volume of a few hundred parsecs at most. With the new \Gaia data, this volume is expanded to more than $1\,\mathrm{kpc}$ with DR1 and will be expanded even further with DR2, allowing the study of gradients or overdensities/groups in the chemodynamical space of the Milky Way disk and beyond.

\subsection{Milky Way disk definitions and membership assignment}

The disk is the most massive stellar component of the Milky Way. Numerous studies have observed the disk in the Solar neighbourhood. The pioneering studies by \citet{Yoshii1982} and \citet{Gilmore1983} found evidence for two thin and thick sub-populations in the disk based on stellar density distributions. Recent studies \citep[e.g.][]{Bovy2012, Bovy2016} find a structural continuity in thickness and kinematics, and the latter property has been shown to be a rather unreliable tracer of the disk sub-populations \citep{Bensby2014}. However, several seminal papers \citep[e.g.][]{Reddy2003, Fuhrmann2011, Adibekyan2012, Bensby2014, Hayden2015} have established that the stellar disk consists of (at least) two major components in chemical space and age, commonly adopted as old, $\upalpha$-enhanced, metal-poor thick and the young thin disk with solar-like $\upalpha$-enhancement at metallicities of $-0.7 < \mathrm{[Fe/H]} < +0.5$. However, the bimodality between these two populations in the high-$\upalpha$ metal-rich regime has been shown to become less or not significant and is still contentious, based on the chosen approaches and population cuts used for the definition of disk populations. \citep{Adibekyan2011} even claimed a third sub-population in this regime. Some of the recent studies using chemistry assume the existence of two distinct populations in $\upalpha$-enhancement up to the most metal-rich stars. In these studies, the metal-rich stars are cut into high and low sequence memberships rather arbitrarily; either by eye or with rather fiducial straight lines (e.g. \citet{Adibekyan2012} Fig.~7, \citep{Recio-Blanco2014} Fig.~12 or \citet{Hayden2017} Fig.~1). Other approaches to separate the $\upalpha$-sequences or stellar populations, e.g. kinematically \citep{Bensby2014} or via age \citep{Haywood2013} result in different separations. A consistent measure or definition to separate both $\upalpha$-sequences especially in the metal-rich regime remains elusive. For a discussion on combining chemistry and kinematics to separate the two $\upalpha$-sequences see e.g. \citet{Haywood2013} and \citet{Duong2018}. For a more detailed overview regarding the definition of the stellar disks, we refer the reader to \citet{Martig2016b} and references therein. 

Although element abundances are easier to determine than stellar ages, their surface abundances and measurements are subject to changes due to processes within the atmosphere of a star. Stellar ages are therefore the most promising tracer of stellar evolution and populations \citep{Haywood2013,Bensby2014, Ness2016, Ho2017b}. We note that, most recently, \citet{Hayden2017} investigated abundance sequences as a function of age, but with a representation as function of age ranges, starting with only old stars and subsequently including more younger ones (see their Fig.~3).

Until now, however, most authors suggested that this issue should be revisited when a larger, homogeneous, and less biased sample is available. With the observations obtained by the GALAH survey, we are now able to investigate the abundance sequences with a significantly larger and homogeneous data set with a rather simple selection function. 

%
\section{Observations} \label{observation} 

The GALAH survey collects data with HERMES, which can observe up to 360 science targets at the same time plus 40 fibres allocated for sky and guide stars \citep{Sheinis2015, Heijmans2012, Brzeski2011,Barden2010}. The selection of targets and observational setup are explained in detail by \citet{DeSilva2015} and \citet{Martell2017}. The observations used in this study were carried out between November 2013 and September 2016 with the lower of the two resolution modes ($\lambda/\Delta \lambda \sim 28000$) with higher throughput, covering the four arms of HERMES, i.e. blue ($4716-4896\,\mathrm{\AA}$ including $\mathrm{H_\upbeta}$), green ($5650-5868\,\mathrm{\AA}$), red ($6480-6734\,\mathrm{\AA}$ including $\mathrm{H_\upalpha}$) as well as the near infrared ($7694-7876\,\mathrm{\AA}$, including the oxygen triplet).

The initial simple selection function of the GALAH survey was achieved with a random selection of stars within the limiting magnitudes $12 < V < 14$ derived from 2MASS photometry \citep{DeSilva2015}. To ensure a large overlap with TGAS \citep{Michalik2015}, our team added special bright fields ($9 < V < 12$) including a large number of stars in the Tycho-2 catalog \citep{Hog2000} which were brighter than the nominal GALAH range \citep{Martell2017}. The exposure times were chosen to achieve a signal-to-noise ratio ($S/N$) of 100 per resolution element in the green channel / arm; 1 hour for main survey targets in optimal observing conditions, often longer in suboptimal observing conditions. The spectra are reduced with the GALAH pipeline \citep{Kos2017}, including initial estimates of $T_\mathrm{eff}$, $\log g$, [Fe/H], and radial velocities ($v_\text{rad}$).

\begin{figure}[!ht]
\centering
\resizebox{\hsize}{!}{\includegraphics{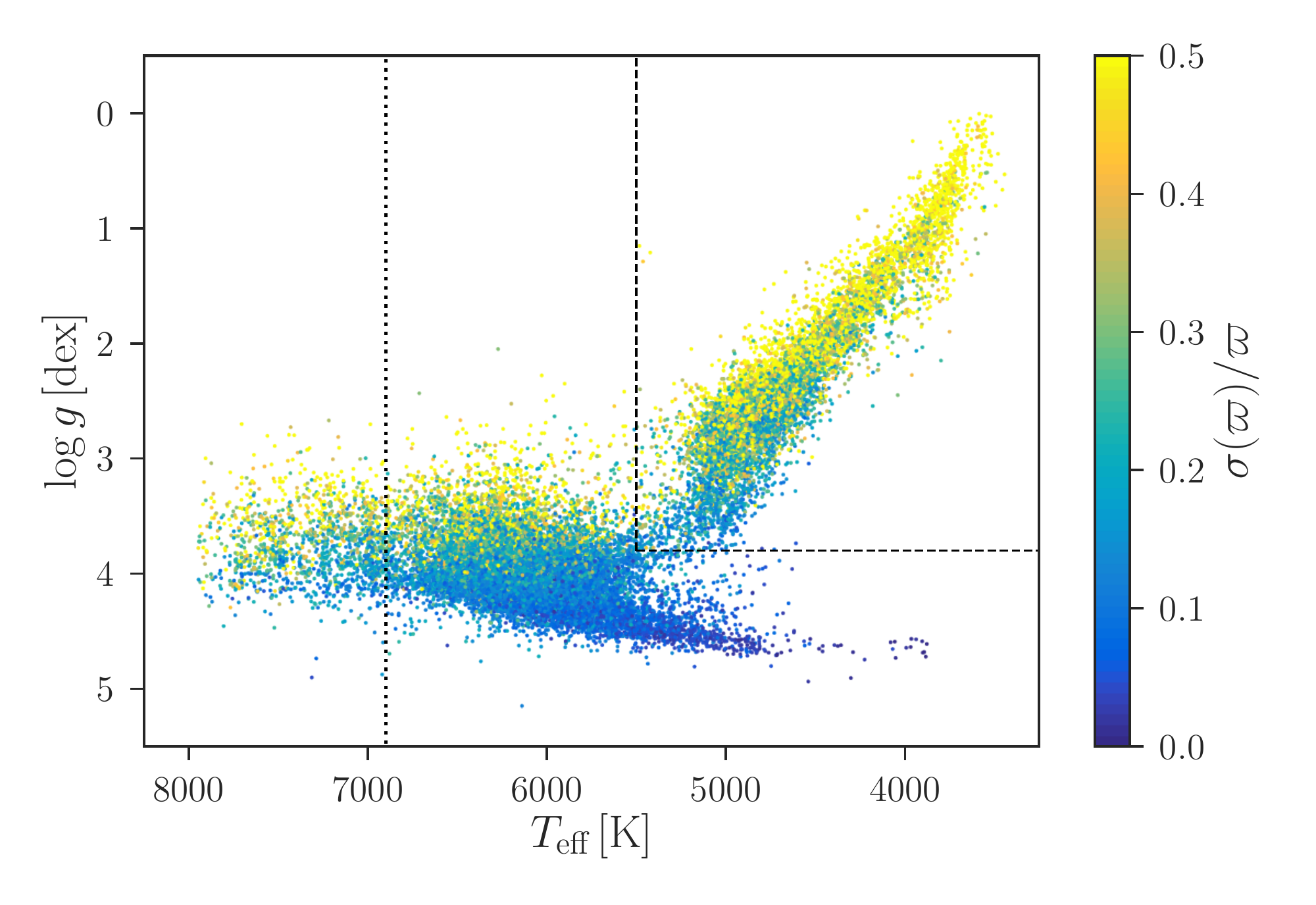}}
\caption{Kiel diagram with effective temperature $T_\text{eff}$ and surface gravity $\log g$ for the complete GALAH+TGAS overlap. The spectroscopic parameters are results of the analysis in Sect.~\ref{pipeline}. Colour indicates the relative parallax  error. Most precise parallaxes are measured for the cool main sequence stars and the parallax precision decreases both towards the turn-off sequence and even more drastically towards evolved giants, which are the most distant stars in the sample. Dotted and dashed black lines indicate the limits to neglect hot stars ($T_\text{eff} > 6900\,\mathrm{K}$) and giants ($T_\text{eff} < 5500\,\mathrm{K}$ and $\log g < 3.8\,\mathrm{dex}$) for the subsequent analysis, respectively. See text for details on the exclusion of stars.}
\label{plx_distribution_hrd}
\end{figure}

\begin{figure}[!ht]
\centering
\resizebox{\hsize}{!}{\includegraphics{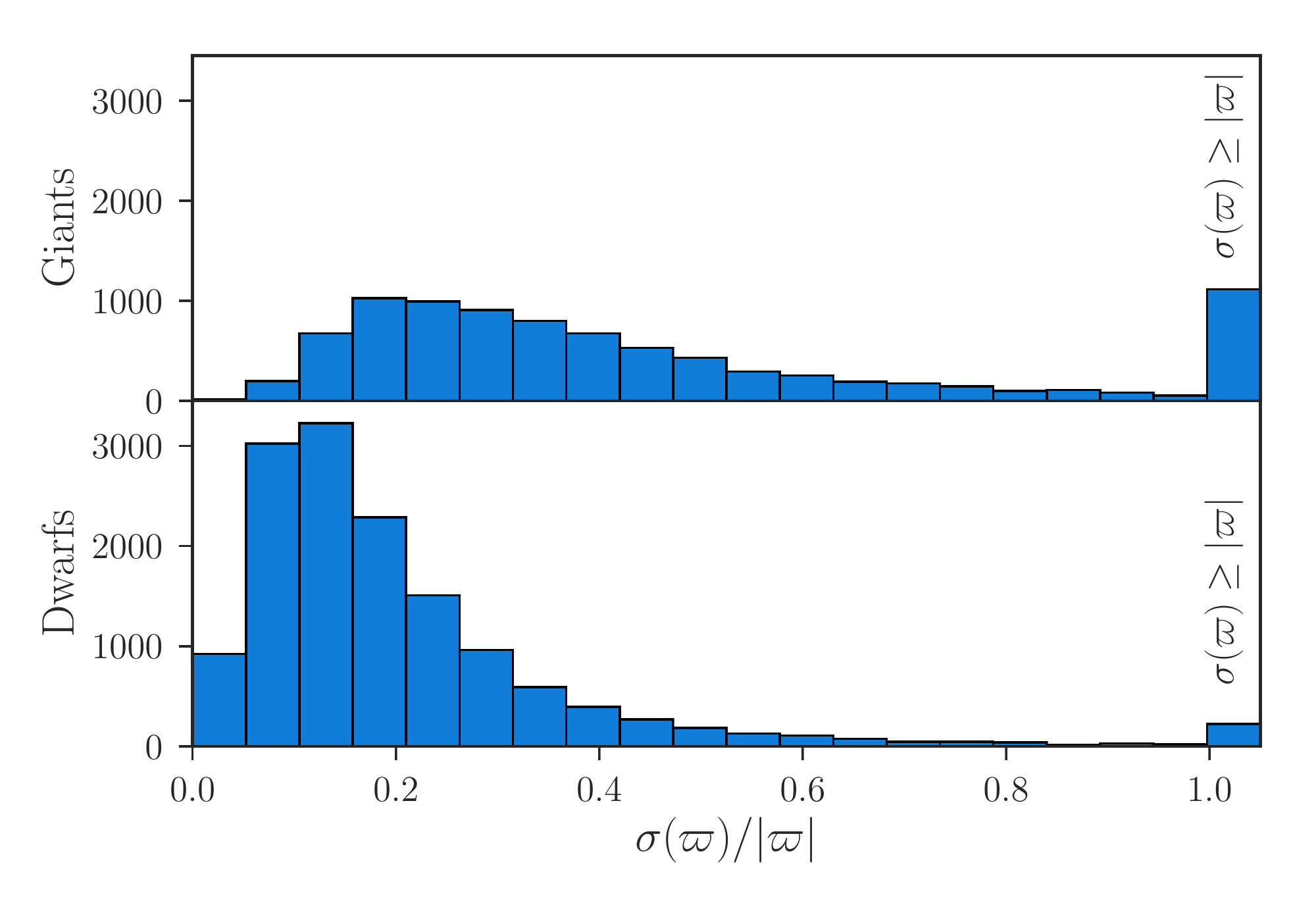}}
\caption{Histograms of relative parallax uncertainties for both giants (top panel with $T_\text{eff} < 5500\,\mathrm{K}$ and $\log g < 3.8\,\mathrm{dex}$) and dwarfs (lower panel, including main-sequence and turn-off stars). The majority of the \GnT~overlap consists of dwarfs. Their mean parallax precision is in the order of $10\%$, while giants parallaxes are less precise with most uncertainties above $20\%$. The histograms are truncated at $\sigma(\varpi) = \varpi$ for readability. All stars with parallax uncertainties larger than the parallax itself are hence contained in the last bin.}
\label{plx_distribution}
\end{figure}

The \GnT~sample, observed until September 2016, consists of \GnTtotal~stars, covering mainly the spectral types F-K from pre-main sequence up to evolved asymptotic giant branch stars. For an overview of the sample, the spectroscopic parameters are depicted in Fig.~\ref{plx_distribution_hrd}, coloured by the parallax precision from TGAS. We note that the shown parameters are estimated as part of this study (see Sect.~\ref{pipeline}). The most precise parallaxes ($\sigma (\varpi)/\varpi \leq 0.05$) are available for main sequence stars cooler than $6000\,\mathrm{K}$, decent parallaxes ($\sigma (\varpi)/\varpi \leq 0.3$) for most dwarfs and the lower luminosity end of the red giant branch. As expected by the magnitude constraints of TGAS as well as GALAH \citep{DeSilva2015}, most of the overlap consists of dwarfs and turn-off stars ($62\,\mathrm{\%}$) which also have smaller relative parallax uncertainties than the more distant giants, see Fig.~\ref{plx_distribution}. Cool evolved giants as well as hot turn-off stars have the least precise parallaxes of the GALAH+TGAS overlap because of their larger distances and are hence not included in the online tables.

In this work, we limit the sample for the analyses to dwarfs and turn-off stars ($T_\text{eff} \geq 5500\,\mathrm{K} \text{~or~} \log g \geq 3.8\,\mathrm{dex}$, see dashed line in Fig.~\ref{plx_distribution_hrd}) with relative parallax uncertainties smaller than $30\,\mathrm{\%}$. This allows the best estimation of ages from isochrones as well as reliable distance and kinematical information and avoids possible systematic differences in the analysis due to the different evolutionary stages of the stars. Evolutionary effects, such as atomic diffusion, have been studied both with observations of clusters (see e.g. studies of the open cluster M67 by \citet{Oenehag2014, BertelliMotta2017, Gao2018}) as well as a theoretical predictions \citep{Dotter2017} and are beyond the scope of this paper.

In addition to removing 8740 giant stars and 7674 stars with parallax uncertainties above $30\%$, we exclude some stars after a visual inspection of the spectra and using our quality analysis (explained in Sect.~\ref{pipeline}). We construct a final sample with reliable stellar parameters and element abundances. We neglect 54 stars with emission lines, 926 stars with bad spectra or reductions, 448 double-lined spectroscopic binary stars, 338 photometric binaries (see Sect.~\ref{binaries}), 3429 stars with broadening velocities above $30\,\mathrm{km/s}$ (mostly hot turn-off stars with unbroken degeneracies of broadening velocity and stellar parameters with the GALAH setup), 1390 stars with $T_\mathrm{eff} > 6900\,\mathrm{K}$ (for which we have not been able to measure element abundances) and 1048 stars with $S/N$ below 25 in the green channel. Note that the different groups of excluded stars defined above are overlapping with each other.

\begin{figure}[!ht]
\centering
\resizebox{\hsize}{!}{\includegraphics{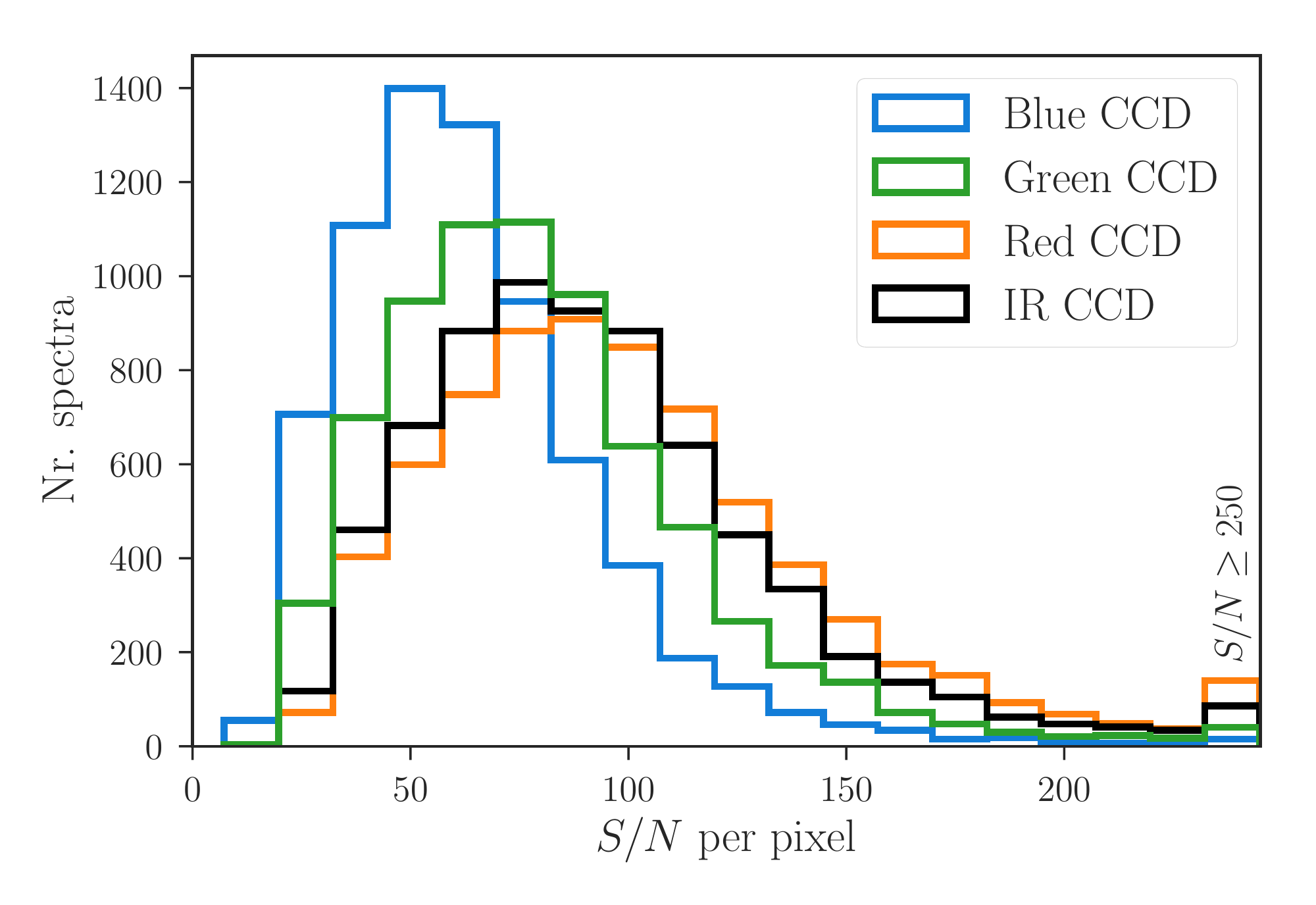}}
\caption{Distribution of $S/N$ per pixel for the different HERMES wavelength bands ($S/N$ per resolution element is about twice as high) for the final sample. The $S/N$ for the green, red and IR channels  are mainly in the range of 50 to 150, i.e. above the nominal survey aim of $S/N$ of 100 per resolution element in the green channel. The $S/N$ in the blue channel is smaller, with typically 25 to 100. The mean values per band are 59/75/94/88. This indicates a smaller influence of the blue band in the parameter estimation with $\chi^2$ minimisation explained in Sect.~\ref{analysis} and lower precision of element abundances measured within this channel.}
\label{SNR_hist}
\end{figure}

For the final selected sample of \GnTused~stars, the majority of the individual $S/N$ vary between 25 to 200, see Fig.~\ref{SNR_hist}. Most of the stars have a higher $S/N$ than the targeted nominal survey value for the green channel. We note that the $S/N$ of the blue arm is lower than in the others. For abundances measured within this arm, like Zn, we also estimate typically lower precision, see Sect.~\ref{ironpeak}.

\section{Analysis}\label{analysis} 

Our analysis combines the use of information derived from our GALAH spectra with additional photometric and astrometric measurements to achieve the best possible parameter estimation. We validate our analysis in a manner similar to other large-scale stellar surveys, such as APOGEE \citep{Majewski2017, SDSSDR14, GarciaPerez2016} or \textit{Gaia}-ESO \citep{Gilmore2012, Randich2013, Smiljanic2014, Pancino2017}, using a set of well-studied stars, including the so called \Gaia FGK benchmark stars \citep[hereafter GBS, see][]{Heiter2015,Jofre2014}. The stellar parameters $T_\text{eff}$ and $\log g $ of the GBS have been derived from direct observables: angular diameters, bolometric fluxes, and parallaxes, and are thus less model-dependent. They therefore provide reference parameters that do not suffer from the same model dependence as isolated spectroscopy. 
Among others, \citet{Schoenrich2014} and \citet{Bensby2014} showed the strength of combining spectroscopy and external information. The latter applied this approach for a sample of 714 nearby dwarfs with high accuracy astrometric parallaxes \citep{vanLeeuwen2007} from the {\sc Hipparcos} mission. We will use their sample as a reference for this study, because the spectral analysis was performed in a similar way, including the anchoring of surface gravity to astrometric information. We stress that their study was performed with higher quality spectra (both regarding the spectral resolution and $S/N$) which allowed a higher precision on measurements to be achieved.

\subsection{Stellar parameter determination} \label{pipeline}

By using the fundamental relation between surface gravity, stellar mass, effective temperature, and bolometric luminosity
\begin{equation}\label{logg}
\log g = \log g_\odot - \log \frac{L_\text{bol}}{L_{\text{bol},\odot}}  + 4 \log \frac{T_\text{eff}}{T_{\text{eff},\odot}} + \log \frac{\mathcal M}{\mathcal  M_\odot}, 
\end{equation}
the degeneracies with $\log g$ and other spectroscopically determined stellar parameters are effectively broken. The thereby improved values of $T_\text{eff}$ and $\log g$ leads to improved estimates of metallicities. Using broadband photometry (apparent magnitudes $K_S$ and inferred bolometric corrections $BC_{K_S}$ as well as extinction $A_{K_S}$) in combination with parallaxes $\varpi$ or distances $D_\varpi$, it is possible to estimate the bolometric magnitudes ($M_\text{bol}$) and luminosities ($L_\text{bol}$) to high precision and accuracy \citep[see e.g.][]{Alonso1995,Nissen1997,Bensby2014}:
\begin{equation}\label{lbol}
- 2.5 \cdot \log \frac{L_\text{bol}}{L_{\text{bol},\odot}} = K_S + BC_{K_S} - 5 \cdot \log \left( D_{\varpi} \right) + 5 - A_{K_S} - M_{\text{Bol},\odot}
\end{equation}

The nominal values for the Sun (used in Eqs. \ref{logg} and \ref{lbol}) of $T_{\text{eff},\odot} = 5772\,\mathrm{K}$, $\log (g_\odot) = 4.438\,\mathrm{dex}$, and $M_{\text{Bol},\odot} = 4.74\,\mathrm{mag}$ are taken from \citet{Prsa2016}.

Any filter with available bolometric corrections can be used for the computation of $L_\mathrm{bol}$; the $V$ band is commonly used for nearby stars. However, our GALAH data set also contains stars with substantial reddening and published catalogues of $V$ band magnitudes, such as APASS \citep{Henden2016}, have multiple input sources, affecting the homogeneity of the data. We therefore decide in favour of using the $K_S$ band as given by 2MASS \citep{Cutri2003}, available for all our targets. Bolometric corrections $BC =BC(T_\text{eff}, \log g, \mathrm{[Fe/H]}, E(B-V))$  were interpolated with the grids from \citet{Casagrande2014}. Distances were taken from \citet{Astraatmadja2016}, using a Milky Way model as Bayesian prior. For attenuation, we use the RJCE method $A_K = A_K (K_S,W2)$ by \citet{Majewski2011,Zasowski2013}. If $K_S$ or $W2$ could not be used, we use the approximation $A_K \sim 0.38 E(B-V)$ estimated by \citet{Savage1979} with $E(B-V)$ from \citet{Schlegel1998}. The reddening of our sample is on average $E(B-V) = 0.12 \pm 0.14\,\mathrm{mag}$. For the nearby dwarfs, however, $A_K$ is very small and thus hard to estimate given the photometric uncertainties; hence it was set to 0 if the RJCE method yielded negative values.

With the exception of $\log g$, stellar parameters and abundances are estimated using the spectrum synthesis code SME \citep{Valenti1996,Piskunov2017}, which uses a Marquardt-Levenberg $\chi^2$-optimisation between the observed spectrum and synthetic spectra that are calculated on-the-fly. As part of the \GnT~pipeline, SME version 360 is used, with {\sc marcs} 1D model atmospheres \citep{Gustafsson2008} and non-LTE-synthesis of iron from \citet{Lind2012}.  The chemical composition is assumed equal to the standard {\sc marcs} composition, including gradual $\upalpha$-enhancement toward lower metallicity\footnote{\label{footnote_alphaenhancement}$\mathrm{[\upalpha/Fe]} = \begin{cases}
 0.4 & \text{for~} \mathrm{[Fe/H]} < -1.0 \\
 0.4 \cdot (-\mathrm{[Fe/H]}) & \text{for~} \mathrm{[Fe/H]} \in [-1.0,0.0] \\
 0.0 &  \text{for~} \mathrm{[Fe/H]} > 0.0 \end{cases}$}. The pipeline is operated in the following way:

\begin{enumerate}
	\item Stellar parameters are initialised from the analysis run used by \citet{Martell2017} if available and unflagged, otherwise the output from the reduction pipeline \citep{Kos2017} is used and if these are flagged, we adopt generic starting values $T_\text{eff} = 5000\,\mathrm{K}$, $\log g = 3.0$, and $\mathrm{[Fe/H]} = -0.5$.
	\item Predefined $3-9\,\textup{\AA}$ wide segments are normalised and unblended and well modelled Fe, Sc, and Ti lines within each segment are identified. Broader segments are used for the Balmer lines. The continuum shape is estimated by SME assuming a linear behaviour for each segment, and based on selected continuum points outside of the line masks.
	\item Stellar parameters are iterated in two SME optimisation loops.
		\begin{enumerate}
		\item  SME parameters $T_\text{eff}$, [Fe/H], $v \sin i \equiv v_\mathrm{broad}$, and $v_\text{rad}$ are optimised by $\chi^2$ minimisation using partial derivatives.
		\item Whenever $T_\text{eff}$ or [Fe/H] change, $\log g$ and $v_\text{mic}$ are updated before the calculation of new model spectra and their $\chi^2$. We adjust $\log g$ according to Eq. \ref{logg} with isochrone-based masses $M = M(T_\text{eff}, \log g, \mathrm{[Fe/H]}, M_{K_S})$ estimated by the \textsc{Elli}~code, see Sect.~\ref{age}. We adjust $v_\text{mic}$ following empirical relations estimated for GALAH\footnote{If $\log g \leq 4.2\text{ or }T_\text{eff} \geq T_0$ with $T_0 = 5500\,\mathrm{K}$: \newline $v_\text{mic} = 1.1 + 1.0 \cdot 10^{-4} \cdot (T_\text{eff} - T_0) + 4\cdot10^{-7} \cdot (T_\text{eff} - T_0)^2$, else: \newline $\,v_\text{mic} = 1.1 + 1.6\cdot10^{-4} \cdot (T_\text{eff} - T_0)$}
		\end{enumerate}
	\item Each segment is re-normalised with a linear function while minimising the $\chi^2$ distance for the chosen continuum points between observation and the synthetic spectrum created from the updated set of parameters \citep{Piskunov2017}
	\item The stellar parameters are iteratively optimised until the relative $\chi^2$-convergence criterium is reached.
\end{enumerate}

\begin{figure}[!ht]
\centering
\resizebox{\hsize}{!}{\includegraphics{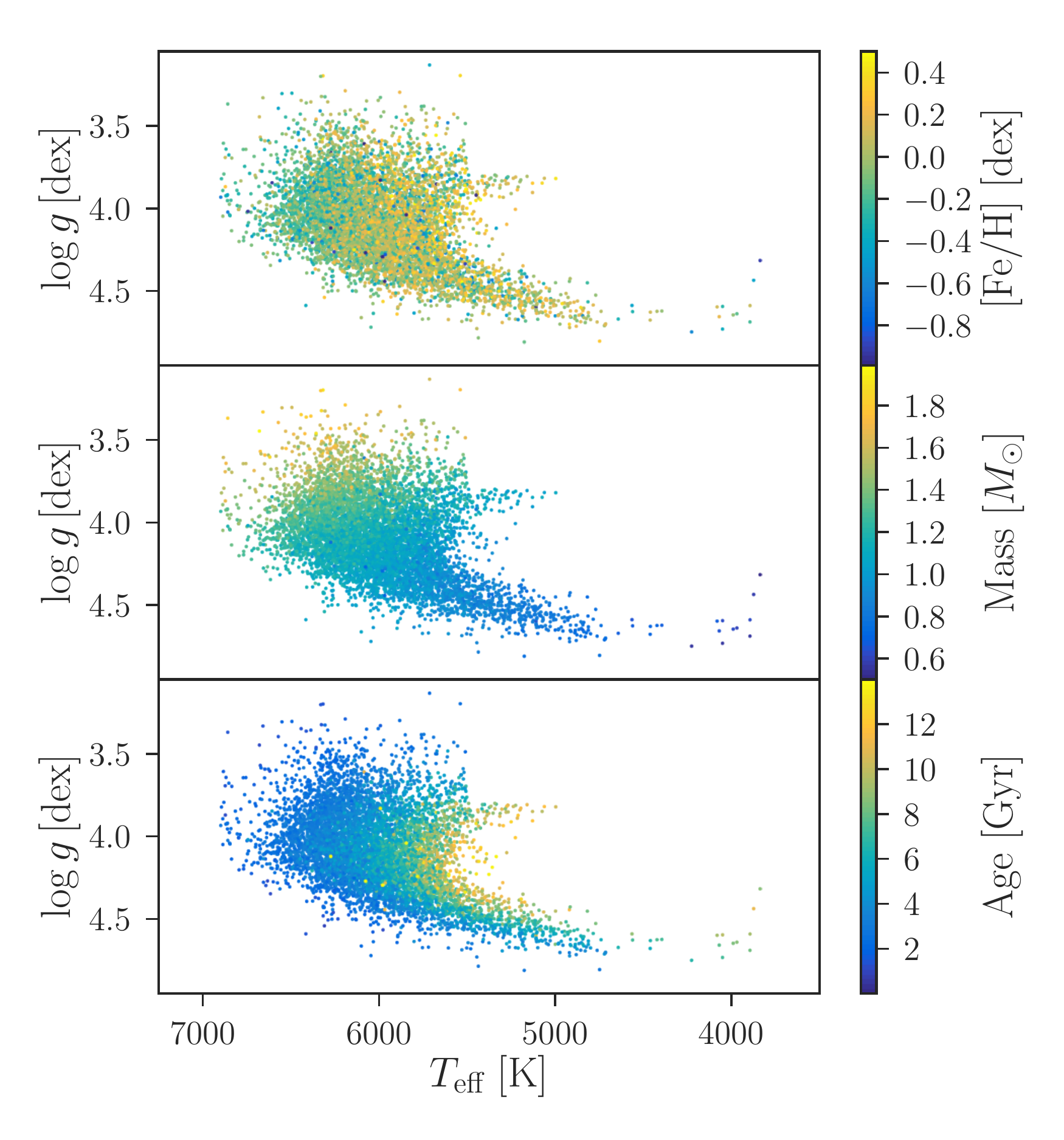}}
\caption{Kiel diagrams ($T_\text{eff}$ and $\log g$) of the \GnT~dwarfs. Colour indicates the metallicity [Fe/H] in the top panel, mass in the middle panel, and age in the bottom panel. The sample is a subset of the clean \GnT~overlap, shown in Fig.~\ref{plx_distribution_hrd} and excludes giants with $T_\text{eff} \geq 5500\,\mathrm{K} \text{~and~} \log g \geq 3.8\,\mathrm{dex}$. Stellar masses increase from the cool main sequence ($\sim 0.8\,\mathrm{\mathcal M_\odot}$) to the hottest turn-off stars ($\sim 2.0\,\mathrm{\mathcal M_\odot}$). Stellar ages decrease towards higher surface gravities on the cool main sequence and towards higher effective temperatures in the turn-off region. In contrast to this rather smooth trend, few metal-poor stars stand out with smaller stellar masses and higher stellar ages also at effective temperatures around $6000\,\mathrm{K}$.}
\label{3hrd}
\end{figure}

During each optimisation iteration, a suite of synthetic spectra based on perturbed parameters and corresponding partial derivatives in $\chi^2$-space are computed to facilitate convergence. The parameters of the synthesis with lowest $\chi^2$ are then either used as final parameters or as starting point of a new optimisation loop. For each synthesis, SME updates the line and continuous opacities and solves the equations of state and radiative transfer based on interpolated stellar model atmospheres \citep{Piskunov2017}. The optimisation has converged, when the fractional change in $\chi^2$ is below 0.001. Non-converged optimisations after maximum 20 iteration are discarded. Figure~\ref{3hrd} shows the final spectroscopic parameters $T_\text{eff}$ vs. $\log g$ of the final sample, colour coded by the fitted metallicities, masses, and ages from the \textsc{Elli}~code, see Sect.~\ref{age}.

We report the metallicity as [Fe/H] and base it on the iron scaling parameter of the best-fit model atmosphere (SME's internal parameter \textit{feh}). It is mainly estimated from iron lines and hence traces to the true iron abundance, as our validation with [Fe/H] from the GBS shows.

\subsection{Validation stars}

\begin{figure}[!ht]
\centering
\resizebox{\hsize}{!}{\includegraphics{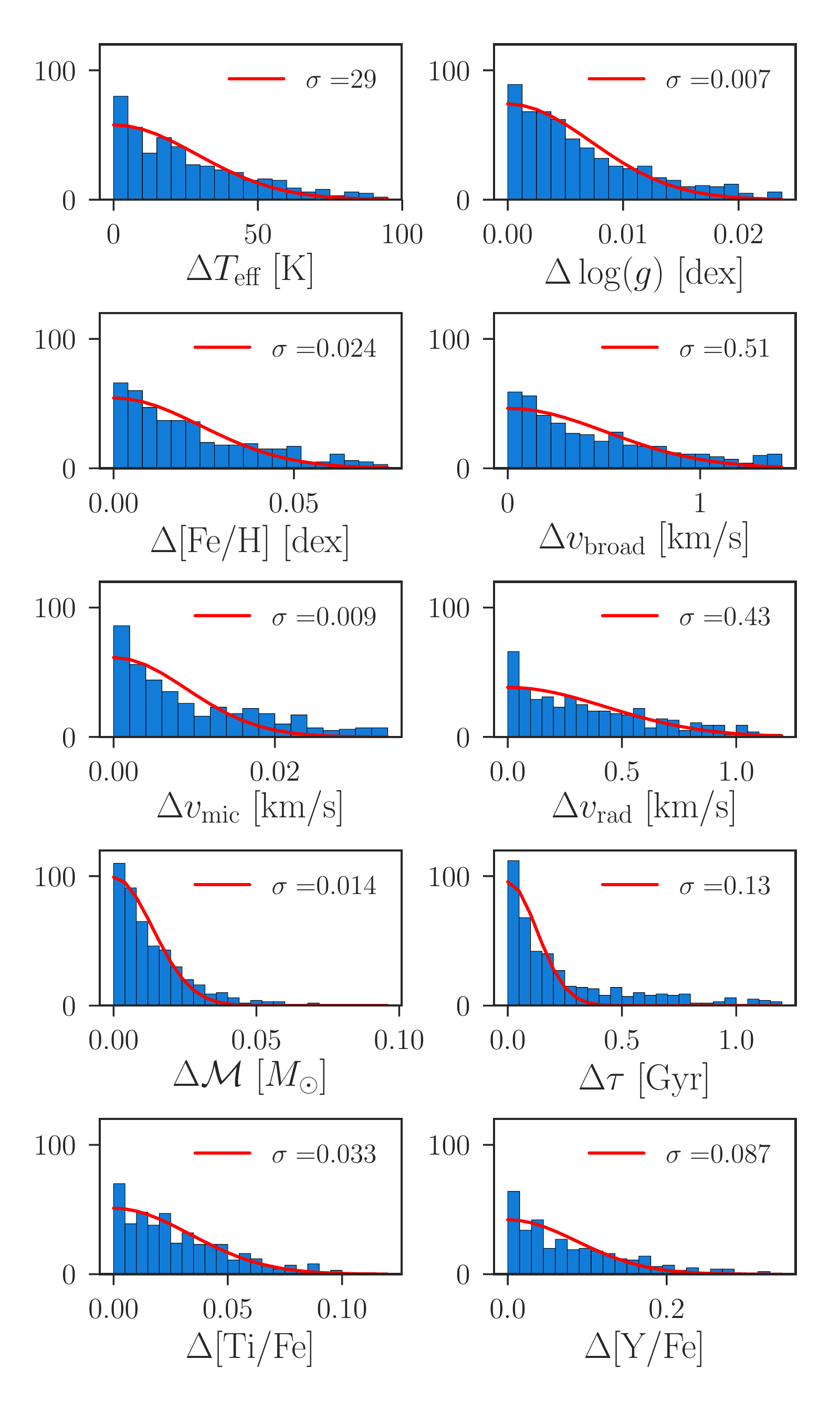}}
\caption{Histograms of parameter and abundance differences obtained from multiple observations of the same star. Shown are the absolute differences from two observations as well as from all three absolute difference combinations for three observations. A Gaussian distribution was fitted to the distributions (red curves). The obtained  standard deviation is indicated in each panel.}
\label{Multivisit_sigma}
\end{figure}

To estimate the precision, we can rely on stars with multiple observations as part of the GALAH+TGAS sample: 334 stars have been observed twice and 44 stars have been observed three times. The individual differences of selected parameters are plotted in Fig.~\ref{Multivisit_sigma}. Because we also use these multiple observations to assess the precision of the abundance estimates (see Sect.~\ref{abundance_determination}), we show the two element abundances Ti and Y as examples. We assume the uncertainties to be Gaussian and estimate the standard deviation of the multiple visits as a measure of precision. The resulting precisions based on the repeated observations are shown in Table~\ref{table:precision}.

\begin{table}[!h]
\caption{Precision and accuracy of the pipeline based on repeated observations and GBS respectively.}              
\label{table:precision}      
\centering                                      
\begin{tabular}{c c c}          
\hline\hline                        
Parameter X & $e_{X,\mathrm{Repeats}}$ &$e_{X,\mathrm{GBS}}$ \\    
\hline                                   
$T_\text{eff}$ & $29\,\mathrm{K}$ & $89\,\mathrm{K}$  \\
$\log g$ & $0.01\,\mathrm{dex}$ & $0.05\,\mathrm{dex}$ \\
$\mathrm{[Fe/H]}$ & $0.024\,\mathrm{dex}$ & $0.07\,\mathrm{dex}$ \\
$v_\text{broad}$ & $0.51\,\mathrm{km/s}$ & $2.0\,\mathrm{km/s}$ \\
$v_\text{mic}$ & $0.009\,\mathrm{km/s}$ & $0.20\,\mathrm{km/s}$ \\
$v_\text{rad}$ & $0.43\,\mathrm{km/s}$ \\
$\tau$ & $0.13\,\mathrm{Gyr}$ \\
$\mathcal M$ & $0.014\,\mathrm{\mathcal M_\odot}$ \\
$\mathrm{[Ti/Fe]}$ & $0.033\,\mathrm{dex}$ \\
$\mathrm{[Y/Fe]}$ & $0.081\,\mathrm{dex}$ \\
\hline                                             
\end{tabular}
\end{table}
 
\begin{figure}[!h]
\centering
\resizebox{\hsize}{!}{\includegraphics{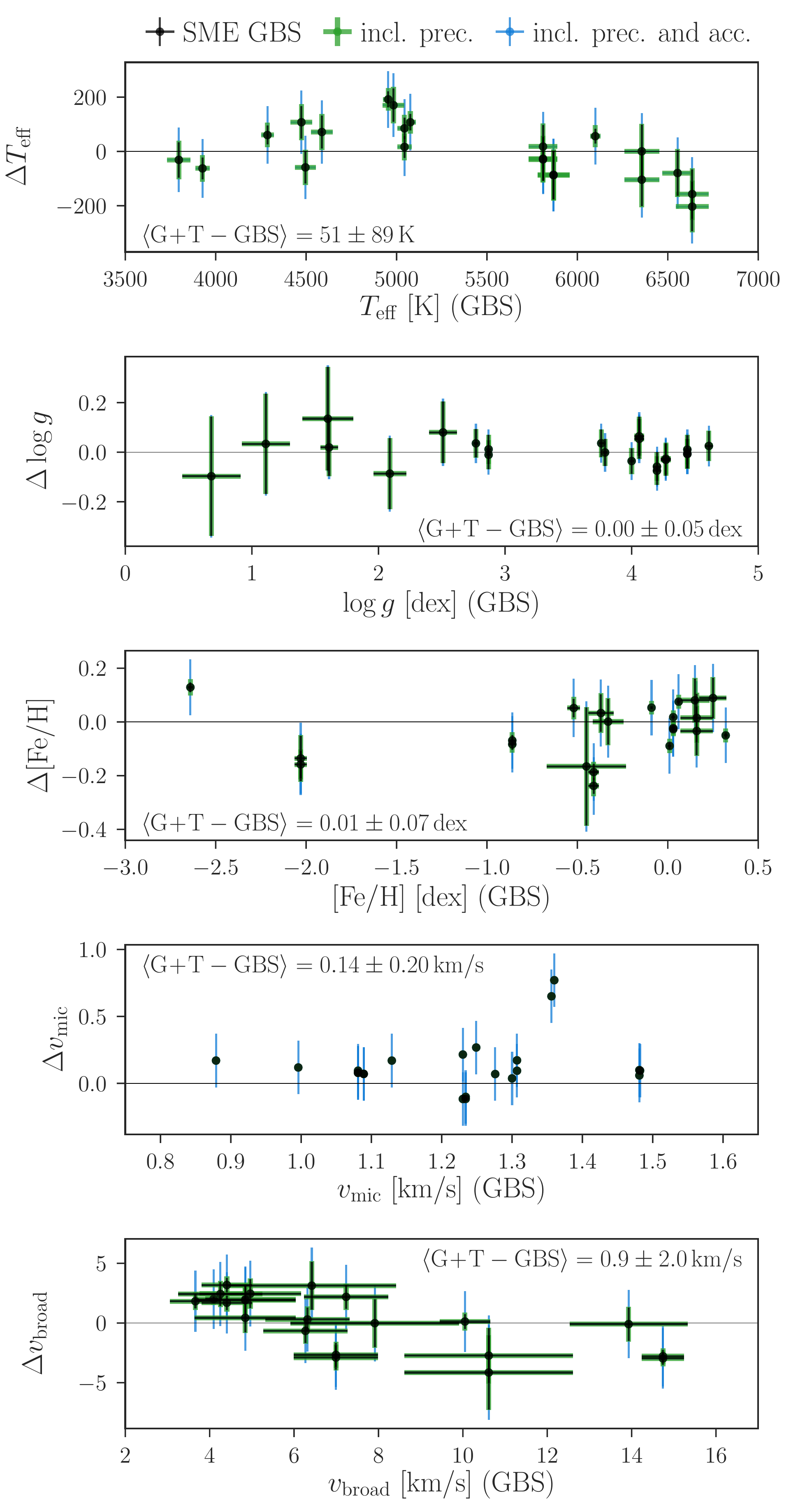}}
\caption{Comparison of the stellar parameters for GBS as estimated by this analysis and \citet{Heiter2015, Jofre2014} (shown as ours theirs versus ours). The fundamental parameters $T_\text{eff}$ and $\log g$ are shown in the two top panels, together with comparisons of metallicity with their recommended iron abundance [Fe/H], microturbulence velocity, and broadening velocity, a convolved parameter of macroturbulence and rotational velocity, in the three bottom panels. Black error bars are the combined uncertainties of GBS as well as the error output of our analysis pipeline (SME). Green error bars include precision uncertainties from repeated observations and blue error bars include both precision and accuracy estimates.}
\label{GBS_errors}
\end{figure}

To estimate the accuracy, we use the GBS. These are, however, typically much brighter and closer than the survey targets and brighter than the bright magnitude limit of \Gaia DR1 TGAS. Hence {\sc Hipparcos} parallaxes \citep{vanLeeuwen2007} are used additionally. New $K_S$ magnitudes are computed for GBS with 2MASS $K_S$ quality flag not equal to `A', following the approach used by \citet{Heiter2015}\footnote{For GBS with bad qualities, we convert $K_{BB}$ magnitudes from \citet{Gezari2000}, using equation (A1) by \citet{Carpenter2001}.}. With this approach 22 GBS observations are analysed and compared to the estimates from the GALAH+TGAS pipeline, as depicted in Fig.~\ref{GBS_errors}. To have a statistically sufficient sample, we also include GBS giants in the analysis. We find a small bias of $51 \pm 89 \,\mathrm{K}$  in comparison to the systematic uncertainties present in both GBS and our parameters. We note, however, temperature-dependent biases of $110 \pm 110 \,\mathrm{K}$ for some stars around $5000\,\mathrm{K}$. Towards higher temperatures, we also note an increasing disagreement, indicating that the temperatures of hotter stars are underestimated by our spectroscopic pipeline ($- 150 \pm 130\,\mathrm{K}$ at $6600\,\mathrm{K}$), a result likely to be caused by the application of 1D LTE atmospheres for hot stars \citep[see e.g.][]{Amarsi2018}, where Balmer lines are the strongest or only contributor for the parameter estimation. For surface gravity, $\log g$, and rotational/macroturbulence broadening, $v_\mathrm{broad}$, we find excellent agreement of $0.00 \pm 0.05\,\mathrm{dex}$ and $0.9  \pm  2.0 \,\mathrm{km/s}$ respectively. The latter is computed as quadratic sum of $v_{\sin i}$ and $v_\text{mac}$ for the GBS. For the metallicity, [Fe/H], we found a significant bias with respect to the GBS. Similar to previous studies of HERMES spectra \citep{Martell2017, Sharma2018} we therefore shift the metallicity by $+0.1\,\mathrm{dex}$ for our sample. The shift is chosen so that the overlap with GBS has consistent [Fe/H] in the solar regime. Two outliers for $v_\mathrm{mic}$ can be seen to drive the bias of $-0.14 \pm 0.20\,\mathrm{km/s}$, which we do not correct for, because the majority of the GBS sample agree well with our estimates and the two outliers are the most luminous giants, which are not representative of the final sample.

With these precision and accuracy estimates (the latter coming from the error-weighted standard deviation between GALAH and GBS estimates), we estimate the overall uncertainties of our parameters X (not mass and age, see Sect.~\ref{age}) by summing them in quadrature to the formal covariance errors of SME ($e_{X,\mathrm{SME}}^2 $):
\begin{equation} \label{overall_error}
\centering
e_{X,\mathrm{final}}^2 = e_{X,\mathrm{SME}}^2 + e_{X,\mathrm{Repeats}}^2 + e_{X,\mathrm{GBS}}^2
\end{equation}

For element abundances, we estimate the overall uncertainties without the GBS term. In the case of $\log g$, we replace $e_{\log g,\mathrm{SME}}^2$ by the standard deviation of 10,000 Monte Carlo samples of Eq. \ref{logg}. For this sampling, we use the uncertainties of $e_{T_\text{eff},\mathrm{final}}$, the maximum likelihood masses as $\mathcal M$ with an error of $6\%$ (based on mean mass uncertainties of an initial \textsc{Elli}~run), $e_{K_S}$ from 2MASS with mean uncertainties of $0.02\,\mathrm{mag}$, and propagate this information to adjust $BC$ (with typical changes below 0.07). Because \citet{Astraatmadja2016} only state the three quantiles, we sample two Gaussians with standard deviations estimated from the 5th and 95th distance percentile respectively. Because there are no Bayesian distance estimates for {\sc Hipparcos}, we choose to sample parallaxes $\varpi$ rather than distances $D_\varpi$. For $e_{A_K}$ we use the quadratically propagated uncertainties from the RJCE method (with mean uncertainties of $0.03\,\mathrm{mag}$) or assume $0.05\,\mathrm{mag}$ for estimates based on $E(B-V)$. We do not use Eq.~\ref{overall_error} for age and mass, because they are estimated with the adjusted stellar parameters.

\subsection{Mass and age determination}\label{age}

\begin{figure*}[!h]
\centering
\includegraphics[width=17cm]{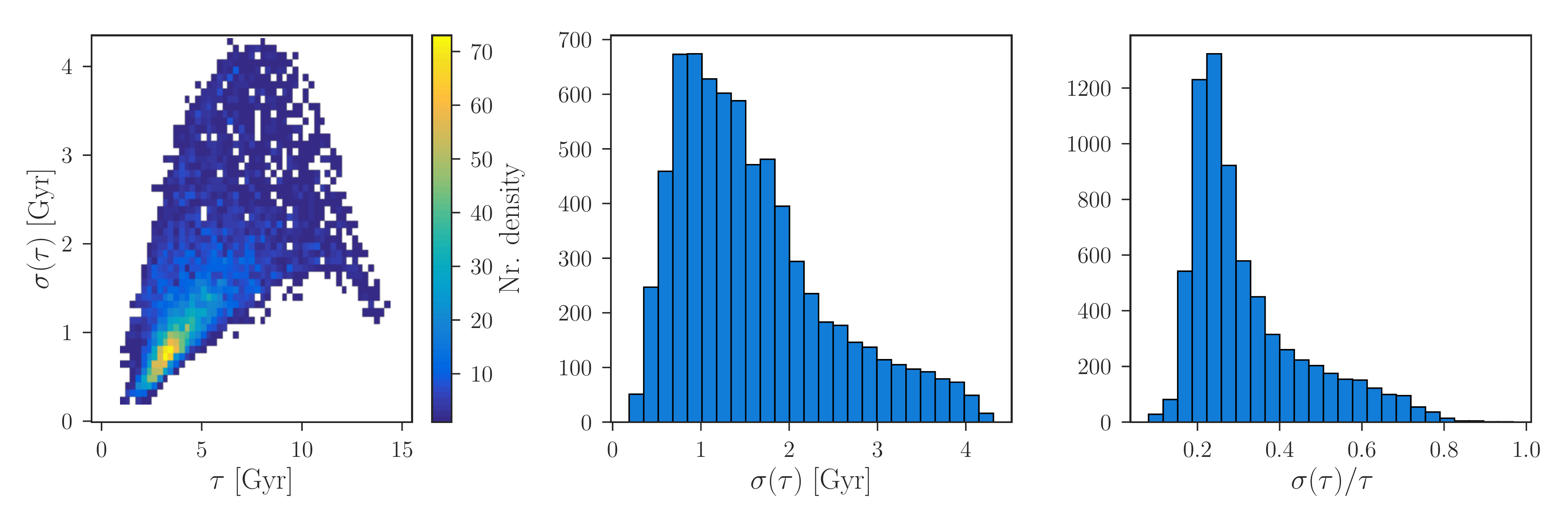}
\caption{Distributions of stellar ages $\tau$ [Gyr] and their uncertainties. The left panel shows the distribution of uncertainties versus ages, the middle panel show the absolute age uncertainties and the right panel shows the relative age uncertainties. The majority of age estimates show uncertainties below $2\,\mathrm{Gyr}$ and relative uncertainties below $30\%$.}\label{age_errors}
\end{figure*}

For the mass and age determination, we use the \textsc{Elli}~code \citep{Lin2018}, employing a Bayesian implementation of fitting Dartmouth isochrones based on $T_\text{eff}$, $\log g$, [Fe/H], and absolute magnitude $M_K$. $M_K$ is based on 2MASS $K_S$, the distance estimates from \citet{Astraatmadja2016} and accounts for extinction $A_K$ (estimated as described in Sect.~\ref{pipeline}). The Dartmouth isochrones span ages from 0.25 to $15\,\mathrm{Gyr}$ and metallicities from $-2.48$ to $+0.56$ with $\upalpha$-enhancement analogous to the {\sc marcs} atmosphere models\footref{footnote_alphaenhancement}. Starting with a maximum likelihood mass and age estimation, MCMC samplers as part of the \textsc{emcee} package \citep{ForemanMackey2013} are used to estimate masses and ages. Stellar ages and their uncertainties are estimated by computing the mean value and standard deviation of the posterior distribution. The stellar ages estimated with the \textsc{Elli}~code have typical uncertainties of $1.6\,\mathrm{Gyr}$ (median of posterior standard deviations), which typically correspond to less than $30\%$, see Fig.~\ref{age_errors}. As pointed out for example by \citet{Feuillet2016}, the posterior distribution does not necessary follow a Gaussian. Although this is the case for the large majority of our stars, we also provide the 5th, 16th, 50th, 84th, and 95th percentiles to the community for follow-up studies. Because the results of this study did not change significantly with quality cuts for stellar ages, we do not apply them.

\subsection{Binarity}\label{binaries}

The observational setup of the GALAH survey allocates one visit per observation (with exception of pilot and validation stars). Therefore, binaries or triples can usually not be identified via radial velocity changes. 

\begin{figure}[!ht]
\centering
\resizebox{\hsize}{!}{\includegraphics{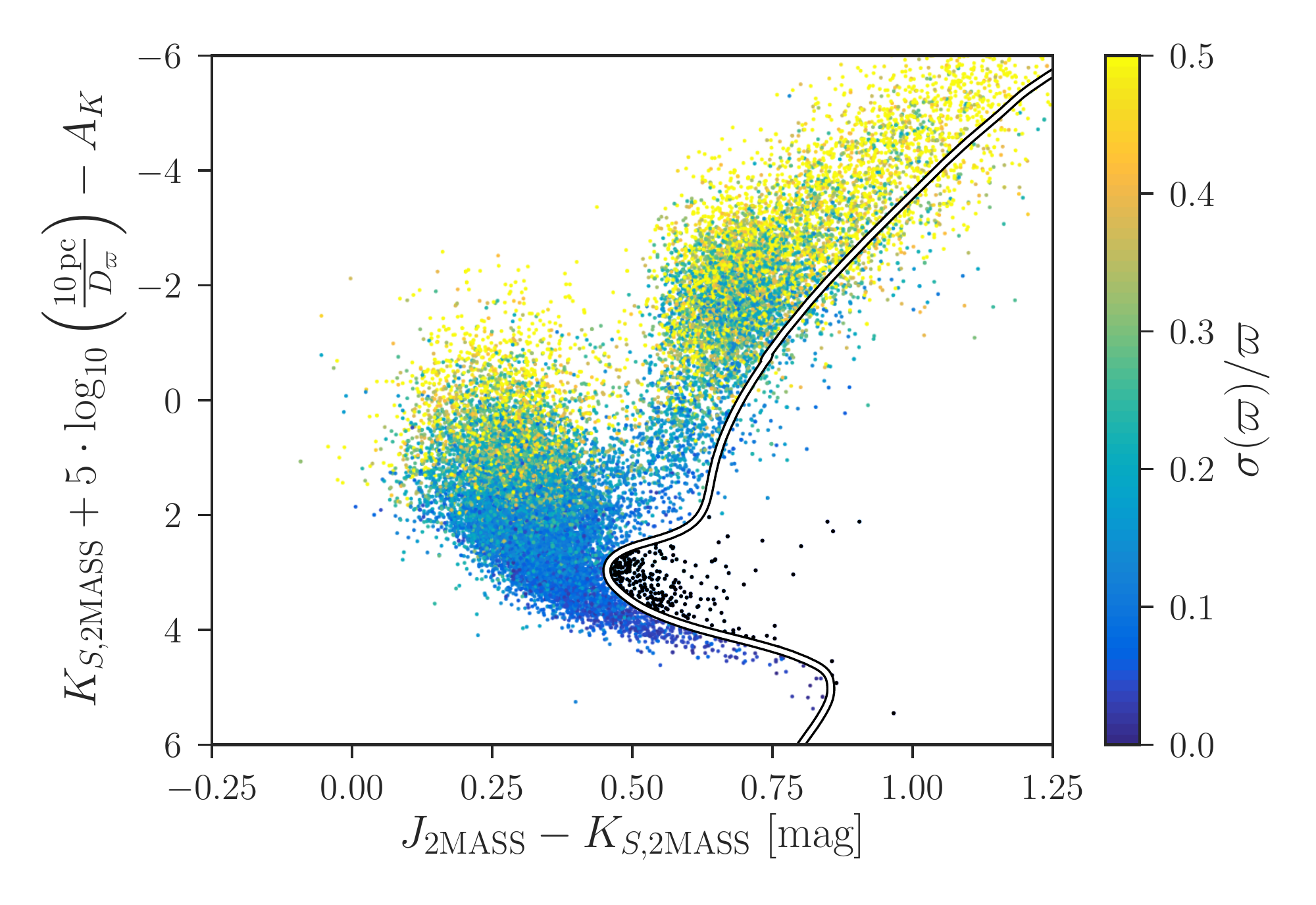}}
\caption{Colour magnitude diagram of the full \GnT~sample coloured by the parallax precision. The colour index is $J-K_S$ from 2MASS photometry and absolute magnitude for $K_S$, inferred from 2MASS as well as distances $D_\varpi$ and extinction $A_K$. A Dartmouth isochrone with age ($15\,\mathrm{Gyr}$) and metallicity ($+0.56$) is shown as white curve. This is used to identify 338 dwarfs with photometry outside of the expected range (above the white curve) for cool single main sequence stars ($M_{K_S} > 2\,\mathrm{mag}$), here shown in black. The identified stars are all nearby and their reddening is negligible, especially in the infrared. We note that for some stars, possibly mis-identified as binaries, the photometry indicates colour excesses or a pre-main-sequence stage, which is still an important reason to eliminate them from the subsequent analysis, as the pipeline is not adjusted for these stars.}\label{cmd}
\end{figure}

Here, we use both the tSNE classifications by \citet{Traven2017}, to identify obvious binaries, as well as visual inspection to identify double-line binaries which are less distinct from the tSNE classification. Within the sample, a binary fraction of $4\%$ has been identified with high confidence from spectral peculiarities. Additionally, 338 probable photometric binaries on the main sequence are identified which show a significant deviation between spectroscopically determined $\log g$ or $L_\mathrm{bol}$ with respect to photometrically determined ones. For these, the suspected secondary contributes significantly to the luminosity of the system without obvious features within the GALAH spectra. These stars lie above the main sequence within a colour-(absolute) magnitude diagram. We have identified the stars with photometric quantities beyond what is expected for a single star on the main sequence (shown as black dots in Fig.~\ref{cmd}) by using a Dartmouth isochrone with the highest age ($15\,\mathrm{Gyr}$) and metallicity ($+0.56\,\mathrm{dex}$). We note that some of these stars show colour excesses. While these might have been mis-identified as binaries, they are definitely peculiar objects (e.g. pre-main-sequence stars), for which the pipeline is not adjusted and have subsequently been neglected. We want to stress again, that identified binaries have been excluded from the cleaned sample.

\subsection{Abundance determination}\label{abundance_determination}

With the stellar parameters estimated in Section \ref{pipeline}, elemental abundances were calculated in the following way:

\begin{enumerate}
	\item Predefined segments of the spectrum are normalised and the element lines chosen with two criteria.  First, the lines have to have a certain depth, i.e., their absorption has to be significant. We use the internal SME parameter depth to assess this, see \citet{Piskunov2017}.
	\item The lines have to be unblended. This is tested by computing a synthetic spectrum of the segment with all lines and one only with the lines of the specific element. The $\chi^2$ difference between the synthetic spectra for each point in the line mask has to be lower than 0.0005 or 0.01 (the latter for blended but indispensable lines), otherwise the specific point is neglected for the final abundance estimation.
	\item The abundance for the measured element is optimised using up to 20 loops with the unblended line masks.
\end{enumerate}

\begin{figure*}[!ht]
\centering
\includegraphics[width=17cm]{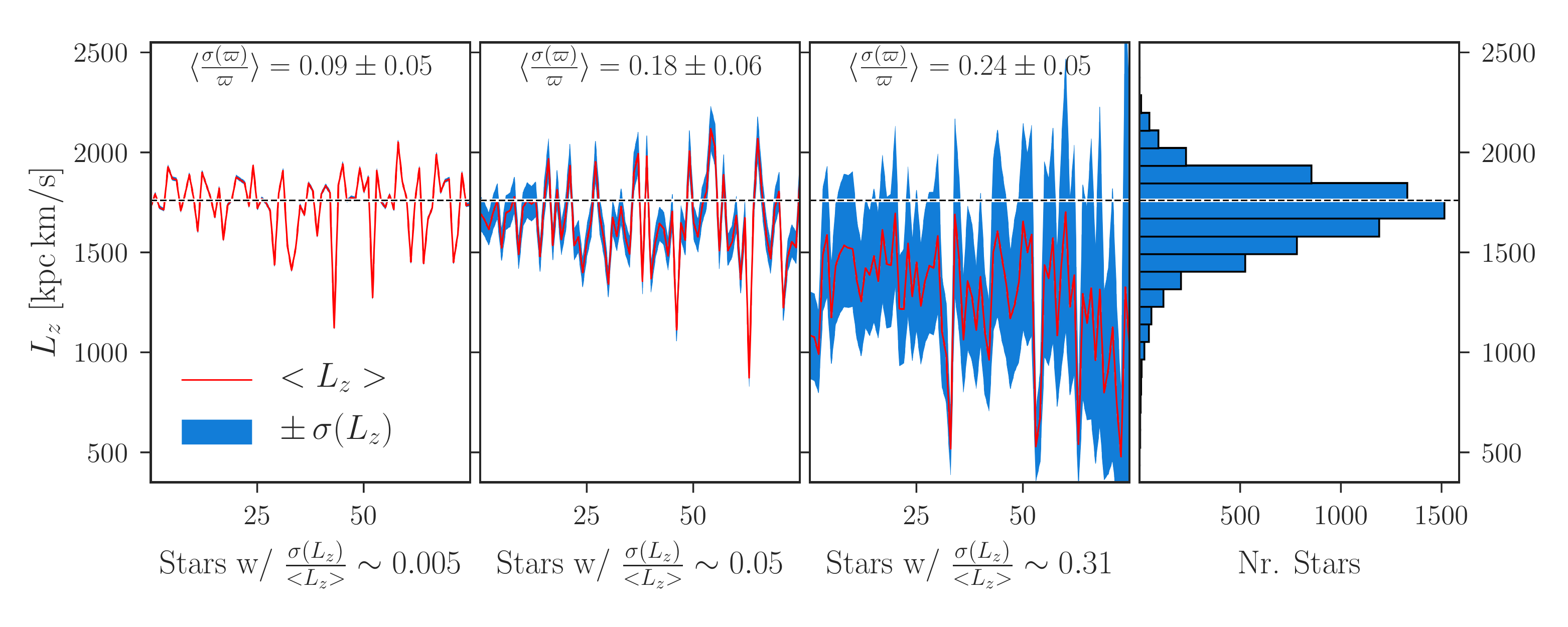}
\caption{The left three panels show the distribution of angular momentum $L_z$, sorted by relative uncertainty and depict three close in views groups of with 75 stars with mean $L_z$ uncertainties of first $0.5\%$, second $5\%$, and third $30\%$ in order to demonstrate the precision reached for different parallax qualities. Red is mean $L_z$ for each star and blue their $1\sigma$ area. A white dashed line indicated the Solar angular momentum. Average parallax precisions are indicated in the top of each panel. The best precision on parallaxes also lead to the most precise $L_z$. The values of the least precise momenta (third panel) are significantly lower than those of the Sun, even when taking the standard deviation of the angular momentum estimates into account. Due to the selection of stars and the density structure of the disk, these stars are statistically further away and are expected to be at larger Galactic heights and closer to the Galactic centre. We note that their angular momenta are also different from the majority of stars, which have a Sun-like angular momenta, as shown in the fourth panel. For a discussion of the angular momenta of the stars in the chemodynamical context see Sect.~\ref{discussion}.}\label{Lz_quality}
\end{figure*}

The selection of lines used for parameter and abundance analysis and their atomic data is a continuation of the work presented by \citet{Heiter2015b} and Heiter et al. (submitted). The complete linelist will be presented in \citet{Buder2018b}.

Abundances are estimated assuming LTE, with the exception of Li, O, Al, and Fe, for which we use corrections by \citet{Lind2009}, \citet{Amarsi2016b}, \citet{Nordlander2017}, and \citet{Amarsi2016a}, respectively, to estimate non-LTE abundances.

Solar abundances are estimated based on a twilight flat in order to estimate the difference to the solar composition by \citet[][G07]{Grevesse2007}. This difference for each element $X$, i.e. $A(X)_\odot - A(X)_\odot^\text{G07}$, is then subtracted from the element abundance of the stars of the sample.

\subsection{Kinematic parameters}\label{analysis_kinematics}

For our target stars, the space velocities $U$, $V$, and $W$ were calculated using the \textsc{galpy} code by \citet{Bovy2015}, assuming $(U_\odot, V_\odot, W_\odot) = (9.58,10.52,7.01)\,\mathrm{km/s}$ \citep{Tian2015} relative to the local standard of rest.

We estimate kinematic probabilities of our sample stars to belong to the thin disk ($D$), thick disk ($TD$), and halo ($H$) following the approach by \citet{Bensby2014} (see their Appendix A) with adjusted solar velocities.

To estimate the Galactocentric coordinates and velocities as well as the action-angle coordinates of the sample, we use \textsc{galpy}. We choose the axisymmetric \textit{MWPotential2014} potential with a focal length of $\delta=0.45$ for the confocal coordinate system and the \textsc{galpy} length and velocity units $8\,\mathrm{kpc}$ and $220\,\mathrm{km/s}$ respectively. We place the Sun at a Galactic radius of $8\,\mathrm{kpc}$ and $25\,\mathrm{pc}$ above the Galactic plane. To speed up computations, we use the \textsc{actionAngleStaeckel} method. We estimate mean values and standard deviations of the action-angles per star from 1000 Monte Carlo samples of the 6D kinematical space randomly drawn within the uncertainties. We neglected the uncertainties of the 2D positions and estimated the standard deviation of the distances from the 5th and 95th percentiles given by \citet{Astraatmadja2016}.

As shown in Fig.~\ref{Lz_quality}, the distance uncertainties are the dominant source of the action uncertainties. While for excellent parallaxes (left panel), the scatter in the action estimates is negligible, it becomes noticeable for parallaxes with uncertainties around $18\%$. For parallax uncertainties above $24\%$, the action uncertainties increase to as high as $31\%$. From the samples depicted in Fig.~\ref{Lz_quality}, one can see that these large uncertainties are particularly common for stars with low angular momentum. Because of the GALAH selection (observing in the Southern hemisphere and leaving out the Galactic plane) as well as the density structure of the disk with more stars towards the Galactic centre, we expect stars with larger distances (and hence larger distance uncertainties) to be situated at larger Galactic heights and smaller Galactic radii than the Sun. The right panels in Fig.~\ref{Lz_quality} confirm this expectation. Stars with angular momenta comparable with the solar value have usually precisely estimated actions. The latter stars are also the majority of stars in the sample, as the histogram in the right panel shows.


\section{Results}\label{results}

In Sect.~\ref{age_distribution}, we describe the stellar age distribution, before presenting abundance and age trends in Sect.~\ref{alpha_feh} and the kinematics of the sample in Sect.~\ref{kinematics}.

\begin{figure}[!h]
\centering
\resizebox{\hsize}{!}{\includegraphics{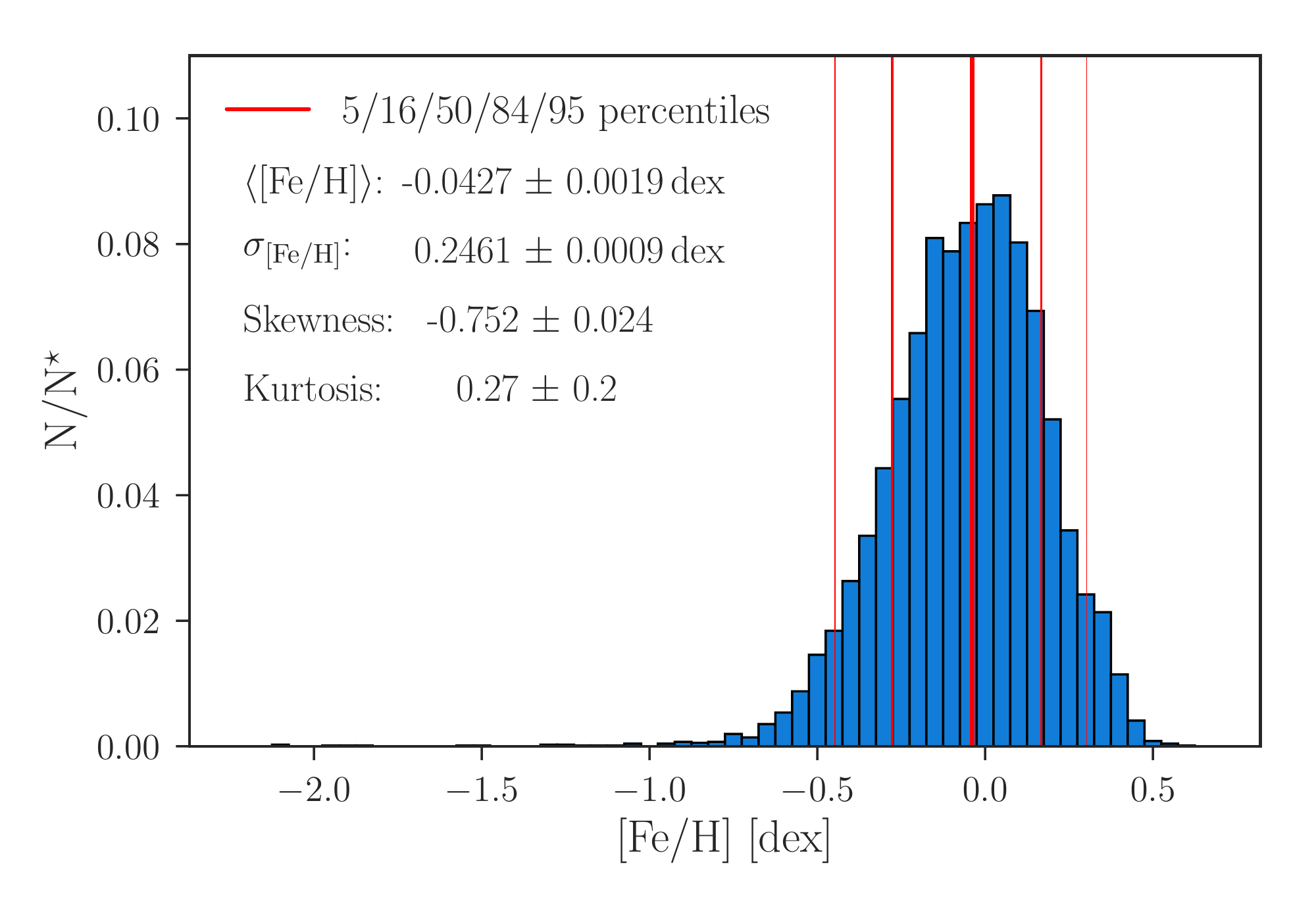}}
\caption{Metallicity distribution function of the GALAH+TGAS sample. The majority of the stars have solar-like metallicity, [Fe/H], within $\pm0.5$. The distribution is skewed towards metal-poor stars between $-2.0\,\mathrm{dex}$ and $-0.5\,\mathrm{dex}$. The 5, 16, 50, 84, and 95 percentiles are $-0.45\,\mathrm{dex}$, $-0.28\,\mathrm{dex}$, $-0.04\,\mathrm{dex}$, $0.17\,\mathrm{dex}$, and $0.30\,\mathrm{dex}$. respectively. Mean metallicity and standard deviation as well as skewness and kurtosis are indicated in the plot and discussed in the text.}
\label{mdf}
\end{figure}

We note that the vast majority of the dwarfs from the GALAH+TGAS sample are more metal-rich than $-0.5\,\mathrm{dex}$, as seen in the metallicity distribution function in Fig.~\ref{mdf}. These stars have no intrinsic selection bias in metallicity or kinematics. 

From 10,000 Monte Carlo samples, we find the parameters of the metallicity distribution to be $\langle\mathrm{[Fe/H]}\rangle = -0.04$, $\sigma_\mathrm{[Fe/H]} = 0.26$, $\mathrm{skewness} = -0.667\pm0.029$, kurtosis\footnote{Here we follow \citet{Hayden2015} and define kurtosis as the fourth standardised moment-3.}$ = -0.21\pm0.23$. The mean of our metallicity distribution is slightly lower but consistent within the uncertainties to the one estimated by \citet{Hayden2015} using APOGEE data for the same (solar) Galactic zone\footnote{We refer to the Solar Galactic zone ($7<R<9\,\mathrm{kpc}$ and $\vert z \vert < 0.5\,\mathrm{kpc}$), which contains $99.5\%$ of the \GnT~sample.}. The APOGEE distribution also shows a narrower standard deviation ($0.2\,\mathrm{dex}$) around a mean value of $+0.01\,\mathrm{dex}$ and is less skewed ($-0.53\pm0.04$) but more extended towards the metal-rich and metal-poor tail of the distribution (with a kurtosis of $0.86\pm0.26$). The kurtosis, a measure for the sharpness of the peak, indicates that the APOGEE distribution has a sharper peak than the GALAH distribution. The skewness indicates that the GALAH sample contains in general also relatively more metal-poor stars compared to the APOGEE sample. This is possibly caused by the different selection functions of the two surveys. GALAH avoids the Galactic plane ($\vert b \vert \leq 10\,\mathrm{deg}$), whereas APOGEE targets the plane where we expect relatively more stars of the low-$\upalpha$-sequence that are more metal-rich than $\mathrm{[Fe/H]} = -0.7$.

\subsection{The age distribution} \label{age_distribution}

\begin{figure}[!ht]
\centering
\resizebox{\hsize}{!}{\includegraphics{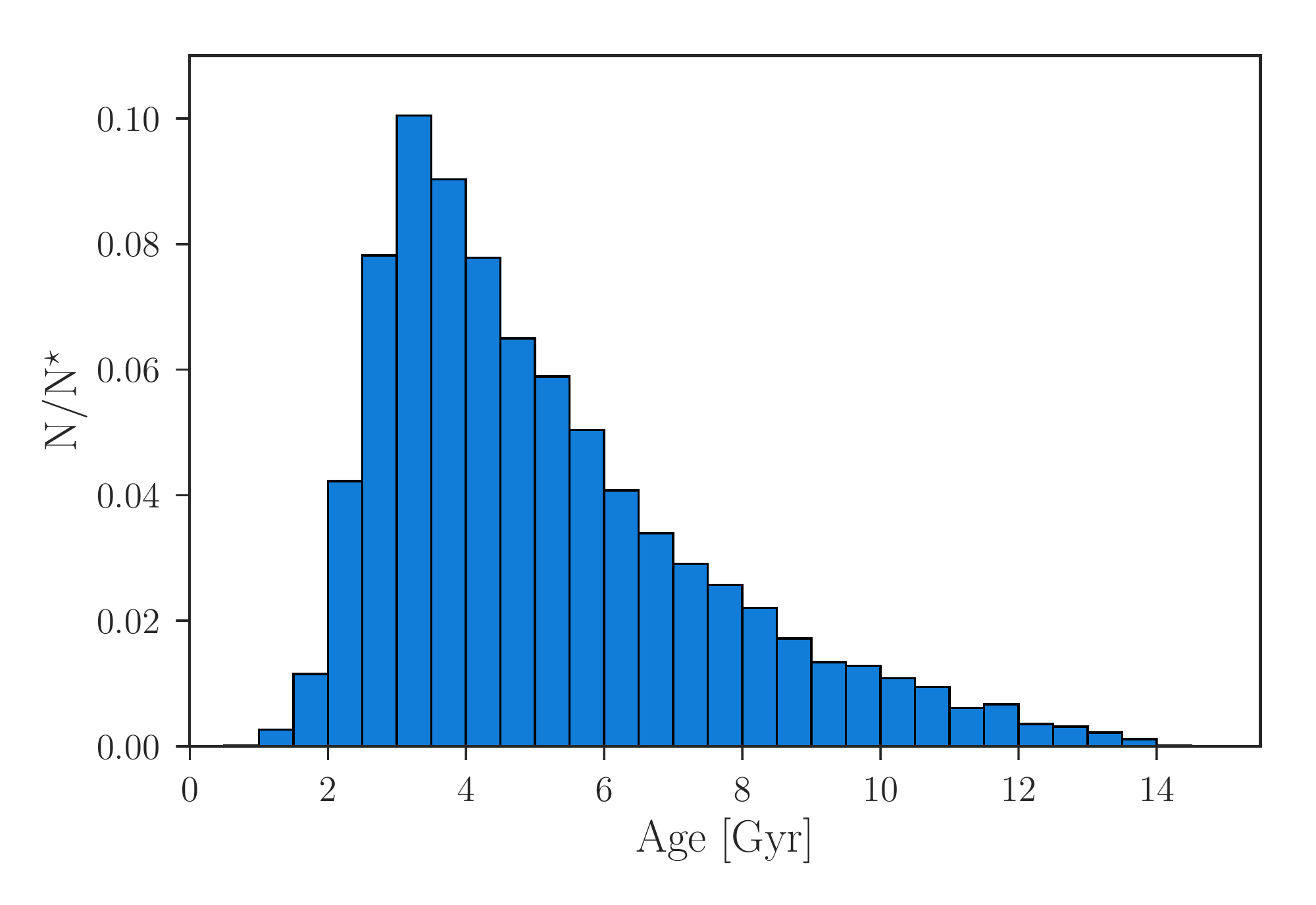}}
\caption{Distribution of stellar ages. The distribution peaks around $3\,\mathrm{Gyr}$ and decreases towards higher ages. We stress that the exclusion of stars with effective temperatures above $6900\,\mathrm{K}$ leads to fewer stars in the clean sample, with ages below $2\,\mathrm{Gyr}$. The peak of the distribution is however not affected by this selection.}\label{age_histogram}
\end{figure}

The age distribution of the \GnT~sample is shown in Fig.~\ref{age_histogram}. It peaks between $3$ to $3.5\,\mathrm{Gyr}$, which is at an older age than estimated by the studies of \citet{Casagrande2011} and \citet{SilvaAguirre2018} who both placed the peak at approximately $2\,\mathrm{Gyr}$. While this might be partially explained by a combination of both selection function, and target selection effects, we note that the exclusion of hot stars with effective temperatures above $6900\,\mathrm{K}$ in our sample, see Sect.~\ref{observation}, affects primarily stars with ages below the peak of the histogram. However, these hot stars have an average maximum likelihood age of $1.5 \pm 0.8\,\mathrm{Gyr}$ and the location of the age peak does not change when including them.

\subsection{The age-[$\upalpha$/Fe]-[Fe/H] distributions}\label{alpha_feh}

\begin{figure*}[!ht]
\centering
\includegraphics[width=17cm]{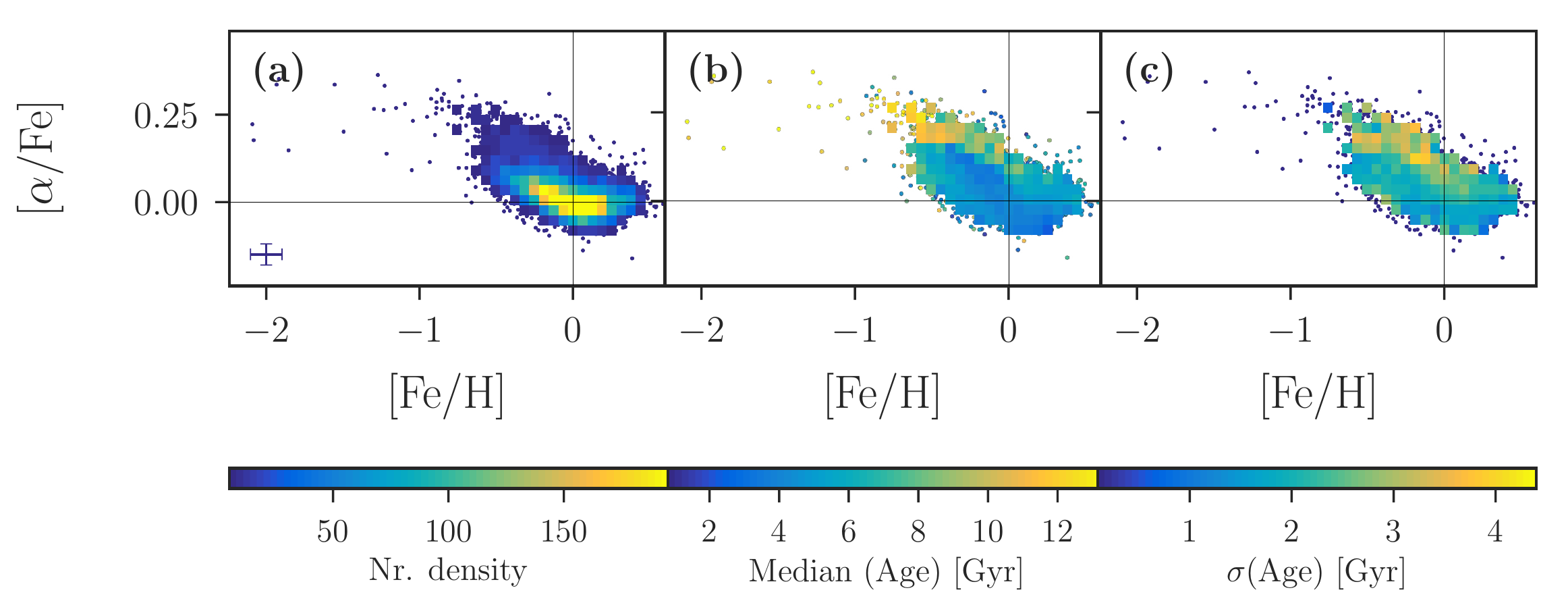} \newline
\includegraphics[width=17cm]{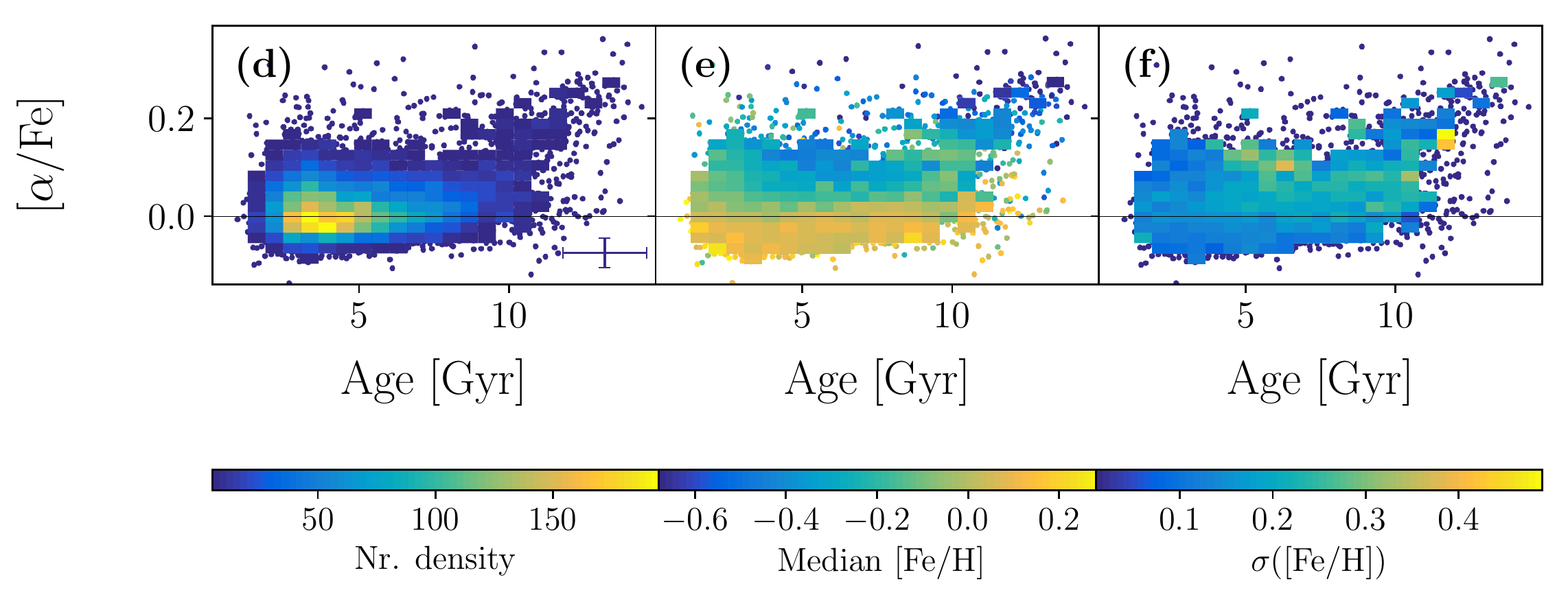} \newline
\includegraphics[width=17cm]{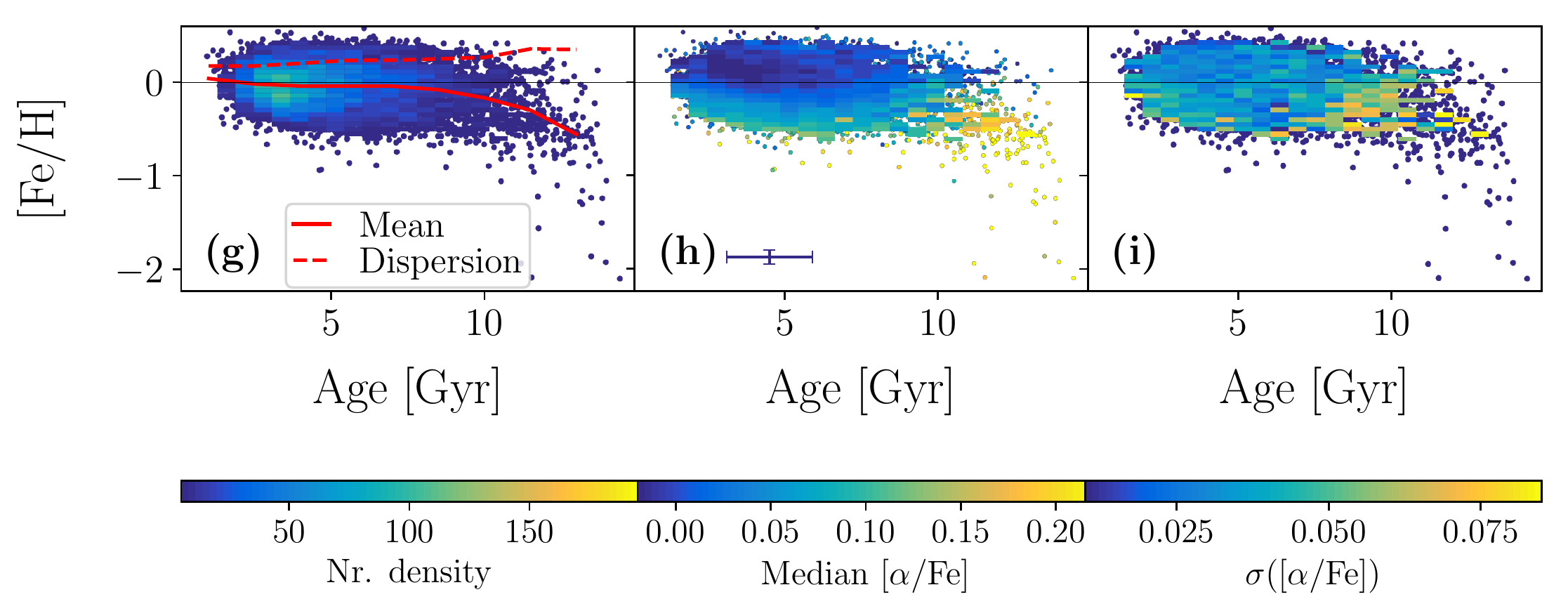}
\caption{Diagrams of the age-[Fe/H]-[$\upalpha$/Fe] distribution in three rotating visualisations (top to bottom). (a) to (c) show [$\upalpha$/Fe] both as a function of [Fe/H]. (d) to (f) and (g) to (i) show [$\upalpha$/Fe] and [Fe/H] as a function of age, respectively. We show the density distributions in the left panels (a), (d), and (g). The same distributions are shown with bins coloured by the median age, [Fe/H], and [$\upalpha$/Fe] in the middle panels (b), (e), and (h), respectively. The right panels show the same distributions coloured by the standard deviation of age, [Fe/H], and [$\upalpha$/Fe] in the middle panels (c), (f), and (i), respectively. Dots are used for individual stars in sparse regimes instead of density bins. In (g), we also show the mean metallicity (red line) and dispersion (red dashed line) as a function stellar age. The mean is decreasing with age from $0.04$ to $-0.56\,\mathrm{dex}$, while the dispersion is increasing with stellar age from $0.17$ to $0.35\,\mathrm{dex}$. See text in Sec. \ref{alpha_feh} for detailed discussion.}
\label{feh_alpha_age}
\end{figure*}

We have detected abundances for up to 20 elements, which are presented in Sect.~\ref{abundancetrends}. For an extended overview of abundance trends for the elements detectable across the whole GALAH range, we refer the reader to \citet{Buder2018b}. For this study, we focus on the $\upalpha$-elements and iron as well as their correlations with stellar age. The combination of these three parameters is shown in Fig.~\ref{feh_alpha_age}. 

The abundance patterns of $\upalpha$-elements in the Galactic disks are expected to follow roughly a similar pattern according to the stellar enrichment history by supernovae type Ia and II \citep[see e.g.][]{Gilmore1989}. While both types of supernovae produce a variety of elements, there is a significant difference in the yields of iron and $\upalpha$-elements and the time in the Galactic evolution, when they each contribute to the chemical enrichment. Early in the chemical evolution of the Galaxy, SN Type II dominate the production of metals and large quantities of $\upalpha$-elements are then produced \citep[e.g.][]{Nomoto2013}. The timescales for SN Ia are larger than those of SN Type II, with estimated intermediate delay times of around $0.42$ to $2.4\,\mathrm{Gyr}$ \citep{Maoz2012}. After this delay time, SN Ia fed material into their environment - but with a larger yield ratio of iron to $\upalpha$-chain elements, therefore decreasing the abundance ratio [$\upalpha$/Fe] while increasing [Fe/H] \citep[e.g.][]{Matteucci1989, Seitenzahl2013}.

\begin{figure*}[!h]
\centering
\includegraphics[width=17cm]{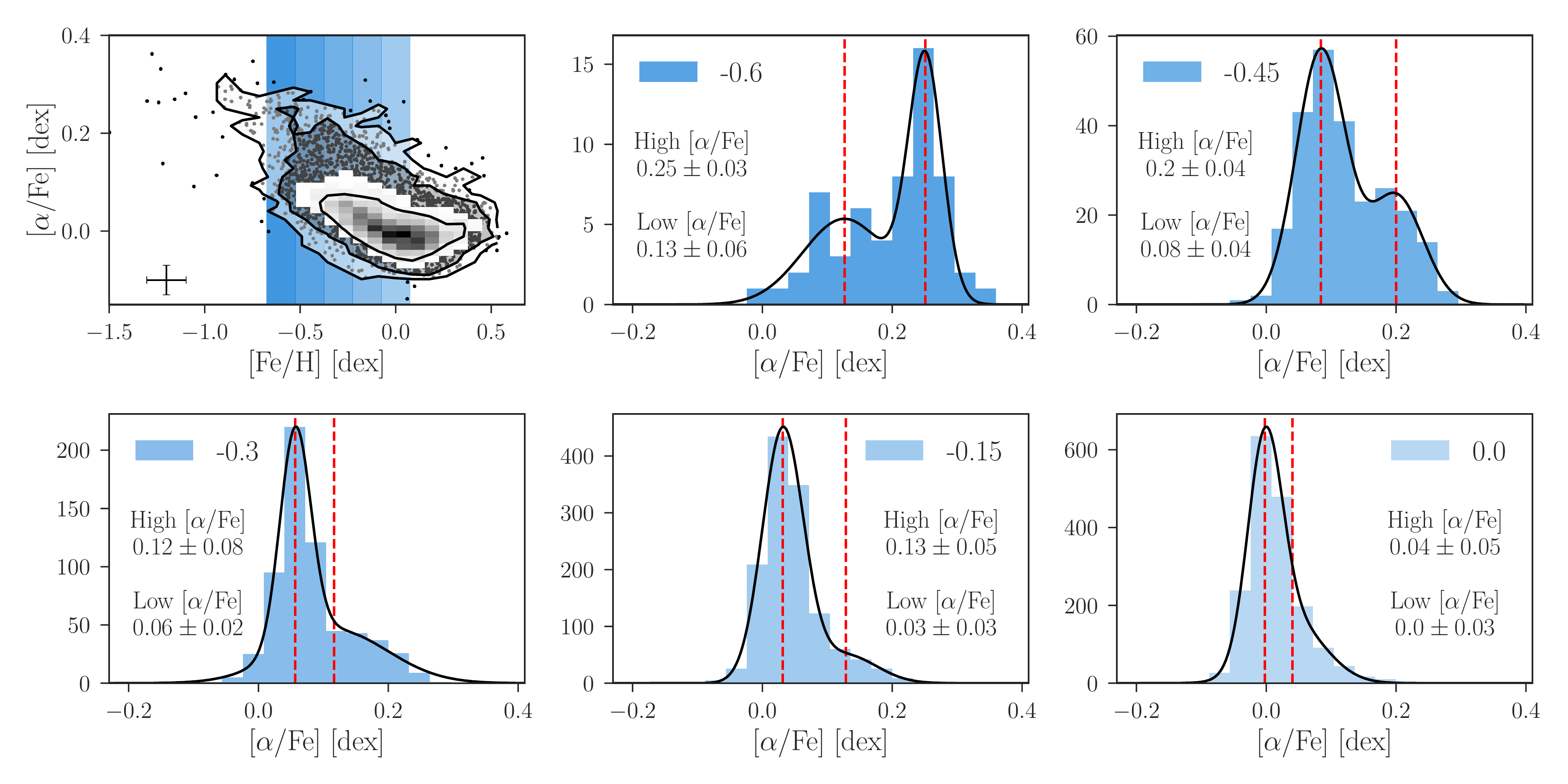}
\caption{$\upalpha$-enhancement [$\upalpha$/Fe] over metallicity [Fe/H] for all metallicities (upper left panel). The blue bars indicate the metallicity bins, which were used to select stars to estimate the $\upalpha$-enhancement distribution in corresponding blue at different metallicities in the other panels, with mean [Fe/H] indicated in the upper left or right corner. Two Gaussian distributions were fitted to the data with mean values indicated by the red lines and their distribution in black. Mean and standard deviation of the two Gaussians are annotated in each panel. See Sect.~\ref{alpha_feh} for detailed discussion.}\label{alpha_histograms}
\end{figure*}

The combined $\upalpha$-element abundance is estimated for 99\% of our stars. For each of these stars, at least one $\upalpha$-process element has been detected  and all significant measurements have been combined with their respective uncertainties as weight. Mg, Si, and Ti are the most precisely measured elements and have the highest weight. Hence, we note that the [$\upalpha$/Fe]-ratio, as defined here, is in practice very similar to the previously used error-weighted combination of Mg, Si, and Ti for GALAH DR1 \citep{Martell2017} and for the study by \citet{Duong2018}.
We see overall good agreement in the [$\upalpha$/Fe] pattern with the stars in the solar vicinity analysed by the APOGEE survey \citep{Hayden2015}, i.e. predominantly solar ratios for $-0.7$ and $+0.5\,\mathrm{dex}$ and fewer stars with increasing $\upalpha$-enhancement towards lower metallicity. We discuss this bimodality further in Sect.~\ref{alpha_feh} by inspecting the quantitative distribution of [$\upalpha$/Fe] in several metallicity bins, see Fig.~\ref{alpha_histograms}.
\importance{We stress that there is no unambiguous or universal definition of $\upalpha$-enhancement, but studies estimate and define this parameter differently, which complicates comparison. Hereinafter we use an average, weighted with the inverse of the errors, of the four $\upalpha$-process elements (Mg, Si, Ca, and Ti) when we refer to [$\upalpha$/Fe] and $\upalpha$-enhancement.}
We note that because our definition of $\upalpha$-enhancement is driven by Ti as the most precisely determined element, our values are comparable with the study by \citet{Bensby2014} based on Ti. \citet{Fuhrmann2011} used only Mg as tracer of the $\upalpha$-process ratio. All different definitions hence induce possible systematic trends.

\paragraph{The [$\upalpha$/Fe]-[Fe/H] distribution} is shown in Fig.~\ref{feh_alpha_age} (a) to (c). 

The pattern of our study is agrees very strongly with the results found by \citet{Ness2016} for APOGEE (see their Fig.~8) and \citet{Ho2017} for LAMOST (see their Fig.~5), all three showing high age for the high-$\upalpha$ sequence and younger ages for stars on the low-$\upalpha$ sequence. We note that stars with larger ages usually have larger absolute age uncertainties. In contrast to the expected rather monotonic trend between $\upalpha$-enhancement and stellar age (especially at constant metallicity), we note that around $-0.4 < \mathrm{[Fe/H]} < 0$, young and fast rotating stars are dominating the interim-[$\upalpha$/Fe] regime. For hotter stars with $v \sin i > 15\,\mathrm{km/s}$, the estimated iron abundances A(Fe) are typically lower than the one of slow rotators. While this could be a trend introduced by the analysis approach that depends on sufficiently deep metal lines, another possibility is an actual correlation between [Fe/H] and rotation. An analysis of this correlation is complex and beyond the scope of this paper. When we neglect such stars (10\% of the sample), the trend of stellar age and [$\upalpha$/Fe] is monotonic.

For the high-$\upalpha$ metal-rich regime, a mix of different ages is noticeable, with an age spread up to $4\,\mathrm{Gyr}$. We take a closer look at this region in Sect.~\ref{discussion}.

To assess how distinct the two $\upalpha$-enhancement sequences are at different metallicities, we plot the histograms for five $0.15\,\mathrm{dex}$-wide metallicity bins in Fig.~\ref{alpha_histograms}. By eye, two clear peaks can only be identified for the three lower metallicities with decreasing separation. However, the fit of two Gaussians recovers the two peaks for all five distributions. For the most metal-poor bin, the $\upalpha$-enhanced stars are more numerous with an enhancement of $0.25\pm0.03\,\mathrm{dex}$, compared to the low-$\upalpha$ stars at $0.13\pm0.06\,\mathrm{dex}$. We note that even the low-$\upalpha$ stars are slightly enhanced at these metallicities. At higher metallicities, the mean enhancement of the low-$\upalpha$ sequence decreases gradually to become solar at solar metallicity. 

The enhancement of the high-$\upalpha$ sequence decreases more steeply down to $0.04\pm0.05\,\mathrm{dex}$ at solar metallicity. The peaks of the two (forced) sequences are thus consistent within one sigma (indistinguishable) at solar metallicity. We stress that in our fit we forced two Gaussian distributions and the actual distribution looks like a positively skewed Gaussian distribution. This means that an assignment to the high- or low-$\upalpha$ sequence based on a given [$\upalpha$/Fe] threshold \citet[e.g.][]{Adibekyan2012} is significantly less accurate or meaningful than in the metal-poor regime \citet[see also][]{Duong2018}.

The widths of the Gaussian fits to the high and low-$\upalpha$ sequence are of order $0.02-0.08\,\mathrm{dex}$ for [$\upalpha$/Fe] and similar to our measurement uncertainties and we note that the separation between the two sequences in the metal-poor regime is larger than this.

\begin{figure*}[!h]
\centering
\includegraphics[width=17cm]{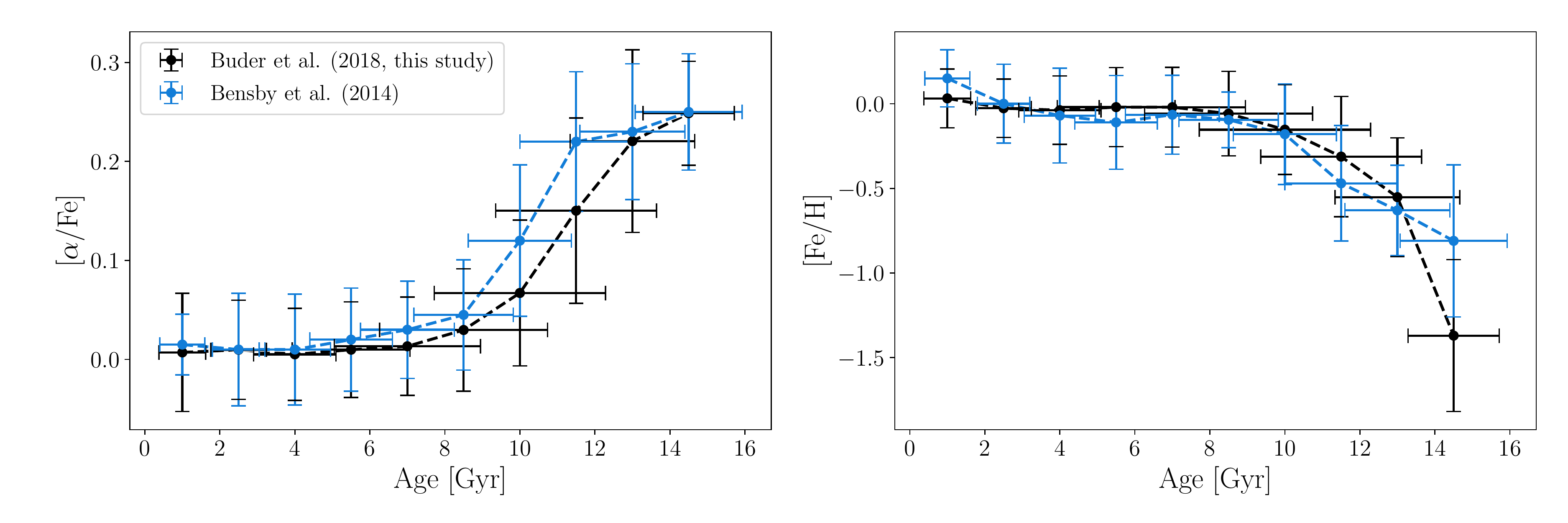}
\caption{Comparison of the relations of stellar age and mean metallicity [Fe/H] (left) as well as mean [$\upalpha$/Fe] (right) for the stars of this study (black) and the stars from \citet[][blue]{Bensby2014}, showing that the two relations agree between the studies within their uncertainties. The data points are calculated in $1.5\,\mathrm{Gyr}$ steps from 1 to $14.5\,\mathrm{Gyr}$) and the errors are the means of the age uncertainties as well as the standard deviations for the mean [Fe/H] and [$\upalpha$/Fe], respectively.}\label{fig:age_feh_alpha_relations_comparison}
\end{figure*}

\paragraph{The [$\upalpha$/Fe]-age distribution} is shown in Fig.~\ref{feh_alpha_age} (d) to (f). The main findings from these panels are:
\begin{enumerate}
\item The mean $\upalpha$-enhancement stays rather constant at $\mathrm{[\upalpha/Fe]} \sim 0.05$ up until $8\,\mathrm{Gyr}$ and then increases with stellar age. A comparison of our relation with the one found for the stars analysed by \citet{Bensby2014}, see left panel of Fig.~\ref{fig:age_feh_alpha_relations_comparison}, shows that the observed relations agree within their measurement uncertainties.
\item We find 6\% of the stars below $8\,\mathrm{Gyr}$ with $\mathrm{[\upalpha/Fe]} > 0.125$. The corresponding fraction for stars older than $11\,\mathrm{Gyr}$ is 60\%. This indicates a small jump around $8$ to $10\,\mathrm{Gyr}$ (from mean $\mathrm{[\upalpha/Fe]} \sim 0.05$ to 0.09) between high and low-$\upalpha$ enhancement, as also found by \citet{Haywood2013}. At around 8 to $12\,\mathrm{Gyr}$, we note a large range of metallicity in these coeval stars.
\item We find 67 stars among the young ones (i.e. 4978 stars with $<6\,\mathrm{Gyr}$), that are significantly $\upalpha$-enhanced ($\mathrm{[\upalpha/Fe]} > 0.13$) with normal rotational velocities (and another 59 with $v \sin i > 15\,\mathrm{km/s}$). With $\sim 0.9\%$ of our sample ($\sim 1.8\%$ when including the fast rotators), their ratio is in agreement with the sample analysed by \citet{Martig2015}, who found 14 out of 1639 stars to be $\upalpha$-rich and similar to most of the ratios of other samples listed by \citet{Chiappini2015}. Looking only at the young stars ($<6\,\mathrm{Gyr}$) our ratio of $1.4-2.5\%$ is however smaller than the one found by \citet{Martig2015} of $5.8\%$, pointing towards a different age distribution of the two different samples (containing either only giants or main sequences/turn-off stars). The 59 (47\%) of the young $\upalpha$-rich stars in our sample with increased broadening are all hotter than $6000\,\mathrm{K}$. We want to stress that for such stars the broadening in addition to the decreasing line strengths due to the atmosphere structure make the parameter estimation more uncertain than for the rest of our sample.
\item Among the old stars ($> 11\,\mathrm{Gyr}$), we find that a significant fraction of the sample (30\%), are low-$\upalpha$ stars ($\mathrm{[\upalpha/Fe]} < 0.125$), in contradiction to \citet{Haywood2013}. These stars are primarily cool main sequence or subgiant stars with metallicities above $-0.5\,\mathrm{dex}$, which causes their ages to have larger error bars. \citet{SilvaAguirre2018} also found such stars among APOGEE giants from the Kepler sample of stars, using asteroseismology to determine precise ages (see their Fig.~10).
\item  \citet{Haywood2013} claimed a rather tight correlation between age and $\upalpha$-enhancement for the old high-$\upalpha$ stars. However, \citet{SilvaAguirre2018} did not see evidence for such a tight relation. Our sample implies a tight trend, but is limited by the small number of these stars and we can not draw strong conclusions regarding the dispersion of chemistry and age.
\end{enumerate}

\paragraph{The [Fe/H]-age distribution} is depicted in Fig.~\ref{feh_alpha_age} (g) to (i) and shows:
\begin{enumerate}
\item With increasing stellar age, the mean metallicity, indicated with a red line in panel (g), decreases steadily and non-linear from $0.04$ at $1\,\mathrm{Gyr}$ to $-0.56$ at $13\,\mathrm{Gyr}$ respectively, but also with increasing dispersion. The recent study by \citet{Feuillet2018} has found that in their sample both the lowest [Fe/H] and highest [Fe/H] stars are older than the solar abundance stars and the iron abundance is hence less useful than [$\upalpha$/Fe] to predict age.
\item The dispersion increases steadily with stellar age from $0.17$ at $2\,\mathrm{Gyr}$ to $0.35$ at $13\,\mathrm{Gyr}$, respectively, as the steadily increasing red dashed line (calculated in steps of $1.5\,\mathrm{Gyr}$ similar to the median age uncertainties and each containing at least 50 stars), which traces the dispersion of [Fe/H], shows in Fig.~\ref{feh_alpha_age} (g).
\item Between ages of $0$ and $8\,\mathrm{Gyr}$, we see a rather flat mean metallicity with a spread of $0.5\,\mathrm{dex}$ around solar metallicity. In this age range, stars with lower metallicities ($<-0.25$) show an increase of $\upalpha$-enhancement of up to around $0.1\,\mathrm{dex}$.
\item Above ages of $8\,\mathrm{Gyr}$, metal-poor stars with $\mathrm{[Fe/H]} < -0.25$ exhibit a decreasing trend of metallicity with increasing stellar age. We can not confirm the tight age-metallicity distribution for the oldest stars due to the small sample of these stars, although we note indications of a tight overdensity. Few old stars around solar metallicity stand out, as also found in previous studies by \citet[][(see their Fig.~16)]{Casagrande2011}. These stars cause an increased dispersion, which is in agreement with the continuously increasing dispersion also seen at lower ages (see Fig.~\ref{feh_alpha_age} (g)). Similar results were found by \citet[][see their Fig.~1]{Haywood2008b}, who interpreted this trend as an observational signature of radial migration. We will follow this up, when including kinematics, in the following section.
\end{enumerate}

We find that the old stars of the sample are more $\upalpha$-enhanced. The oldest stars (above $11\,\mathrm{Gyr}$) are most metal-poor and $\upalpha$-enhanced (around $0.25\,\mathrm{dex}$ at metallicity [Fe/H]$\sim-0.5$), while slightly younger stars (still above $8\,\mathrm{Gyr}$) are less $\upalpha$-enhanced and more metal-rich (around $0.15\,\mathrm{dex}$ at solar metallicity). Most of the stars of the sample exhibit slightly increased or solar [$\upalpha$/Fe] and are on average younger than 6 to $8\,\mathrm{Gyr}$, but the sample also contains a minority of old stars (around $8\,\mathrm{Gyr}$) with low metallicity (around $-0.6\,\mathrm{dex}$) and only slight $\upalpha$-enhancement. We assess and discuss these results in more detail in Sect.~\ref{discussion}

\subsection{Kinematics}\label{kinematics}

To get an overview of the kinematical content of the GALAH+TGAS overlap, we first examine the 3D velocities. We then use these velocities to assign membership probabilities to different Galactic components. 

\begin{figure}[!h]
\centering
\resizebox{\hsize}{!}{\includegraphics{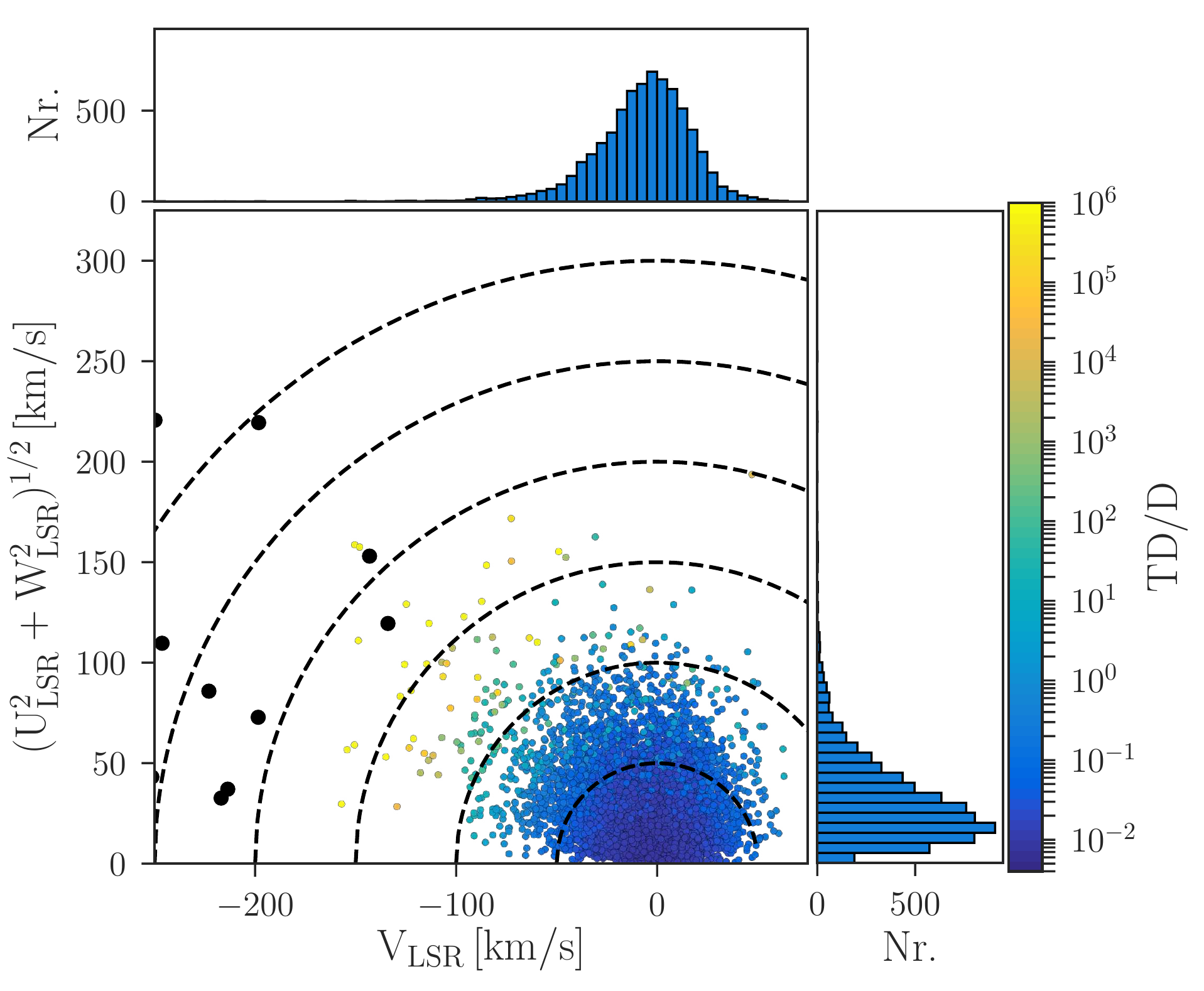}}
\caption{Toomre diagram of the sample from the perspective of the local standard of rest (LSR). Colour indicates the kinematic probability ratio of thick-to-thin disk membership, estimated in Sect.~\ref{analysis_kinematics}.  For 12 stars (marked as enlarged black dots), the kinematic membership probabilities point to neither thick or thin disk membership, but actually the halo. Dashed circles indicate total velocities in steps of $50\,\mathrm{km/s}$. The majority of the stars shares similar velocities to the local standard of rest and their kinematic membership ratio $TD/D$ points towards the thin disk. Fewer stars are seen with total velocities above $100\,\mathrm{km/s}$ and they typically show lower space velocities $V$. These stars have a significantly higher $TD/D$ ratio, characterising them as thick disk stars.}
\label{toomre}
\end{figure}

From the Toomre diagram in Fig.~\ref{toomre}, we can deduce that most stars belong to the disk, because their total velocities are lower than $180\,\mathrm{km/s}$, which is typically adopted as a good limit of halo kinematics \citep{Venn2004, Nissen2010}. This was expected from the target selection \citep{DeSilva2015}. Most of the sample shows an azimuthal relative velocity close to the local standard of rest, i.e. $\vert V \vert < 50\,\mathrm{km/s}$ and also $U$ and $W$ are close to the local standard of rest, indicating a solar-like motion in the thin disk. However, there are stars that also show large deviations from the thin disk kinematics, with total velocities larger than $100\,\mathrm{km/s}$ relative to the local standard of rest. In previous studies, such stars have been identified as thick disk stars, based on a hard limit of $70\,\mathrm{km/s}$ \citep{Fuhrmann2004}. 

Adopting hard limits to separate populations is however not appropriate when it comes to kinematics, as both thin and thick disc populations show significant dispersions in their characteristic space velocities \citep{Nordstroem2004}. Several more sophisticated approaches have been implemented \citep[see e.g.][]{Reddy2006, Ruchti2011, Bensby2014}. To begin with, we follow the approach by \citet{Bensby2014} to estimate the probability of each star to belong to one of the Milky Way components thin disk (D), thick disk (TD) or halo (H) by their kinematical information including the population velocity dispersions. The probability is influenced by the velocity distribution (assumed to be Gaussian), rotation velocities, as well as the expected ratio of stars among the components \citep{Bensby2003}. Similar to \citet{Bensby2014} we subsequently use the ratios of the membership probabilities.

\begin{figure}[!h]
\centering
\resizebox{\hsize}{!}{\includegraphics{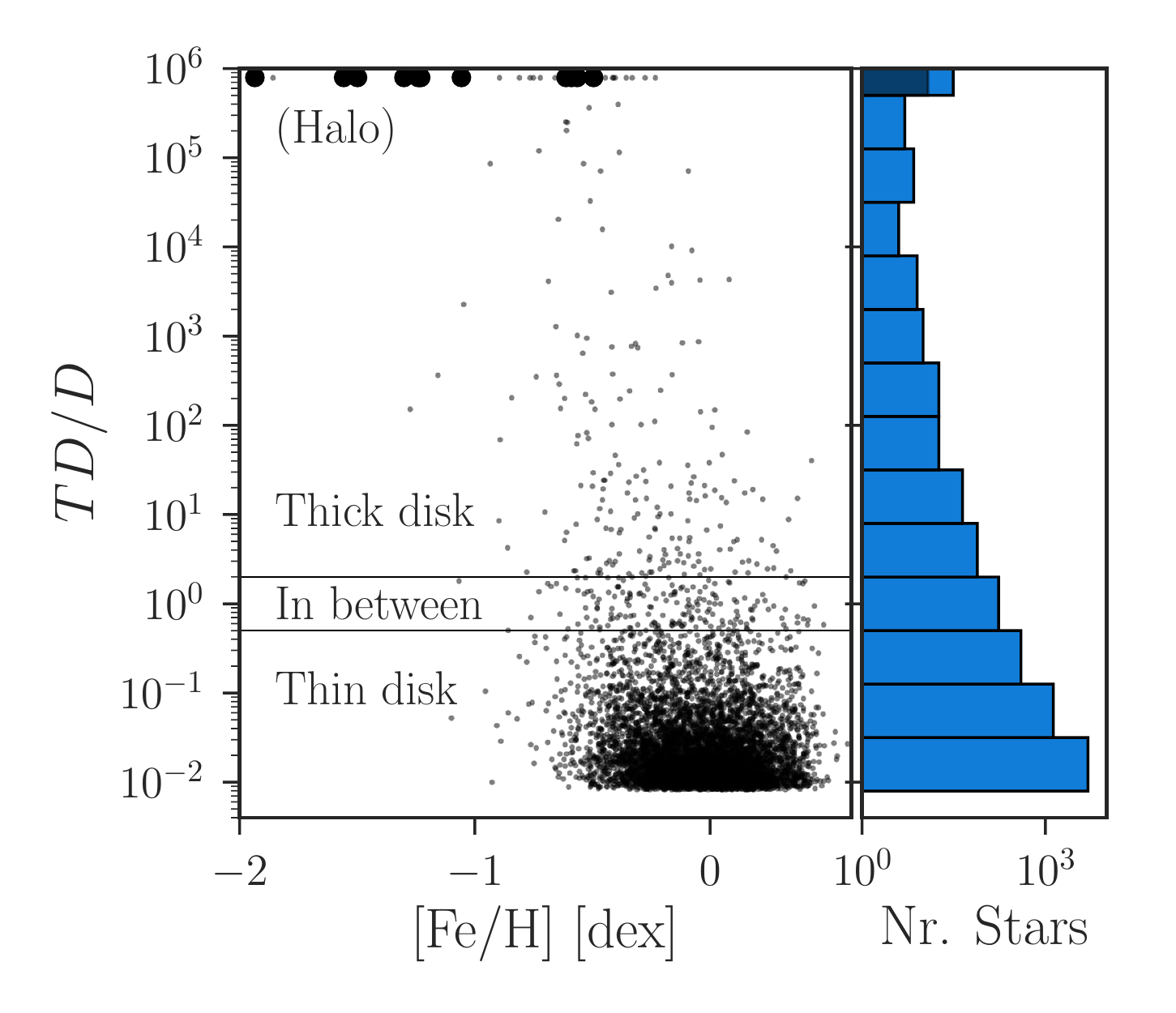}}
\caption{Ratio of kinematic membership probabilities of thick to thin disk $TD/D$ of the sample, estimated in Sect.~\ref{analysis_kinematics} following the approach by \citet{Bensby2014} and comparable with their Fig.~1. Most stars are assigned to the thin disk with this criterion. 12 stars, marked as enlarged black dots, are likely to be halo stars, based on their membership ratios with respect to the halo, $D_H$ and $TD_H$ respectively.}
\label{td_d_distribution}
\end{figure}

For each component, it is possible to separate the sample into most likely thick disk stars ($TD/D > 10$) and most likely thin disk stars ($TD/D < 0.1$), as shown in Fig.~\ref{td_d_distribution}. The 12 stars, that fit the kinematics of the halo best, are marked as big black circles and contribute $0.16$ \% to the sample. Almost all of these stars show a total space velocity of more than $180\,\mathrm{km/s}$ and would also be identified as halo stars with the simplified velocity criteria.

Most of the \GnT~dwarfs are most likely to be affiliated with the thin disk, while only a small fraction belongs to either thick disk or halo. For a very large fraction of the stars however, the probability ratio is indecisive. 

While the analysis of kinematics with the classical Toomre diagram is a powerful tool for a small volume, the approximations (assuming similar positions and velocities) are less appropriate for larger volumes. We therefore include another way to interpret the kinematical information by calculating action-angle coordinates and characterise orbits with integrals of motion as proposed by \citet{McMillan2008}. 

\begin{figure*}[!h]
\centering
\includegraphics[width=17cm]{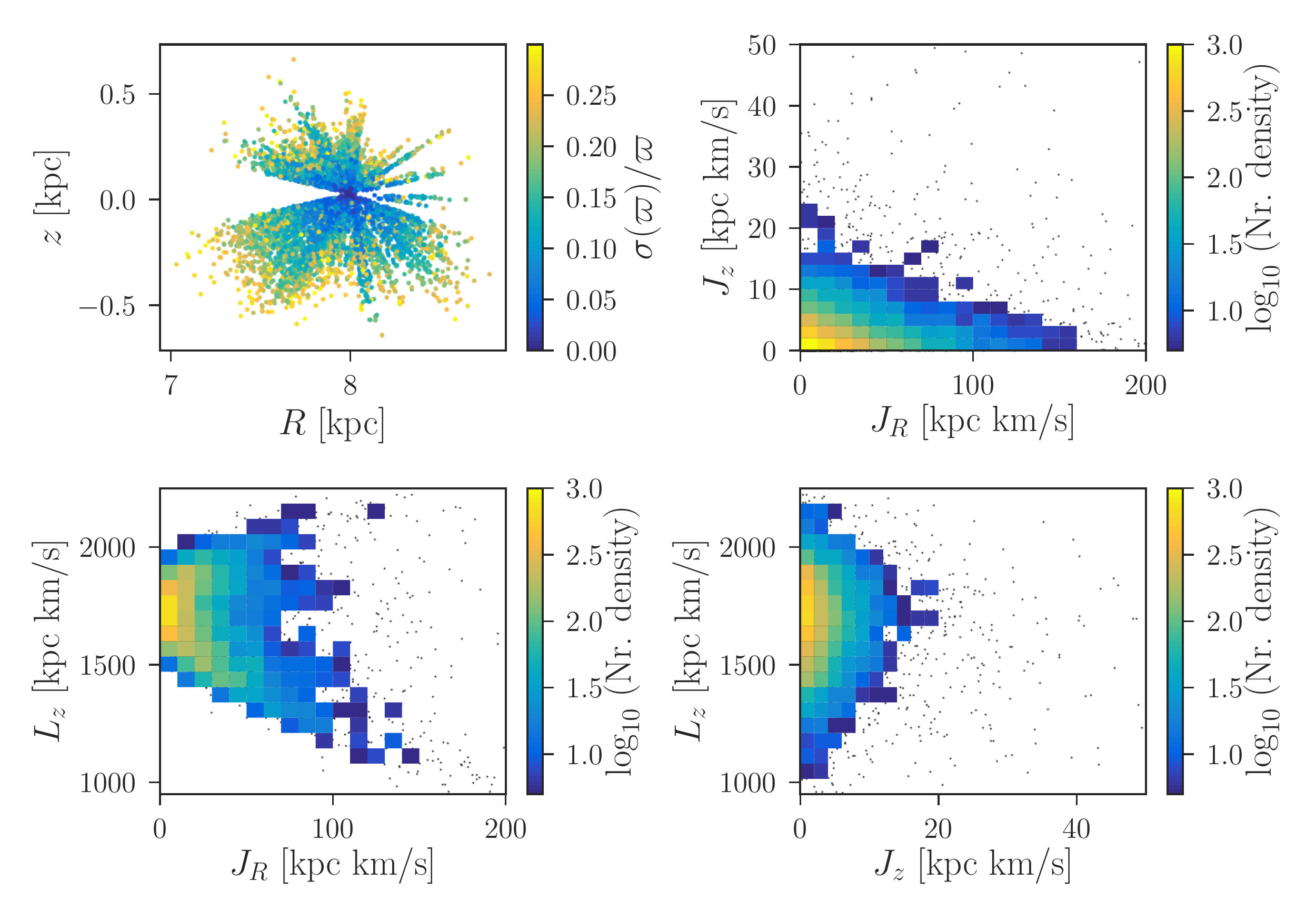}
\caption{Distribution of the sample stars in the $R$-$z$ plane (top left) as well as the three actions $J_R$-$J_z$ (top right), $J_R$-$J_\phi = L_z$ (bottom left), and $J_z$-$L_z$ (bottom right). With exception of the top left panel (coloured by parallax precision), colour indicates the number of stars per bin in each panel, with a lower limit of 5 stars per bin. Individual stars outside the bins are shown as black dots. The top left panel shows that increasing distance from the Sun, the parallax precision decreases. It also shows that with the exception of two special pointings, the Galactic plane ($\vert b \vert < 10\,\mathrm{deg}$) is neglected by GALAH. The individual action angle plots show that most of the stars move on circular orbits with solar Galactocentric orbit and angular momentum. The top right corner shows decreasing density with a diagonal pattern up to a line which intercepts with $J_R$ at $150\,\mathrm{kpc\,km/s}$ and $J_z$ at $20\,\mathrm{kpc\,km/s}$ and with only few stars with larger actions. The  $J_R$-$L_z$ shows trends of a minority of stars to be on eccentric orbits which have their apocentre in the solar neighborhood (sub-solar $L_z$ and increased $J_R$) or eccentric orbits with pericentres in the solar vicinity (super-solar $L_z$ and increased $J_R$). While stars do not show increased vertical actions in the $J_z$-$L_z$ plane (bottom right), an increase vertical actions can be seen for stars with solar angular momentum.}
\label{actions}
\end{figure*}

Contrary to $U$, $V$, and $W$, the three orbit labels, the actions $J_R$, $J_\Phi = L_z$, and $J_z$, allow us to quantify and compare orbits of stars independent of their position relative to the Sun or the local standard of rest. The distribution of the three actions is shown in Fig.~\ref{actions} and shows that most of the stars have similar orbits to the Sun, meaning low eccentricities and radial actions $J_R$, low vertical oscillations and vertical actions $J_z$, and azimuthal actions similar to the one of the Sun (with $L_{Z,\odot} = R_\odot \cdot v_{\Phi,\odot} = 8\,\mathrm{kpc} \cdot 220\,\mathrm{km/s} = 1760\,\mathrm{kpc\,km/s}$). However there are several stars with $J_R > 75\,\mathrm{kpc\,km/s}$ on more eccentric orbits, which manifests in mean stellar radii either significantly closer to ($L_z \ll L_{z,\odot}$) or further away from ($L_z \gg L_{z,\odot}$) the Galactic center. We will follow this up in Sect.~\ref{discussion}, when analysing angular momenta of the individual stars with different ages and chemistry, see Fig.~\ref{age_dissection}.

\section{Discussion of chemodynamics of the Galactic disks}\label{discussion}

\begin{figure*}[p]
\centering
\includegraphics[width=17cm]{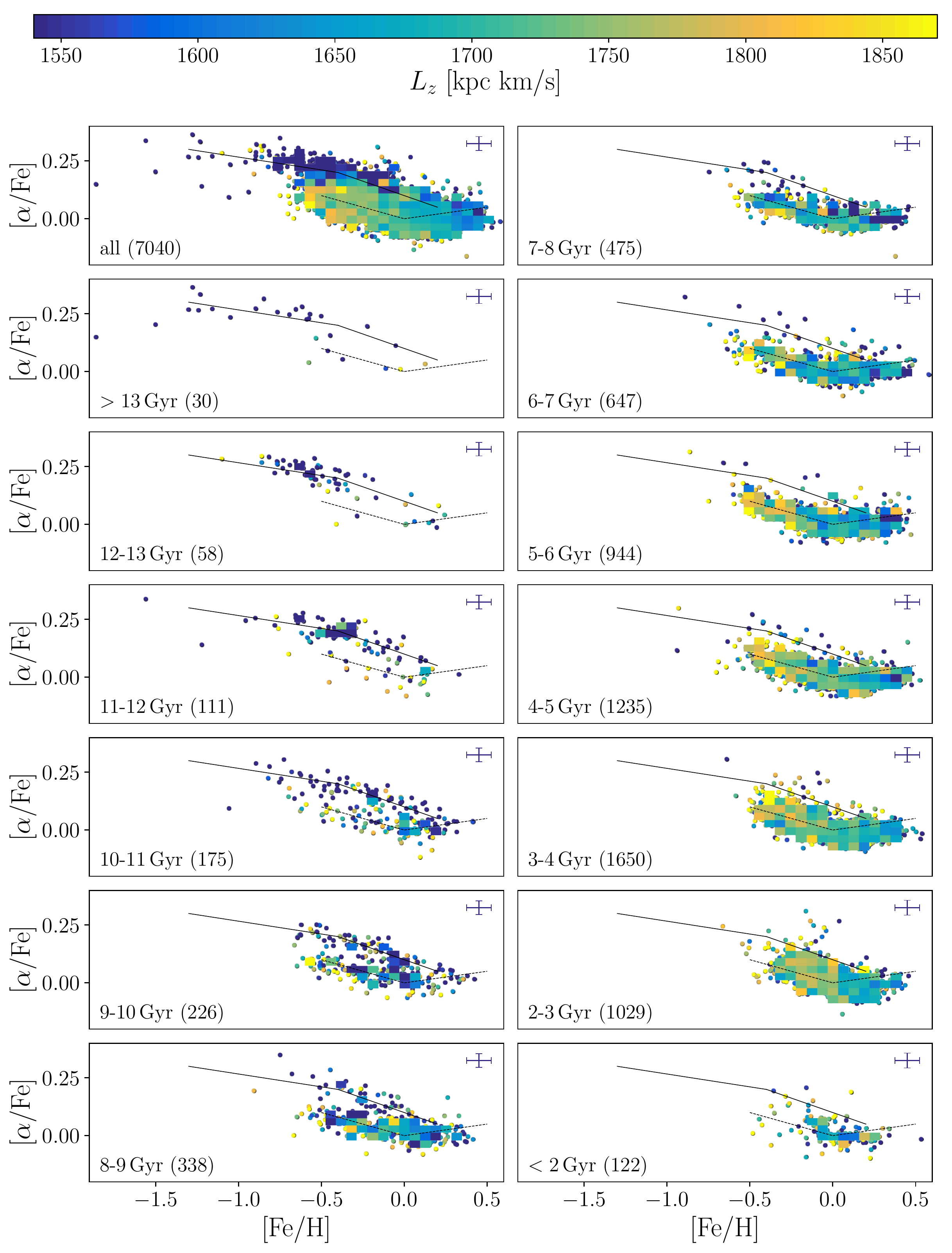}
\caption{$\upalpha$-enhancement, [$\upalpha$/Fe] with [Fe/H] in different age bins. The top left panel shows the distribution of all stars and mean angular momenta $L_z$ are shown in bins with more than five stars. The other panels show the stars bins of stellar age, ranging from above $13\,\mathrm{Gyr}$ (second left panel) to below $2\,\mathrm{Gyr}$ (bottom right panel). Colour indicates the angular momentum of each star, estimated in Sect.~\ref{analysis_kinematics}. The text in each panel indicates the age and respective number of stars. Fiducial lines indicate the low-$\upalpha$ and high-$\upalpha$ sequences to guide the eye. This plot shows, that stars with high-$\upalpha$ enhancement have generally lower angular momentum, while stars in the low-$\upalpha$ sequence, have mostly solar momentum. We note however that the metal-rich low-$\upalpha$ stars have in general lower angular momenta than metal-poor stars with comparable $\upalpha$-enhancement. For a detailed discussion of the panels see Sect.~\ref{discussion}.}\label{age_dissection}
\end{figure*}

The previous discussion highlighted that we report a significant change of chemistry around the stellar age of $8$ to $10\,\mathrm{Gyr}$. \citet{Haywood2013} and \citet{Bensby2014} used this to establish population assignments based on ages alone or ages and chemistry combined. \citet{Haywood2013} made a seemingly more arbitrary cut in the [$\upalpha$/Fe]-age. By assessing the [$\upalpha$/Fe]-[Fe/H] relation at different stellar ages, we will take a closer look at the transition phase in Fig.~\ref{age_dissection}. This slicing into mono-age populations has already been performed on output from numerical simulations \citet[see e.g.][]{Bird2013, Martig2014}. However it has, to our knowledge, not yet been applied to observational data, especially chemical composition, beyond the analysis of the age-metallicity structure \citep[see e.g.][]{Mackereth2017}.

\begin{figure}[!h]
\centering
\resizebox{\hsize}{!}{\includegraphics{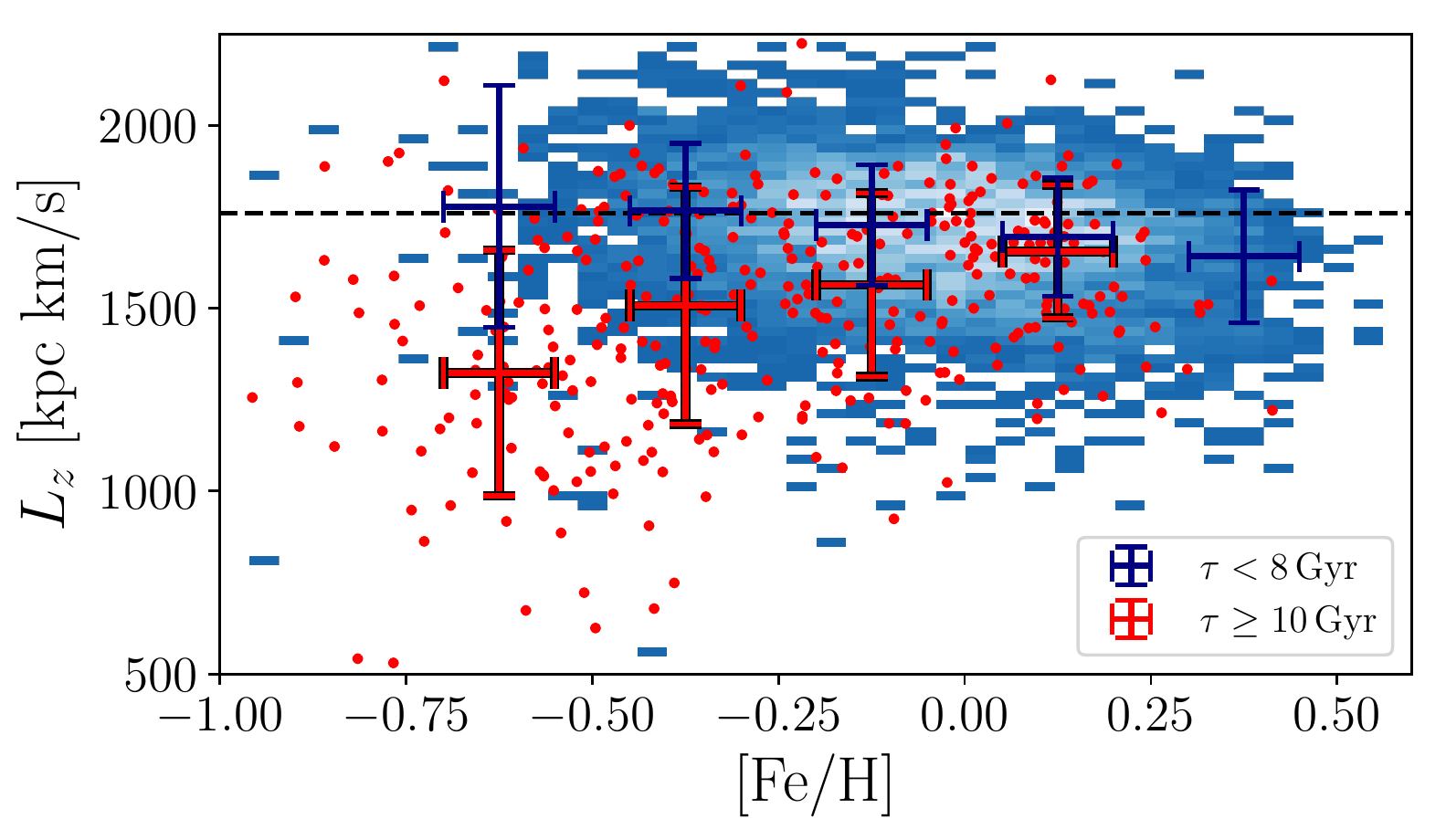}}
\caption{Angular momentum $L_z$ as a function of metallicity [Fe/H]. Stars with ages below $8\,\mathrm{Gyr}$ are shown in a blue density plot and those with ages above $10\,\mathrm{Gyr}$ as red dots. For both sets, mean angular momenta have been calculated in $0.25\,\mathrm{dex}$ steps in [Fe/H] for bins with more than 50 entries and are shown with mean [Fe/H] error and a combination of $L_Z$ uncertainty and standard deviation of the mean $L_z$ as error bars. The solar angular momentum is indicated with a dashed line.}
\label{fig:Feh_Lz}
\end{figure}

In Fig.~\ref{age_dissection}, we plot age bins; we can therefore focus on specific ages rather than age sequences. For each of the $1\,\mathrm{Gyr}$ bins, we show the [$\upalpha$/Fe]-[Fe/H] plane and colour the stars by their angular momentum, which is a function of Galactic radii. . Our main findings are:
\begin{enumerate}
\item The oldest stars ($>13\,\mathrm{Gyr}$) are mostly $\upalpha$-enhanced and metal-poor and have not yet been significantly enriched by SN Ia because they were born before the delay time of SN Ia was reached. Their spread in metallicities can be explained by different SN II masses and and frequencies, as well as gas mixing efficiencies, in the progenitor clouds. Their angular momentum shows that stars older than $13\,\mathrm{Gyr}$ are usually located closer to the Galactic centre with mean $L_z = 1060 \pm 600\,\mathrm{km/s}$, which corresponds to an average Galactocentric distance of $4.8 \pm 2.7\,\mathrm{kpc}$.
\item We note however, four stars with roughly solar metallicities older than $13\,\mathrm{Gyr}$, which could be unidentified binaries because of their proximity to the main sequence binary sequence. Such stars have also been found by \citet[][see their Fig.~16]{Casagrande2011}, \citet[][see their Fig.~21]{Bensby2014} , and \citet[][see their Fig.~10]{SilvaAguirre2018}. Their angular momenta and actions point towards solar-like orbits for half of them or eccentric orbits with significantly lower angular momenta than the Sun for the other half. The presence of these stars could be explained via chemical evolution, radial migration, but also influence of the bar. The recent study by \citet{Spitoni2018} showed, that a revision of the "two-infall" model can explain the presence of old stars with $-0.5 < \mathrm{[Fe/H]} < 0.25$ and $\mathrm{[\upalpha/Fe]} < 0.05$. Such stars have however also been found in the inner disk \citep{Hayden2015} and even the bulge \citep[see e.g.][]{Bensby2017}. The latter found even a high fraction of one-third of low-$\upalpha$ stars among the old stars. However the models by \citet{Spitoni2018} are not able to fully recover e.g. the age distribution and additional dynamic process are needed to explain the observed data. If we assume radial migration of such stars, we expect also them to arrive in our solar neighborhood. Models for radial migration \citep[see e.g.][]{Frankel2018} however usually aim to model the secular evolution of stars and hence currently focus on stars younger than $8\,\mathrm{Gyr}$. We do however not see a reason why the oldest stars, which clearly exist in the inner Galaxy \citep{Hayden2015, Bensby2017} should not have migrated in a similar manor as the younger stars within the lookback time of the migration models. Further analyses beyond the scope of this paper will hopefully help to pin down the more likely reason for the presence of the old stars with solar [Fe/H] and [$\upalpha$/Fe]. Another explanation that can be tested with more extended data sets is the radial migration due to the influence of the Milky Way bar \citep[see e.g.][]{Grenon1999, Minchev2010}.
\item Stars between $12$ and $13\,\mathrm{Gyr}$ exhibit more iron than the oldest stars although at similar [$\upalpha$/Fe]. In the metal-poor regime ($\mathrm{[Fe/H] < 0}$), the high-$\upalpha$ stars have lower angular momentum than the Sun ($\langle L_z \rangle = 1267 \pm 389\,\mathrm{kpc\,km/s}$ with mean $L_z$ uncertainties of $63\,\mathrm{kpc\,km/s}$), while more metal-rich stars ($\mathrm{[Fe/H] > 0}$) have angular momenta ($\langle L_z \rangle = 1513 \pm 262\,\mathrm{kpc\,km/s}$ with mean $L_z$ uncertainties of $46\,\mathrm{kpc\,km/s}$ closer to the solar one ($L_{z,\odot} = 1760\,\mathrm{kpc\,km/s}$).
\item Below $12\,\mathrm{Gyr}$, all stars except three outliers have metallicities above $-1.0\,\mathrm{dex}$. Between $12$ and $9\,\mathrm{Gyr}$, the stars on the high-$\upalpha$ sequence become gradually less $\upalpha$ enhanced and show increasing metallicities, hence indicating a continuous evolution of high-alpha stars along this sequence. This is consistent with the increasing enrichment of the ISM by SN Ia with a delay time distribution \citep[see e.g.][]{Maoz2012}, producing significantly more iron than SN II. Their angular momenta are on average still significantly lower than the solar one. This indicates a continuous evolution of high-$\upalpha$ stars along this sequence. We want to stress that these stars are still slightly $\upalpha$-enhanced, even at solar metallicities. Below $9\,\mathrm{Gyr}$, only a few stars are on the high-$\upalpha$ sequence and almost none below $7\,\mathrm{Gyr}$.
\item Similar to previous studies \citep{Lee2011, Haywood2013} we find a gradual increase of angular momentum and rotational velocity with metallicity among the $alpha$-enhanced metal-poor stars, which are is also correlated with age above $10\,\mathrm{Gyr}$. For stars below $10\,\mathrm{Gyr}$ (but even more pronounced for stars below $8\,\mathrm{Gyr}$), the angular momentum decreases with metallicity, as shown in Fig.~\ref{fig:Feh_Lz}.
\item Around $10\,\mathrm{Gyr}$ and at younger times, stars at the metal-poor end of the low-$\upalpha$ sequence appear. This finding is consistent with previous results by \citet[][see their Fig.~8]{Haywood2013}. The angular momenta of many of them indicate an origin at larger Galactic radii, i.e. they are only visitors to the solar neighbourhood, in agreement with findings by \citet{Haywood2008b} and \citet{Bovy2012}.
\item Between $3$ and $9\,\mathrm{Gyr}$, the full range of metallicities from $-0.7$ up to $0.5$ at the low-$\upalpha$ sequence is covered. The stars at the low-$\upalpha$ metal-poor end show on average significantly larger angular momenta than the Sun, an opposite trend with respect to the high-$\upalpha$ stars. The angular momenta of metal-rich stars ($1640 \pm 180\,\mathrm{kpc\,km/s}$) are on average lower than the solar one.
\item At younger times than $3\,\mathrm{Gyr}$, the spread in metallicities decreases and stars younger than $2\,\mathrm{Gyr}$ only cover metallicities between $-0.3$ and $0.3$. Note that our cut for hot stars, see Sect.~\ref{observation}) has cut out most of the stars with ages below $2\,\mathrm{Gyr}$, see Sect.~\ref{age_distribution}. \citet{Casagrande2011} have found, however, that nearby stars younger than $1\,\mathrm{Gyr}$ also only cover the range of $-0.3 < \mathrm{[Fe/H]} < 0.2$ (see their Fig.~16).
\end{enumerate}

Similar conclusions of a gradual chemical enrichment of the high-$\upalpha$ sequence with time can be drawn from the data by \citet[][see their Fig.~3]{Hayden2017}, as well as from the analysis of Auriga simulations \citep{Grand2018}. 

\begin{figure}[!ht]
\centering
\resizebox{\hsize}{!}{\includegraphics{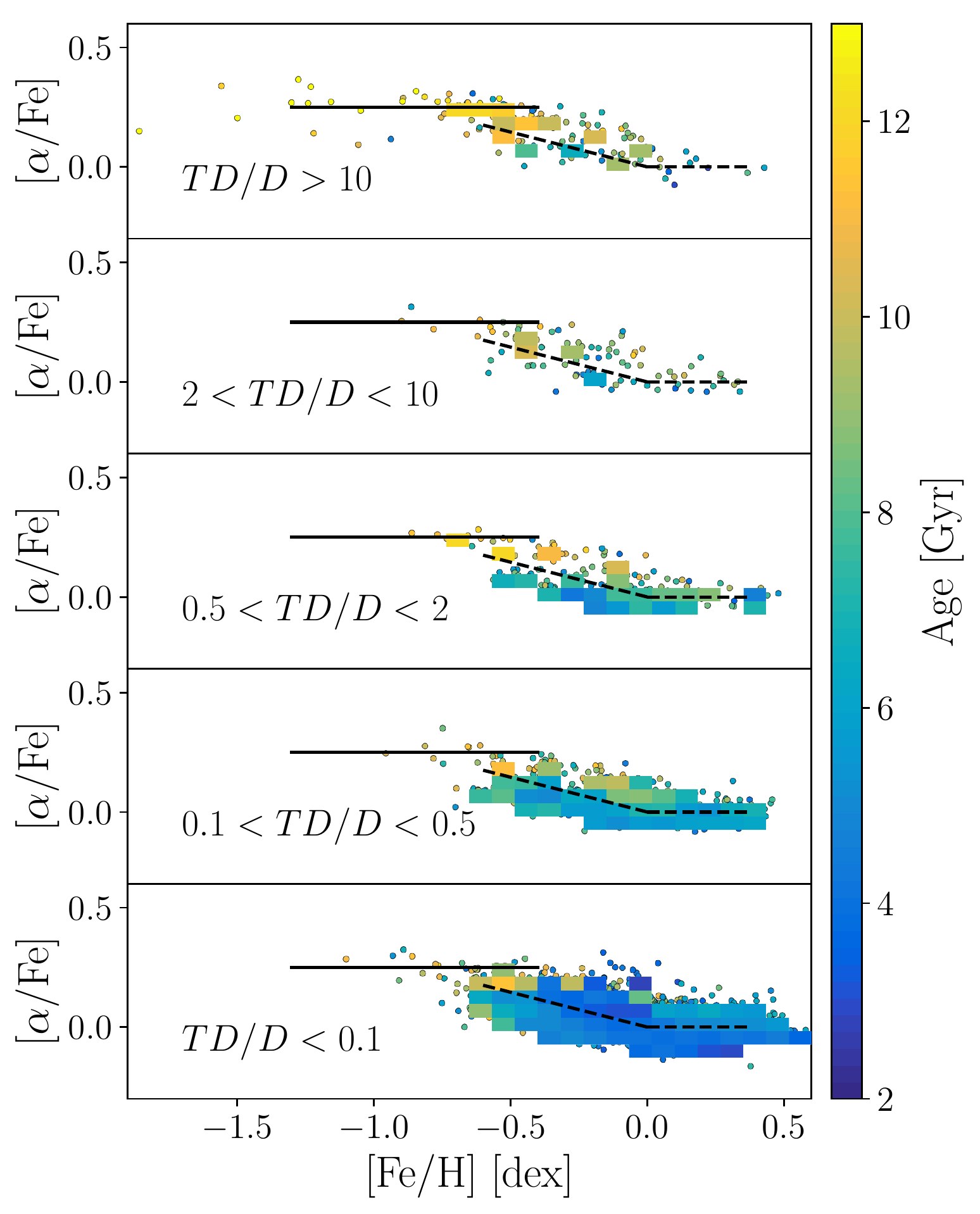}}
\caption{[$\upalpha$/Fe] as a function of [Fe/H] selected based on the kinematic probabilities of belonging to the thin or thick disk $TD/D$ from kinematics. Colour indicates the estimated age. Analogous to Fig.~19 of \citet{Bensby2014}, the abundance plateau of the thick disk and the decrease for thin disk have been plotted in black to guide the eye. While stars with highly thick-disk like kinematics (two upper panels) follow the high-$\upalpha$ sequence, numerous stars with thin-disk like kinematics are also seen in the high-$\upalpha$ metal-rich regime. However, most stars with thin disk like kinematics ($TD_D < 0.5$ two lower panels) follow the low-$\upalpha$ sequence. See Sect.~\ref{discussion} for further discussion.}
\label{alpha_td_d_split}
\end{figure}

Finally, we revisit the $\upalpha$-enhancement of the sample, but dissect the sample by the kinematic probability of belonging to the thick or thin disk. These probabilities were estimated in Sect.~\ref{analysis_kinematics} and we use them to recreate Fig.~19 from \citet{Bensby2014} with the \GnT~sample. From this dissection in Fig.~\ref{alpha_td_d_split} we conclude:
\begin{enumerate}
\item Stars that are kinematically $>10$ times more likely to be part of the thick disc population (top panel), are mostly on the high-$\upalpha$ sequence. We note that in contrast to \citet{Bensby2014}, these stars are not all metal-poor, but also cover the high-$\upalpha$ metal-rich regime. These stars are however almost exclusively older than $8\,\mathrm{Gyr}$. The sample from \citet{Bensby2014} is expected to cover relatively more metal-poor stars, because they selected their sample specifically with the aim to trace the metal-poor limit of the thin disk, the metal-rich limit of the thick disk, the metal-poor limit of the thick disk.
\item Stars that are $2-10$ times as likely to belong to the thick disk ($2 < TD/D < 10$) and those with inconclusive kinematics ($0.5 < TD/D < 2$) cover both the high- and low-$\upalpha$ metal-rich regime as well as a larger range of ages.
\item At the lowest probabilities of belonging to the thick disk ($TD/D < 0.5$), the stars are on average young ($5.0 \pm 2.2\,\mathrm{Gyr}$) and cover the low-$\upalpha$ sequence. However, there are still noticeable amounts of stars in the high-$\upalpha$ metal-rich regime ($\mathrm{[Fe/H]} > -0.1$ and above the dashed line, including stars older than $8\,\mathrm{Gyr}$. This implies that there is no distinct kinematical separation of populations in the metal-rich regime and a significant overlap in kinematic properties is present.
\end{enumerate}

%
\section{Conclusions}

The combination of spectroscopic data from the high-resolution GALAH survey and the astrometric data from the Tycho-\Gaia Astrometric Solution (TGAS) spans a high-dimensional space of chemodynamical information. In this study, we have analysed \GnTused~dwarf and turn-off stars of this overlap. Our main results are summarised as follows:

\subsection{Abundance and age trends}
\begin{enumerate}
\item We show that our parameter and abundance estimates measured with the GALAH pipeline are accurate and precise. 
\item Our selected stars are all within the solar vicinity and consist mostly of young stars with metallicities from $-0.7\,\mathrm{dex}$ up to $+0.5\,\mathrm{dex}$ as shown by the age and metallicity distributions.
\item We report stellar parameters (including iron in non-LTE), 18 element abundances in LTE, an error-weighted $\upalpha$-process element abundance in LTE, and Li as well as O abundances in non-LTE. We show that non-LTE corrections for the O triplet are a vital requirement for accurate abundance estimation of stars in different evolutionary stages.
\item We show in Fig.~\ref{Xfeall1}, that the abundance trends estimated for the \GnT~sample agree well with previous studies with higher spectral quality. Due to the larger number statistics and the selection of our stars, we are able to assess abundance trends of the low-$\upalpha$ sequence in more detail than previous studies. 
\item Among the studied elements, we find the expected similarities of abundance trends among the $\upalpha$-elements Mg, Si, Ca, and Ti. O, however, shows a significantly steeper decrease with metallicity and no flattening at super-solar metallicities and was hence not treated as an $\upalpha$-element. The odd-Z elements Na and Al show similar increasing abundances towards the super-solar metallicities. We note that these trends also agree with those observed for the iron-peak elements Ni, Cu, and Zn.
\item We find significant trends of abundances with stellar age for $\upalpha$- and s-process elements, see Fig.~\ref{Xfeall2}, similar to previous studies of solar twins \citep{Nissen2015, Spina2016} and dwarfs in general \citep{Bensby2014}.
\end{enumerate}

\subsection{The age-[$\upalpha$/Fe]-[Fe/H] relationship}\label{sec:conclusion_2}

\begin{enumerate}
\item We recover the same pattern of the [$\upalpha$/Fe]-[Fe/H] in combination with age as previous studies \citep[e.g.][]{Ness2016, Ho2017}, namely that the high-$\upalpha$sequence is mainly consisting of old stars, while the low-$\upalpha$ sequence covers ages usually below $8\,\mathrm{Gyr}$. We note however, that at the the low-$\upalpha$ metal-rich and low-$\upalpha$ metal-poor regime, we also find stars with ages between $8$ to $10\,\mathrm{Gyr}$.
\item When using age and chemistry together, the high-$\upalpha$ metal-rich stars can be well explained to be part of a population formed from the same material as the canonical (lower metallicity) thick disk, although significantly enriched by both SN II and Ia. This is in contradiction to the claim by \citet{Adibekyan2012}, but backed up by stellar kinematics. The angular momenta of the high-$\upalpha$ metal-poor and high-$\upalpha$ metal-rich stars are both significantly and consistently lower than the solar value. We conclude that these stars belong to the high-$\upalpha$ sequence rather than representing a distinct population.
\item At solar and super-solar metallicity, both old and young stars as well as slightly $\upalpha$-enhanced and solar [$\upalpha$/Fe] stars are observed. We have shown in Fig.~\ref{alpha_histograms}, that in this regime, the two distinct $\upalpha$-sequences (as seen for $\mathrm{[Fe/H] < 0}$) become indistinguishable in [$\upalpha$/Fe] vs. [Fe/H]. We conclude, similarly to \citet{Haywood2013}, that stellar age is a better identifier and discriminator of formation origin than metallicity, especially at high metallicity. Stellar age when available is more fundamental than an arbitrary cut in [$\upalpha$/Fe] for the purpose of understanding Milky Way populations.
\item The decreasing spread of metallicities for stars younger than $4\,\mathrm{Gyr}$ is strong support for radial migration. Radial migration predicts that stars will change their orbits after they have been born. The panel of Fig.~\ref{age_dissection} with stars below $2\,\mathrm{Gyr}$ shows a narrower range in metallicity compared to the panel for stars with ages of $3-4\,\mathrm{Gyr}$. With increasing age, stars are observed at lower and higher metallicity ranges. Yet, there are no stars with metallicities at $-0.5$ and $0.5\,\mathrm{dex}$ for the youngest ages, although the GALAH selection is not biased against these stars. At increasingly younger ages we preferentially see stars formed in the solar neighbourhood, whereas older stars did not have time to migrate to this location. Therefore, we only see stars formed in the solar neighborhood. This observational support for radial migration in the thin disk has also been found in other studies \citet[see e.g.]{Haywood2008, Feuillet2016} and tested by comparison with models including radial migration by \citet{Feuillet2018}. We stress however, that we excluded stars with $T_\text{eff} > 6900\,\mathrm{K}$ from this study, which are mostly younger than $3\,\mathrm{Gyr}$. While we advice caution that our result might be biased, \citet{Casagrande2011} have found a similar spread of [Fe/H] for stars below $1\,\mathrm{Gyr}$ in their analysis.
\end{enumerate}

\subsection{Chemodynamics of the disk in the Solar vicinity}

\begin{enumerate}
\item Our sample includes a small fraction of stars of the distinctly old ($> 10\,\mathrm{Gyr}$) and $\upalpha$-enhanced ($\mathrm{[\upalpha/Fe]} > 0.3$) thick disk. These stars are also kinematically consistent with the chemically defined thick disk.
\item Stellar age does not change after the birth of a star. However, kinematic properties of stars from different populations overlap and might also change. Thus, age should be a more reliable definition of a population.
\item Independent of the population assignment of the high-$\upalpha$-metal-rich stars (see Sect.~\ref{sec:conclusion_2}), their kinematics (e.g. their lower angular momentum compared to the Sun) indicates, that they have migrated towards us via blurring, meaning they are on eccentric orbits with mean radii closer to the Galactic center.
\item Around ages of $8$ to $10,\mathrm{Gyr}$, stars at the metal-poor end of the thin disk are identified. When including kinematical information, we find a large number of these stars to be from outside the solar neighborhood (on eccentric orbits with mean radii further out). At the metal-rich end of our sample, the kinematical information points to a significant number of stars from the inner radii on eccentric orbits. This result is in agreement with the seminal paper by \citet[][see their Fig.~22]{Edvardsson1993} and confirms blurring as reason for a broad age-metallicity relation at different radii.
\end{enumerate}

\subsection{Future investigations}

The follow-up of our study with more detailed studies focusing of certain aspects of the high-dimensional chemodynamical space is promising. \citet{Anders2018} have shown that the chemical space can be dissected by using t-distributed stochastic neighbour embedding. For the high-$\upalpha$ metal-rich regime they find that these stars are different from the classical thick disk. They find however, that the stars in this regime are spread within the t-SNE distribution, indicating an evolution within this regime (see their Fig. 1). The application of their approach onto our data is promising but beyond the scope of this paper.

In this study, we have shown that non-LTE corrections play a crucial role for several elements. We suggest a detailed further investigation of 3D and non-LTE effects for other elements. We are working on extending the non-LTE implementation and will apply them to 7 elements in GALAH DR2 \citep{Buder2018b}.

A quantitative study of the correlation of element abundances and stellar ages for the stars of our sample could explore the influence of other parameters on the [Y/Mg]-age correlation beyond the solar twins \citep{Nissen2015, Spina2016}, to test the hypothesis by \citet{Feltzing2017}, that the [Y/Mg]-age relation is unique to solar analogues. Such analyses can be performed on all different elements, especially when using the second public Data Release from the GALAH survey \citep{Buder2018b} and the second \Gaia Data Release. With this large set of abundances, ages, and kinematics, it will also be possible to identify clumps in chemodynamical space or stellar streams \citep[see e.g.][]{Quillen2018}. A central point of Galactic Archaeology in the future will also be the improvement of stellar age estimation.

\begin{acknowledgements}

S.B. and K.L. acknowledge funds from the Alexander von Humboldt Foundation in the framework of the Sofja Kovalevskaja Award endowed by the Federal Ministry of Education and Research. S.B. acknowledges support from the Hunstead Fund for Astrophysics at the University of Sydney and the National Science Foundation under Grant No. PHY-1430152 (JINA Center for the Evolution of the Elements). K.L. acknowledges funds from the Swedish Research Council (Grant nr. 2015-00415\_3) and Marie Sklodowska Curie Actions (Cofund Project INCA 600398). L.D. gratefully acknowledges a scholarship from Zonta International District 24 and support from ARC grant DP160103747. L.C. is the recipient of an ARC Future Fellowship (project number FT160100402). S.L.M. acknowledges financial support from the Australian Research Council through grant DE1401000598. K.\v{C}., G.T., and T.Z. acknowledge support by the core funding grant P1-0188 from the Slovenian Research Agency. D.M.N. was supported by the Allan C. and Dorothy H. Davis Fellowship. Parts of this research were conducted by the Australian Research Council Centre of Excellence for All Sky Astrophysics in 3 Dimensions (ASTRO 3D), through project number CE170100013.

We thank A.~M.~Amarsi for providing the grid of abundance corrections for oxygen. We thank U. Heiter for providing adjusted $K_S$ band magnitudes for the \Gaia FGK benchmark stars. We thank H.-W.~Rix, M.~Fouesneau, A.~M.~Amarsi,  M.~Bergemann, and A.~Quillen for useful discussions and comments.

We would like to thank the anonymous referee for carefully reading our manuscript and for giving such constructive comments which substantially helped improving the quality of the paper.

The computations were performed on resources provided by the Swedish National Infrastructure for Computing (SNIC) at UPPMAX under project 2015/1-309 and 2016/1-400.

This project was developed in part at the 2016 NYC \Gaia Sprint, hosted by the Center for Computational Astrophysics at the Simons Foundation in New York City, and the 2017 Heidelberg \Gaia Sprint, hosted by the Max-Planck-Institut f\"ur Astronomie, Heidelberg. This work has made use of data from the European Space Agency (ESA) mission \Gaia (http://www.cosmos.esa.int/gaia), processed by the \Gaia Data Processing and Analysis Consortium (DPAC, http://www.cosmos.esa.int/web/gaia/dpac/consortium). Funding for the DPAC has been provided by national institutions, in particular the institutions participating in the \Gaia Multilateral Agreement. This research has made use of the VizieR catalogue access tool, CDS, Strasbourg, France. The original description of the VizieR service was published in A\&AS 143, 23. This research mad use of the TOPCAT tool, described in \citet{Taylor2005}. This publication makes use of data products from the Two Micron All Sky Survey, which is a joint project of the University of Massachusetts and the Infrared Processing and Analysis Center/California Institute of Technology, funded by the National Aeronautics and Space Administration and the National Science Foundation.

\end{acknowledgements}


\end{CJK*}

\bibstyle{aa} 
\bibliographystyle{aa} 
\bibliography{bib.bib} 

\begin{appendix} 

\section{Abundance trends}\label{abundancetrends}

In total, the GALAH survey wavelength range includes detectable atomic absorption lines of up to 30 elements of FKG stars. A subset of 20 of these can be measured in the dwarf and turn-off star spectra, which we examine in this work: Li, C, O, $\upalpha$-process elements (Mg, Si, Ca, and Ti), light odd-Z elements (Na, Al, K), iron-peak elements (Sc, V, Cr, Mn, Co, Ni, Cu, and Zn), and s-process neutron capture elements (Y and Ba). We list the mean precision (estimated from repeated observations), accuracy (inferred from the uncertainties of the oscillator strengths), measured fraction of the clean sample in percent, and numbers of measured lines in Table~\ref{table:elements}.

\begin{table}[!h]
\caption{Elements by atomic numbers and their precision and accuracy as well as number of measured lines}              
\label{table:elements}      
\centering                                      
\begin{tabular}{c c c c c c}          
\hline\hline                        
Z & Elem. X & Precision & Accuracy & Measured & Lines \\    
\hline                                   
 & $\upalpha$ & 0.03 &  & 99\% & \\
3  &  Li  &  0.12  & 0.01 & 42\% & 1 \\
6  &  C  &  0.11  &  0.05 & 26\% & 1 \\
8  &  O  &  0.12  & 0.01 & 97\% & 3 \\
11  &  Na  &  0.08  & 0.01 & 94\% & 4 \\
12  &  Mg  &  0.07  & 0.05 & 99\% & 3 \\
13  &  Al  &  0.06  & 0.08 & 56\% & 4 \\
14  &  Si  &  0.06  & 0.03 & 83\% & 4 \\
19  &  K  &  0.17  & 0.01 & 95\% & 1 \\
20  &  Ca  &  0.17  & 0.02 &  95\% & 2  \\
21  &  Sc  &  0.06  & 0.05 &  77\% & 10  \\
22  &  Ti  &  0.04  & 0.05 & 24\% &  20 \\
23  &  V  &  0.12  &  0.05 & 25\% & 17 \\
24  &  Cr  &  0.08  & 0.15 & 81\% & 9 \\
25  &  Mn  &  0.08  & 0.02 & 69\% &  4  \\
27  &  Co  &  0.09  & 0.09 & 4\% & 3   \\
28  &  Ni  &  0.08  &  0.07 & 74\% & 7  \\
29  &  Cu  &  0.10  & 0.08 & 52\% & 1  \\
30  &  Zn  &  0.11  & 0.05 &81\% &   2  \\
39  &  Y  &  0.12  &   0.05 &82\% & 4  \\
56  &  Ba  &  0.08  & 0.05 & 34\% &  2  \\
\hline                                             
\end{tabular}
\end{table}

\begin{figure*}[h]
\centering
\includegraphics[width=16.6cm]{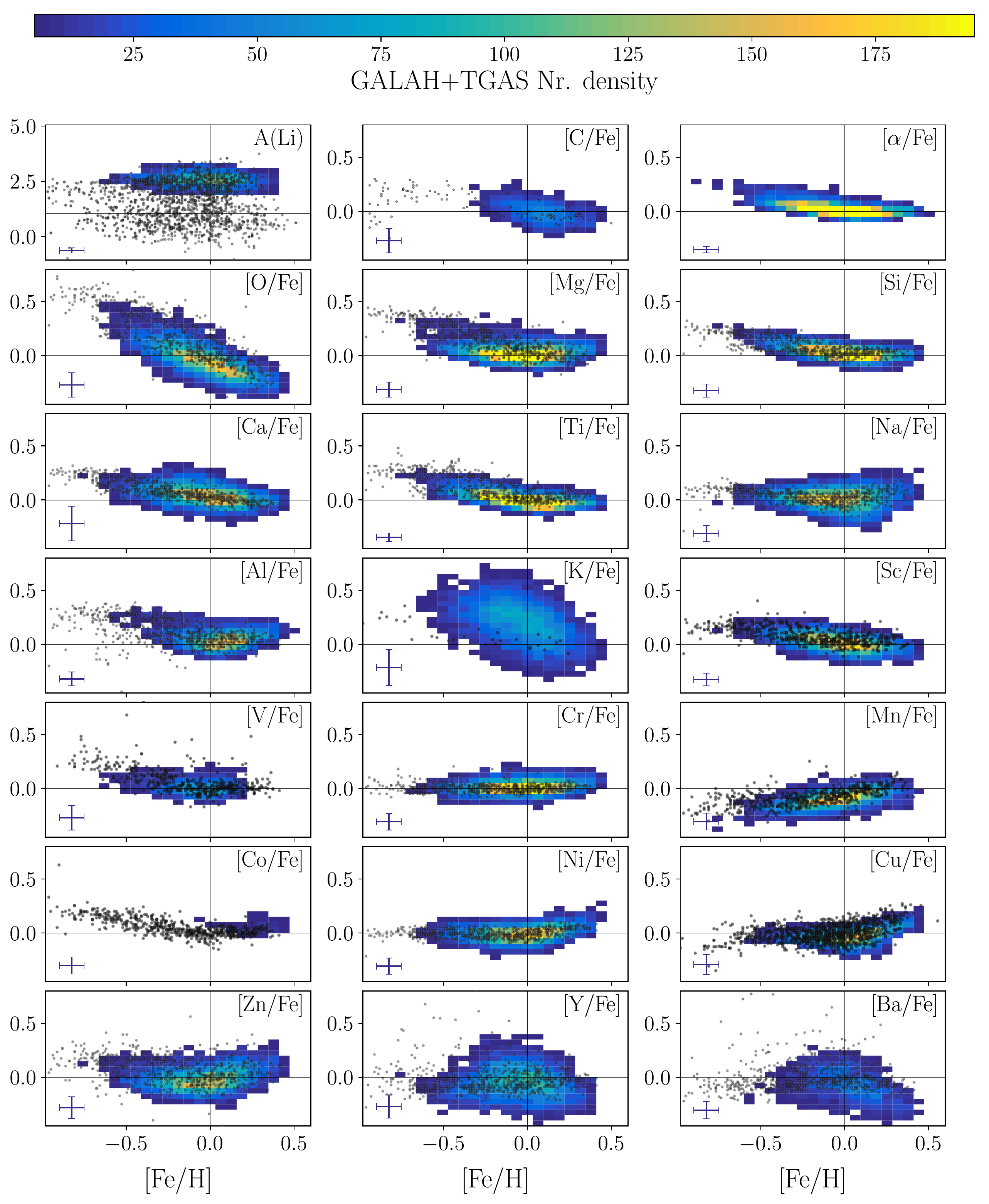}
\caption{Distribution of elemental abundances of the measured elements as a function of metallicity, coloured by density (with a minimum of 5 stars per bin). All elements are shown relative to the iron abundance, except for Li (shown as absolute abundance). The elements are indicated in the upper right corner of each panel. Measurements from the literature are overlaid as grey dots, i.e. results by \citet{Bensby2014} for O, Na, Mg, Al, Si, Ca, Ti, Cr, Ni, Zn, Y, and Ba, \citet{Battistini2015} for Sc, V, Mn, \citet{Nissen2014} for C, \citet{Zhao2016} for K, and \citet{DelgadoMena2017} for Cu. For details regarding the individual elements, see respective paragraphs in Sect.~\ref{abundancetrends}.}\label{Xfeall1}
\end{figure*}

\begin{figure*}[h]
\centering
\includegraphics[width=16.6cm]{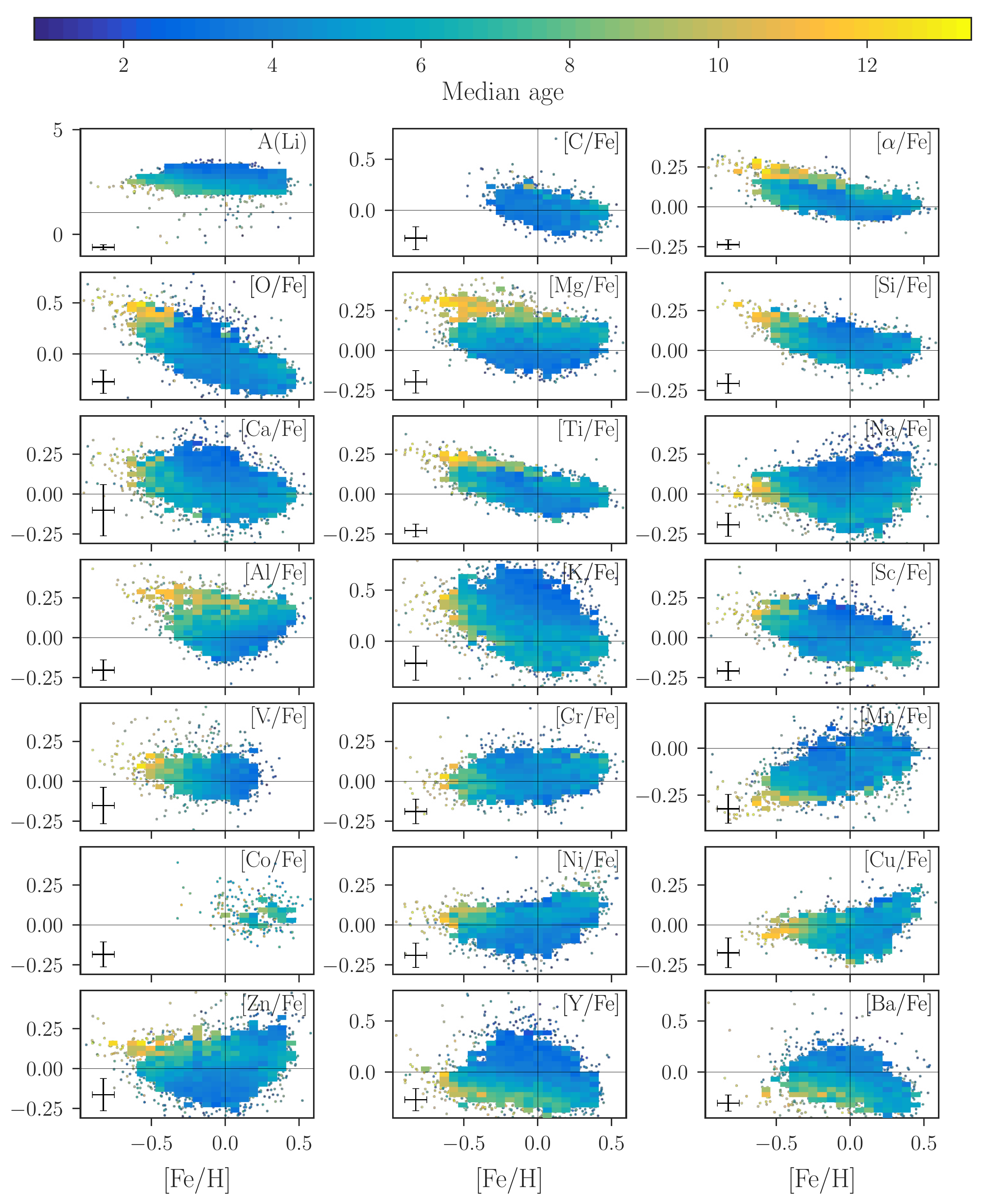}
\caption{Distribution of elemental abundances of the measured elements as a function of metallicity, coloured by median age per bin (with a minimum of 5 stars per bin). Contrary to Fig.~\ref{Xfeall1}, individual stars outside the bins are shown as small dots and are also coloured by age. No literature samples are overlaid. All elements are shown relative to the iron abundance, except for Li (shown as absolute abundance). For details regarding the individual elements, see respective paragraphs in Sect.~\ref{abundancetrends}.}\label{Xfeall2}
\end{figure*}

In Figures \ref{Xfeall1} and \ref{Xfeall2}, we show our measured element abundances as a function of metallicity. We present the density distribution for these in Fig.~\ref{Xfeall1}, coloured by stellar counts per bin. In Fig.~\ref{Xfeall2}, we plot the same distribution, but colour each bin by the median age of its stars. When available, we include results from previous studies \citep{Bensby2014, Battistini2015, Nissen2014, Zhao2016, DelgadoMena2017}. These studies use spectra of higher quality (higher resolution and $S/N$), but are much smaller in sample size. Similar to the GALAH survey, these literature studies include dwarfs of both the low- and high-$\upalpha$ sequences of the disk and span a large metallicity range ($-2.8 < \mathrm{[Fe/H]} < 0.4$). We note that the selection function for these studies is different to that of GALAH and they contain much higher relative numbers of metal-poor stars. 

For a quantitative discussion of age trends we refer the reader to other studies, e.g. for [Y/Mg] \citep{Nissen2015, Spina2016}, [C/N] \citep{Masseron2015, Martig2016, Ness2016} or the study of 17 abundance-age trends for APOGEE by \citet{Feuillet2018}. 

\subsection{Lithium}

Li is measured using 1D non-LTE corrections by \citet{Lind2009}. In our sample, we can only detect Li in stars with a relatively large absolute Li abundance, of $\mathrm{A(Li)} > 2.0$. In the metal-poor regime, the warm stars are situated on the Spite plateau \citep{Spite1982}, around $\mathrm{A(Li)} = 2.3$, as expected. Towards higher metallicities, A(Li) is mostly measured between 2.0 and 3.3. The latter value is close to the meteoritic A(Li) of $3.26 \pm 0.05$ \citep{Asplund2009}.
Li is expected to be a good tracer of the evolution of the star, because the initial composition is depleted by the proton bombardment processes at temperatures higher than $2.5\cdot10^6\,\mathrm{K}$ \citep{Pinsonneault1997}.
\challenge{From the respective panel in Fig.~\ref{trends1a}, we see a strong correlation of higher Li for higher effective temperatures. The lines of Li in colder and evolved stars are below the $3\sigma$ detection limit. The strong anti-correlation between Li and effective temperature (or mass) is well known and previously depicted for thin/thick disk stars e.g. by \citet{Ramirez2012}. This is due to the larger surface convective envelopes of cooler stars, extending to hotter layers in the stellar interior.} \agetrend{that among the stars with significant Li detections (all above the solar value of $\mathrm{A(Li)} = 1.05\,\mathrm{dex}$), we see a tendency of lower Li with increasing age for a given metallicity, as expected (see e.g. the work for solar twins by \citet{Carlos2016} and references therein). The upper envelope for stars at $-0.5 < \mathrm{[Fe/H]} < 0.5$ goes significantly above the Spite plateau due to interstellar and stellar production \citep[see e.g.][]{Prantzos2017} and even reaches the meteoritic values estimated by \citet{Asplund2009}.}
\conclusion{While we see indications that Li is a good tracer of age for the younger dwarfs in our sample, we do not draw any conclusions because of the uncertain temperature-age causality as well as the influence of the metallicity, detection limits, and other potentially important factors such as rotational velocity.}

\subsection{Carbon}

In the spectra, only atomic C lines with high excitation energies could be identified, which are strongest in hot and metal-rich stars. Due to the high excitation energy of the C lines, in our sample only hot stars with metallicities above $-0.3\,\mathrm{dex}$ have detectable line strengths.
Starting from enhanced abundances at metallicities of $-0.75\,\mathrm{dex}$, we see a decreasing trend towards solar metallicities, which flattens at super-solar metallicities. Our results are consistent with the study of \citet{Nissen2014} who demonstrated a linear C-enhancement trend from $\mathrm{[Fe/H]}=0$ to $-0.75\,\mathrm{dex}$.
\importance{Although it shows a behaviour like an $\upalpha$ element, it is expected to follow the iron abundance more closely than these elements and the origin of C is still debated \citep[see][and references therein]{Nissen2014}.}
\agetrend{no significant correlation between C and stellar age for our sample of dwarfs and turn-off stars. Several recent studies, e.g. \citet{Masseron2015}, have discovered significant age-trends for evolved stars, which is an observational manifestation of mass-dependent mixing during the dredge-up phase of the stellar evolution.\citep[e.g.][]{Feuillet2018}. Since our study is limited to dwarf stars, we do not detect such correlations.}
\conclusion{Further, because of the detection limit, we can not draw conclusions regarding the stars with $\mathrm{[Fe/H]} < 0$.}

\newpage
\subsection{Oxygen}

\begin{figure}[!h]
\centering
\resizebox{\hsize}{!}{\includegraphics{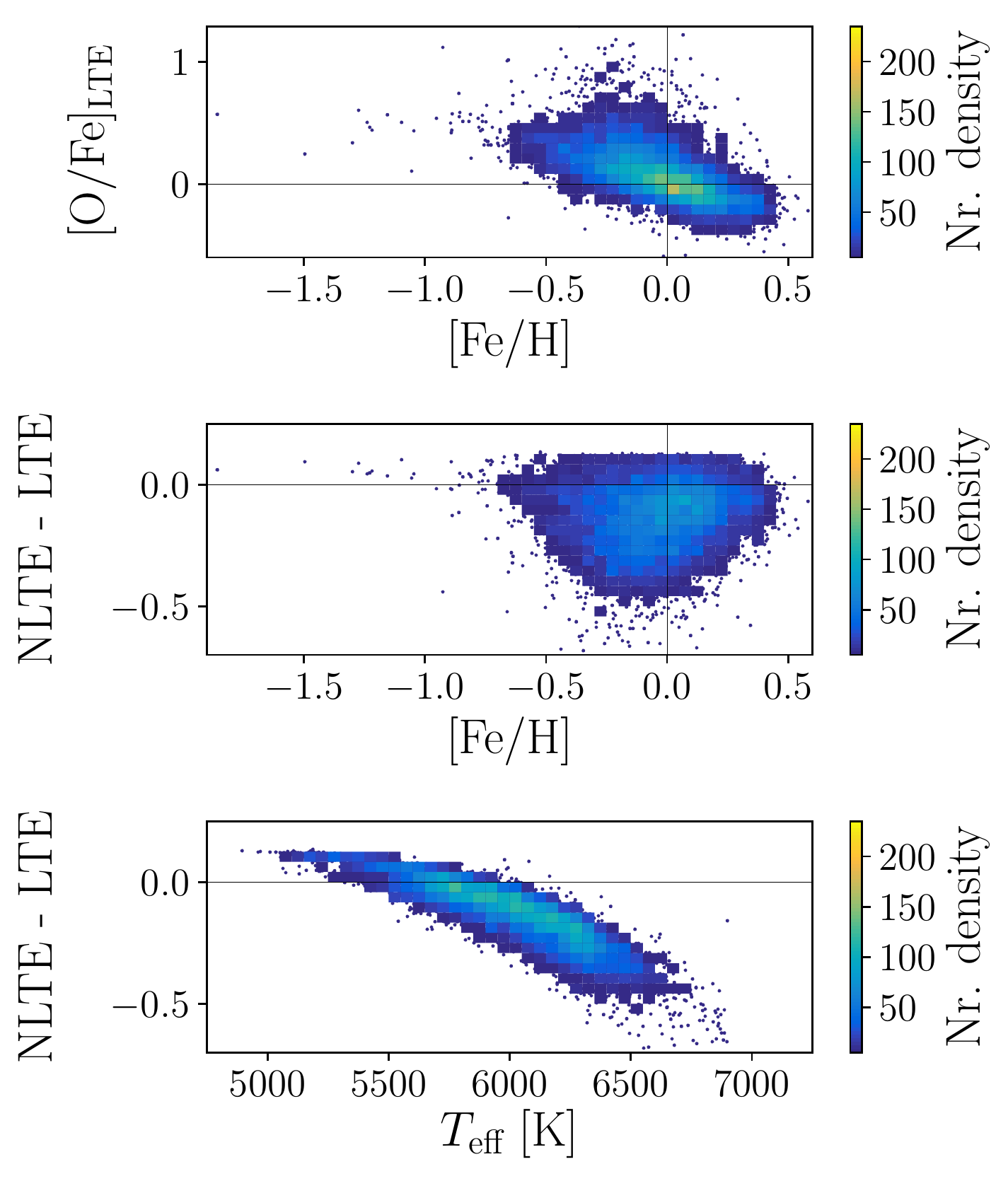}}
\caption{Visualisation of the change of O abundance with respect to the assumption of LTE. Top panel shows [O/Fe] assuming LTE, while middle and bottom panel show difference with respect to non-LTE abundances as a function of metallicity (middle) and effective temperature (bottom). Stars are shown as blue dots or in bins (containing a minimum of 5 stars). The plots indicate that the O abundance of hot stars need to be significantly corrected downwards by up to $0.7\,\mathrm{dex}$, when estimated under the LTE assumption with [O/Fe] up to 1.0.}\label{O_Fe_LTE_NLTE}
\end{figure}

For O, we apply 1D non-LTE corrections based on the model atom and non-LTE radiative transfer code described in \citet{Amarsi2015,Amarsi2016b}, but using 1D {\sc marcs} model atmospheres. These corrections are vital for the O abundances estimated from the O triplet (O {\sc I} $7772\,\mathrm{\AA}$, O {\sc I} $7774\,\mathrm{\AA}$, and O {\sc I} $7775\,\mathrm{\AA}$), as shown in Fig.~\ref{O_Fe_LTE_NLTE}. The corrections are significant for hotter stars and those at the turn-off region. While the global trend of [O/Fe] with metallicity is similar for LTE and non-LTE measurements, the attributed corrections can be as large as $-0.5\,\mathrm{dex}$ for the hottest stars of the sample, hence shifting them down to a similar level as cooler main-sequence stars. We want to stress, that non-LTE corrections for O play a particularly important role when it comes to magnitude limited selections of stars, especially for dwarfs, as more distant and luminous stars tend to be more evolved and in the turn-off region, hence being more affected by departures from LTE.

When comparing with the study by \citet{Bensby2014}, who also analysed the O triplet, we see a strong agreement, i.e. a steep and quite tight linear decrease of [O/Fe] around $0.5\,\mathrm{dex}$ for metal-poor stars to around $-0.25\,\mathrm{dex}$ for the most metal-rich stars.
\importance{The different behaviour of O with respect to the combined $\upalpha$-enhancement is an important nuance. Numerous simulations \citep[e.g. ][]{MCM2013} use O as main or sole tracer of $\upalpha$-enhancement. Accurate estimates of O in stars are vital for scaling measurements of nebulae and galaxies in general, because it is the most abundant element after hydrogen and helium.}
\challenge{We note that for hot and young stars we measure large abundances of O also in non-LTE. For these stars, the stellar parameters are hard to estimate and their non-LTE corrections are hence also less certain, because they are strongly changing with stellar parameters. We can not exclude atmospheric 3D effects for these stars. Although these effects are not expected to be as large as the LTE to non-LTE correction, they are still non-negligible as demonstrated by \citet[][see their Fig.~9]{Amarsi2016b}.}
We note that hot and young stars scatter the age trend at intermediate metallicities, hence blurring the age trend in this regime.

\subsection{$\upalpha$ elements}

Because of the significantly different behaviour of O with respect to the other $\upalpha$-process elements Mg, Si, Ca, and Ti, in the GALAH range, as well as the strong non-LTE corrections (and possibly 3D corrections) needed for O, we have decided to not define O as an $\upalpha$-process element for this study.

Magnesium agrees with the combined $\upalpha$-enhancement trend, but with a larger spread and scatter at all metallicities.
In agreement with \citet{Adibekyan2012}, we find a flatter trend ([Mg/Fe] between 0.0 and $0.1\,\mathrm{dex}$) at the metal-poor end of the low Mg regime than \citet{Bensby2014} (up to $0.2\,\mathrm{dex}$). At the metal-rich end of the distribution, we find a larger relative fraction of Mg-enhanced stars than \citet{Bensby2014}. \citet{Adibekyan2012} also found stars with these abundances, which they assigned to the high-$\upalpha$metal-rich population. We note that \citet{Fuhrmann2011} did not find these stars in his volume-complete study based on Mg.
Mg shows a strong proportionality with age and on average, the trend seems to be metallicity- and temperature-independent.

Silicon measurements in GALAH spectra follow the expected $\upalpha$-enhancement trends. In agreement with the studies by \citet{Adibekyan2012} and \citet{Bensby2014}, Si is in general closer to the solar value, i.e. also less enhanced in the metal-poor regime than Mg and O.

Calcium is measured with lower precision and although a tight trend of Ca abundances is expected for this $\upalpha$ element, the abundances derived from GALAH spectra are very scattered, but agree in general with \citet{Bensby2014} and \citet{Adibekyan2012}.

Titanium agrees well with previous studies by \citet{Adibekyan2012} and \citet{Bensby2014}.
\importance{Ti can be very well measured in optical dwarf spectra, because numerous clean Ti lines are available with good line data. Hence Ti is one of the most precise elements for studies in the optical.}
\challenge{Ti is however not well reproduced by chemical evolution models \citep{Kobayashi2011,Kobayashi2011b}.}

\subsection{Light odd-Z elements}

Sodium shows a large abundance spread ($0.15\,\mathrm{dex}$) compared to the median measurement uncertainties, centered around the solar value for solar metallicities and trends towards more enhancement both towards the sub- and super-solar regime, with different steepness. In both metal-poor and metal-rich stars [Na/Fe] is elevated by up to $0.15\,\mathrm{dex}$ with respect to solar. The correlations with stellar parameter, see Fig.~\ref{trends1a}, indicate that the scatter is caused by non-LTE effects \citep{Lind2011}. The substantial star-to-star scatter persists in the super-solar metallicity regime.
The estimated trends agree well the previous LTE studies by \citet{Adibekyan2012} and \citet{Bensby2014}.
\importance{We note that the abundance trend of this element is very similar to those of Ni and Cu, although with a larger spread.}
The age-correlation indicates a slight Na-enhancement for the younger stars and Na-depletion for the intermediately old stars around solar metallicities, hence the opposite behaviour to that shown by $\upalpha$ elements. The oldest stars of the sample tend to show solar [Na/Fe] at low metallicities, in agreement with Ni and Cu.
We note that the similar behaviour of Na and Ni was already found for solar twins by \citet{Nissen2015}, who identified a very tight correlation of [Na/Fe] and [Ni/Fe] of their sample (see their Fig.~12). They also found that neither element correlates tightly with stellar age.

Aluminium is measured using 1D non-LTE corrections by \citet{Nordlander2017}\footnote{Using the grid available at \url{https://www.mso.anu.edu.au/~thomasn/NLTE/}.}. [Al/Fe] shows a significant spread of $0.25$ around solar abundance ratios at solar metallicities and increasing abundances towards both the metal-rich and metal-poor regimes.
The oldest and most metal-poor stars show Al enhancement up to $0.3\,\mathrm{dex}$, but old stars with solar metallicity are still Al enhanced, while younger stars are closer to solar [Al/Fe] at sub- and solar metallicities, in agreement with \citet{Bensby2014} and \citet{Adibekyan2012}. Contrary to \citet{Bensby2014}, but in agreement with \citet{Adibekyan2012}, the super-solar metallicity stars show an increasing trend similar to the odd-Z element Na. Our Al measurement behave in general very similar to the Mg measurements, including in the metal-rich regime, where a gradual increase with age and with [Al/Fe] can be noticed.%

Potassium shows, similar to O abundance estimates in LTE, an increasing trend of [K/Fe] with effective temperature, most prominent for turn-off stars, see Fig.~\ref{trends1a}, indicating a large influence of non-LTE for the measured atomic resonance line K~{\sc I}~$7699\,\mathrm{\AA}$. These effects are estimated to be of the order of $-0.4$ to $-0.7\,\mathrm{dex}$ \citep{Ivanova2000}. This line suffers from an interstellar contribution, which increases the line depth as a function of a reddening-dependent component, which has not been corrected for. 
It is therefore expected that the LTE trend differs from the non-LTE study by \citet{Zhao2016}. K behaves similar to O both when assuming LTE and non-LTE, which indicates that K behaves like an $\upalpha$ element.
\conclusion{Due to the observational difficulties and strong expected non-LTE effects, we do not draw strong conclusions for this element from our results.}

\subsection{Iron-peak elements}\label{ironpeak}

Scandium shows a similar behaviour as the $\upalpha$ elements Si and Ti, i.e. a flat trend for super-solar metallicities and an increase of Sc abundances towards metal-poor stars, in agreement with \citet{Battistini2015}. 
\citet{Adibekyan2012} found Sc trends similar to Al, with a significant increase of Al towards super-solar metallicities. All three studies have at least three Sc lines in common, hence it is unlikely that the difference originate in the chosen lines themselves. It is worth mentioning that \citet{Adibekyan2012}, using the linelist by \citet{Neves2009}, did not include hyper-fine structure splitting, contrary to this study. The difference in abundance measurements with or without the use of hyper-fine structure splitting is of a very complex nature, but has been shown to play an equally important role for abundance estimations as blending, microturbulence velocities, line choices, line centres, and oscillator strengths \citep{Jofre2017}. 

Vanadium stays flat over most of the metallicity range, but shows a slight increase towards lower metallicities. A large number of lines are however too blended or weak to be used for the V abundance estimation in the metal-rich and poor regime. The estimated trend is consistent with those estimated by \citet{Adibekyan2012} and \citet{Bensby2014}. Where we can detect V, [V/Fe] seems to be rather independent of stellar age, i.e. young and old stars span a range in [V/Fe] around $\pm 0.15$.

Chromium traces Fe along most of the metallicity range and our trends agree in general with \citet{Bensby2014} given our lower precision. In the metal-rich regime, super-solar [Cr/Fe] seems to be favoured. We note that \citet{Adibekyan2012} estimated a very slight anti-correlation of Cr with metallicity (around 0.03 dex [Cr/Fe] within 1 dex metallicity), which our measurements do not suggest.

Manganese measurements agree with the decreasing trend of Mn with metallicity found by \citet{Battistini2015} and \citet{Adibekyan2012} under the assumption of LTE.
\challenge{\citet{Battistini2015} also discuss the significant influence on this trend when considering non-LTE. We only use LTE in our study, but will include non-LTE in future GALAH analysis}

Cobalt was only detected for a very small number of stars with strong lines (4\%), typically metal-rich stars in the sample. We therefore make no conclusions for Co.

Nickel traces Fe for $\mathrm{[Fe/H] < 0}$ and increases towards super-solar metallicities, in agreement with \citet{Adibekyan2012} and \citet{Bensby2014}.

Copper was not detected in the metal-poor regime; we observe solar [Cu/Fe] with a small spread at $\mathrm{[Fe/H] < 0}$, an increasing spread of [Cu/Fe] towards solar [Fe/H], and increasing Cu abundance with metallicity in the super-solar regime. This trend agrees with previous studies by \citet{DelgadoMena2017}, which was performed on a significantly smaller sample.

Zinc measurements in our sample follow the same trend as Ni and Cu, i.e. showing a rather flat behaviour at low metallicities and stronger increase at super-solar metallicities. Zn is however more scattered than Ni and Cu, although the mean uncertainties are comparable. Zn is not created in SN Type Ia according both to theoretical yields \citep{Iwamoto1999, Kobayashi2006} and observations \citep{Nissen2011, Mikolaitis2017, Skuladottir2017}. The yields of Zn are very metallicity-dependent \citep{Kobayashi2006}. We therefore do not expect Zn to behave exactly like $\upalpha$-elements. In our sample, stars with high [Zn/Fe] also show high [$\upalpha$/Fe] at and below solar metallicities. At super-solar metallicities, where the high- and low-$\upalpha$ sequence can not be distinguished, our measurements suggest a linear increase of Zn with a large spread or scatter, as also found by \citet{Bensby2014}.
\challenge{Strong blending (especially towards cooler temperatures) decreases the precision of our measurements, but we note that \citet{Bensby2014, DelgadoMena2017} also estimated a significant spread of Zn with their high-resolution and high $S/N$ data.}

\begin{figure*}[!h]
\centering
\includegraphics[width=8.33cm]{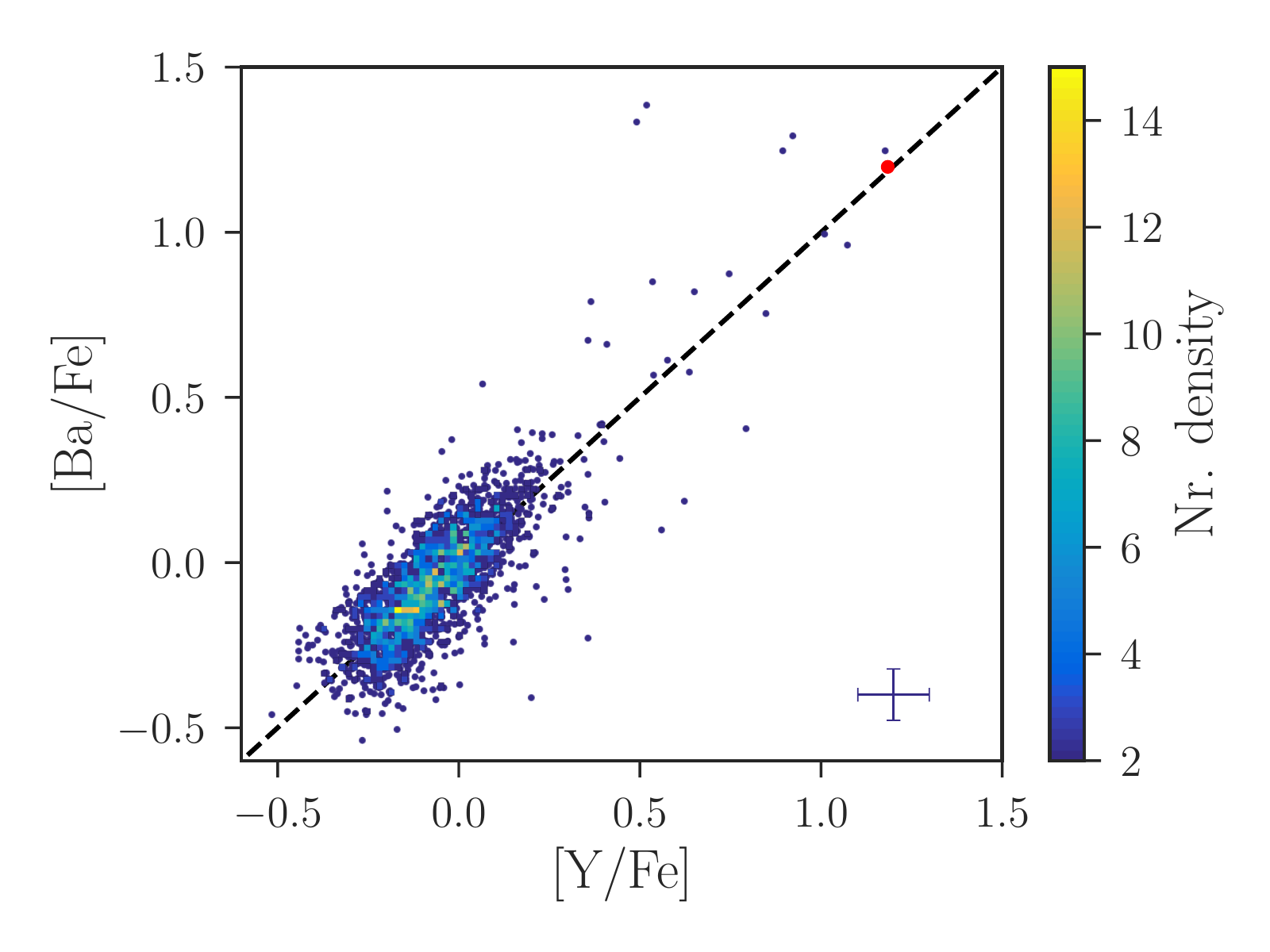}
\includegraphics[width=8.33cm]{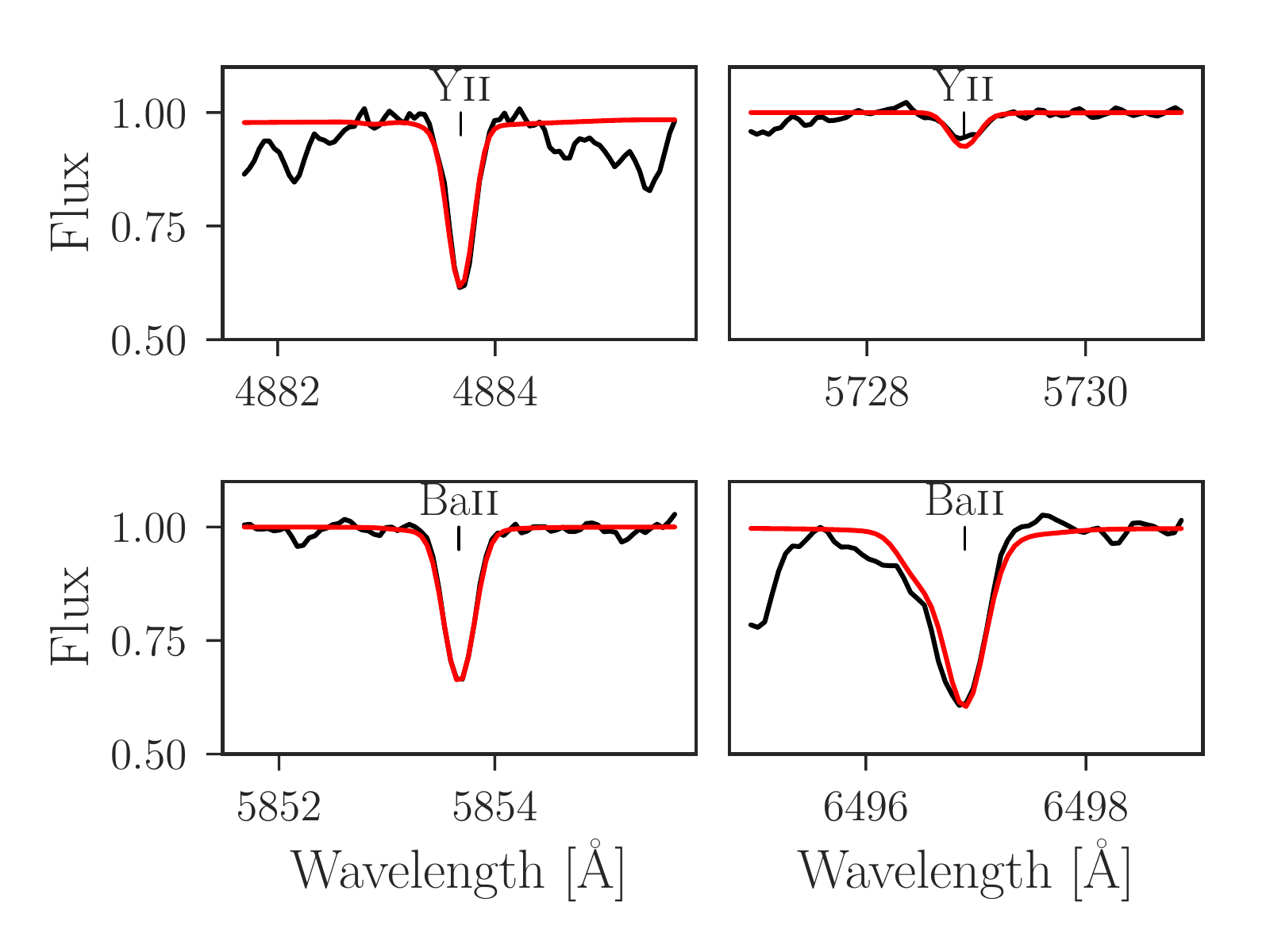}
\caption{The left panel shows the distribution of Ba and Y abundances when measured for both elements. The elements behave close to a 1:1 correlation indicated by a dashed black line in the background. For higher abundances however, Ba is slightly more abundant than Y. Right panels show the observation (black) and synthesis (red) of Ba and Y lines of the s-process enhanced 2MASS J20120895-4129341 (indicated by a red dot in the left panel).}\label{high_YBa}
\end{figure*}

\subsection{The neutron-capture elements}

Several neutron-capture elements have lines that are detectable in GALAH spectra only for giant stars. For this sample of dwarfs and subgiants, the line strength is too small to be significant at the survey, $S/N$. However, Y and Ba have strong singly ionised lines readily detectable also in the majority of our unevolved stars. We therefore only report the abundance of the s-process neutron capture elements Y and Ba relative to iron.

Yttrium shows a lens-like shape, with a lower almost flat lower limit at about $-0.4\,\mathrm{dex}$ and a convex upper limit between metallicities of $-0.7$ and $ 0.4\,\mathrm{dex}$ with most abundances below $+0.4\,\mathrm{dex}$. 
While \citet{Bensby2014} also found some Y-enhanced stars, the vast majority of their stars showed roughly solar Y abundances relative to iron.
High abundance measurements in stars could be a result of non-LTE effects. However, our sample shows such elevated Y values across a wide range of temperatures and surface gravities, see Fig.~\ref{trends1b}. We can not confirm a behaviour similar to O or K, for which LTE-based abundances are strongly overestimated in some stars. Additionally, a strong correlation between [Y/Fe] and age was found for solar twins by \citet{Nissen2015} and \citet{Spina2016}, which are insensitive to differential non-LTE effects because of the similar stellar parameters among solar twins.
We find old stars of our sample to show depleted Y, while young stars are Y-enhanced. From both theory \citep{Travaglio2004} and cluster observations \citep{DOrazi2009,Maiorca2011}, this can be explained with the increasing contribution of s-process material from low-mass asymptotic giant branch stars to the Y and Ba abundances over time.

Barium was measured less frequently than Y, especially for metal-poor stars, but the overall lens-shaped trend of Ba with metallicity is similar and consistent with \citet{Bensby2014}, who showed that a large fraction of their sample with $T_\mathrm{eff} > 6100\,\mathrm{K}$ are Ba-enhanced (see their Fig.~16). We note that similar to \citet{DelgadoMena2017}, the most metal-rich stars show lower [Ba/Fe] than [Y/Fe] by around $0.1\,\mathrm{dex}$. Investigating the flagged stars, this trend seems to be caused by detection limits for Ba.
\agetrend{these hotter stars are however also younger and their Ba abundance correlates, similar to Y, with the stellar age, as shown by \citet{Nissen2015} and \citet{Spina2016}.}

When comparing [Y/Fe] and [Ba/Fe] in Fig.~\ref{high_YBa}, we find a reassuringly strong correlation of both, with a slightly steeper slope of Ba abundance compared to Y. In the right panels of Fig.~\ref{high_YBa}, we depict the s-process enhanced star 2MASS J20120895-4129341, indicated by a red dot in the left panel, with strong Y and Ba lines.

\subsection{Correlations with stellar parameters}

Correlations of abundances with effective temperature, surface gravity, and microturbulence velocity are shown in Figs.~\ref{trends1a} and \ref{trends1b}. If the actual element abundances do not change with evolutionary stage, we expect flat trends for all these parameters. An element, for which our measurements are systematically off, is K, which suffers from strong non-LTE effect especially for the turn-off stars. We note however, that some elements are subject to changes in the evolutionary stage (e.g. Li, which shows a steep trend with $T_\mathrm{eff}$).

\begin{figure*}[p!]
\centering
\includegraphics[width=17cm]{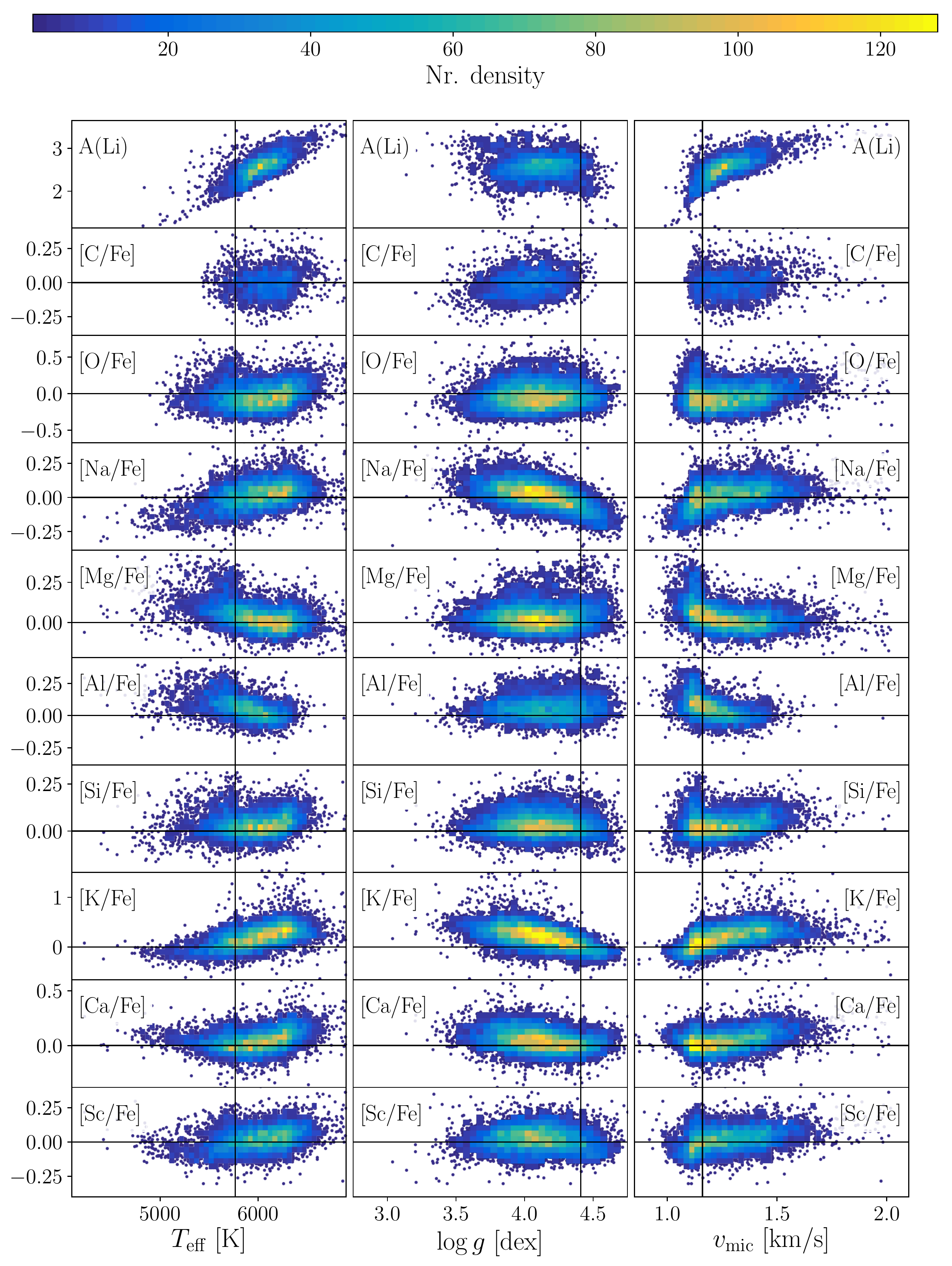}
\caption{Element abundances ordered by atomic number for the clean sample of dwarf, turn-off and subgiant stars (elements X indicated in top right corner of each panel) as a function of $T_\text{eff}$ (left panels), $\log g$ (middle panels), and $v_\text{mic}$ (right panels). All elements are shown relative to the iron abundance, except for Li (shown as absolute abundance). Colour indicates the density of stars with a minimum of 5 star per bin. Individual stars outside the bins are shown as small dots. See individual paragraphs in Sect.~\ref{abundancetrends} for further descriptions.}\label{trends1a}
\end{figure*}

\begin{figure*}[p!]
\centering
\includegraphics[width=17cm]{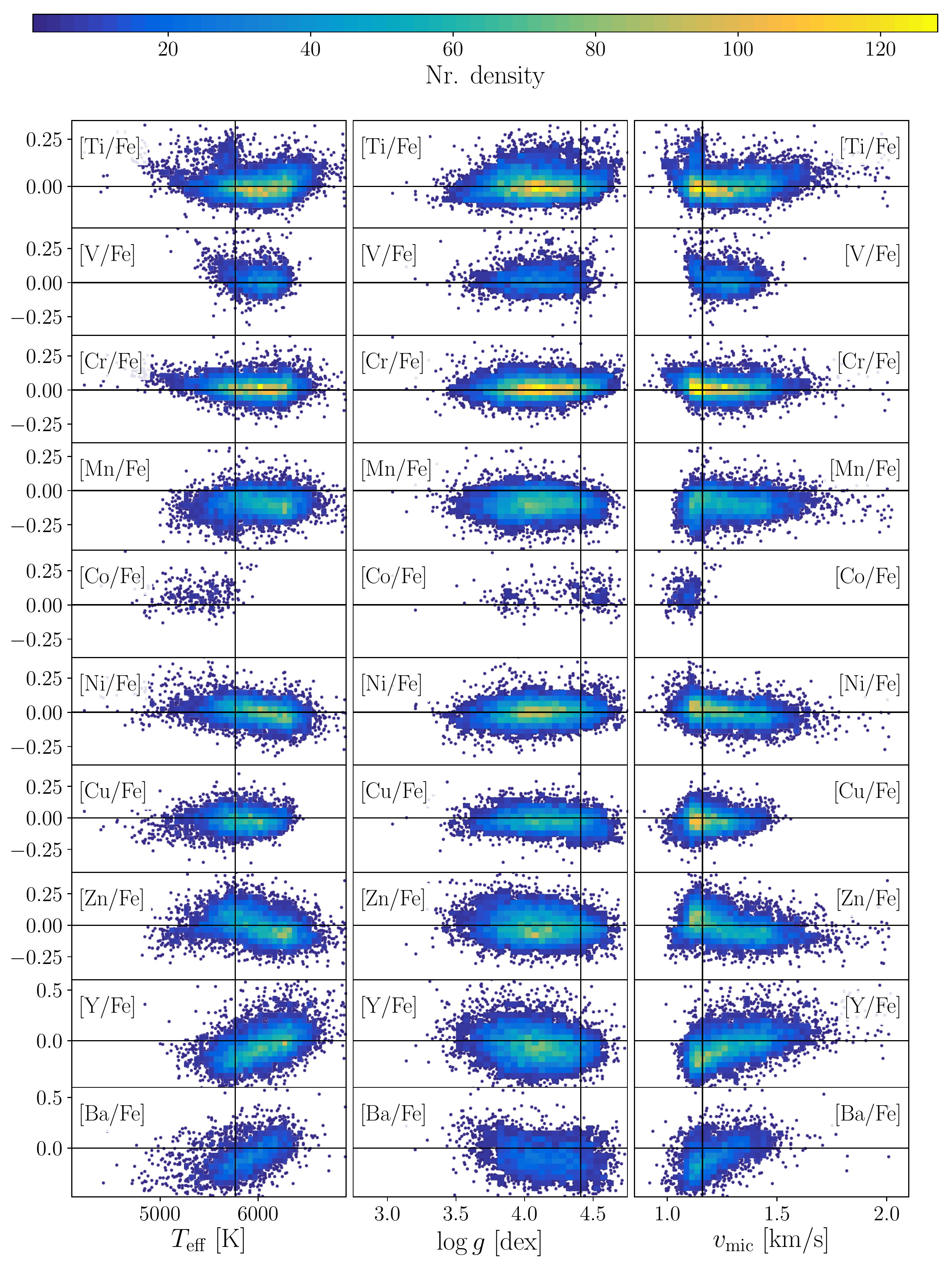}
\caption{continuation of Fig.~\ref{trends1a}}\label{trends1b}
\end{figure*}

\section{Description of the online tables}\label{table_description}

In Table~\ref{catalog} we list the content of each column in the chemodynamical table. The catalog data can be accessed online via {\sc{URL FOR VIZIER HERE}}.

\longtab[1]{
\begin{longtable}{l l l l l}
\caption{Catalog Description}\label{catalog}\\
\hline
\hline
Name      & Description                   & Units       & Datatype & Source \\
\hline
\endfirsthead
\caption{Continued.} \\
\hline
Name      & Description                   & Units       & Datatype & Source \\
\hline
\endhead
\hline
\endfoot
\hline
\endlastfoot
source\_id  			& Gaia identifier        												&             		& int64    		& Gaia\\      
tmass\_id 				& 2MASS identifier												&          		& char[16] 	& 2MASS \\      
sobject\_id  			& GALAH Spectrum identifier										&  			& int64    		& GALAH \\      
ra        				& Right Ascension (ICRS,Epoch=J2000)								& deg		& float64  		& UCAC4 \\
dec       				& Declination     (ICRS,Epoch=J2000)								& deg		& float64  		& UCAC4 \\
snr\_c1   			& $S/N$ per pixel for ccd-1 (blue channel)   							&             		& float64 		& Sect.~2 \\      
snr\_c2   			& $S/N$ per pixel for ccd-2 (green channel)  							&             		& float64 		& Sect.~2 \\         
snr\_c3   			& $S/N$ per pixel for ccd-3 (red channel)    							&             		& float64 		& Sect.~2 \\      
snr\_c4   			& $S/N$ per pixel for ccd-4 (infrared channel)							&            		& float64 		& Sect.~2 \\      
Bmag      				& APASS $B$ magnitude											& mag 		& float64 		& APASS \\
e\_Bmag				& Uncertainty in APASS $B$ mag  									& mag 		& float64 		& APASS \\
Vmag      				& APASS $V$ magnitude											& mag 		& float64 		& APASS \\
e\_Vmag				& Uncertainty in APASS $V$ mag  									& mag 		& float64 		& APASS \\
Jmag      				& 2MASS $J$ magnitude											& mag 		& float64 		& 2MASS \\
e\_Jmag				& Uncertainty in 2MASS $J$ mag  									& mag 		& float64 		& 2MASS \\
Hmag      				& 2MASS $H$ magnitude           									& mag 		& float64 		& 2MASS \\
e\_Hmag				& Uncertainty in 2MASS $H$ mag  									& mag 		& float64 		& 2MASS \\
Kmag      				& 2MASS $Ks$ magnitude          									& mag 		& float64 		& 2MASS \\
e\_Kmag				& Uncertainty in 2MASS $K$ mag  									& mag 		& float64 		& 2MASS \\
Qfl					& Uncertainty in 2MASS $K$ mag  									& mag 		& char[3] 		& 2MASS \\
W1mag     			& WISE $W1$ magnitude											& mag 		& float64 		& WISE \\
e\_W1mag			& Uncertainty in WISE $W1$ mag  									& mag 		& float64 		& WISE \\
W2mag     			& WISE $W2$ magnitude											& mag 		& float64 		& WISE \\
e\_W2mag			& Uncertainty in WISE $W2$ mag  									& mag 		& float64 		& WISE \\
W3mag     			& WISE $W3$ magnitude											& mag 		& float64 		& WISE \\
e\_W3mag			& Uncertainty in WISE $W3$ mag  									& mag 		& float64 		& WISE \\
W4mag     			& WISE $W4$ magnitude											& mag 		& float64 		& WISE \\
e\_W4mag			& Uncertainty in WISE $W4$ mag  									& mag 		& float64 		& WISE \\
ebv       				& Reddening $E(B-V)$           										& mag       	& float64 		& SFD+98 \\     
A\_K       				& Attenuation in $Ks$				        							& mag   		& float64 		& GALAH, Sect.~3.1 \\     
BC\_K      				& Bolometric corrections for $Ks$									& mag   		& float64 		& GALAH, Sect.~3.1 \\     
parallax				& Parallax														& mas		& float64		& Gaia \\
parallax\_error			& Parallax	error													& mas		& float64		& Gaia \\
pmra					& Proper motion in Right Ascension									& mas/yr		& float64		& Gaia \\
pmra\_error			& Proper motion error in Right Ascension								& mas/yr		& float64		& Gaia \\
pmdec				& Proper motion in Declination										& mas/yr		& float64		& Gaia \\
pmdec\_error			& Proper motion error in Declination									& mas/yr		& float64		& Gaia \\
rMo\_3				& Mode distance of the posterior, using Milky Way Prior					& pc			& float64		& AB+16 \\
r5\_3					& 5th percentile of the posterior, using Milky Way Prior					& pc			& float64		& AB+16 \\
r50\_3				& 50th percentile of the posterior, using Milky Way Prior 					& pc			& float64		& AB+16 \\
r95\_3				& 95th percentile of the posterior, using Milky Way Prior					& pc			& float64		& AB+16 \\
sigmaR\_3			& Distance standard error, using Milky Way Prior						& pc			& float64		& AB+16 \\
comb\_sp				& Spectra used for the combination of stellar parameters					&			& int64		& Sect.~3.1 \\
teff\_sme  				& Effective temperature  $T_\mathrm{eff}$							& K           		& float64 		& Sect.~3.1 \\  
e\_teff\_sme  			& Uncertainty in effective temperature  $T_\mathrm{eff}$					& K           		& float64 		& Sect.~3.1 \\
logg\_sme  			& Surface gravity  $\log g$										& dex		& float64		& Sect.~3.1 \\
e\_logg\_sme  			& Uncertainty in surface gravity  $\log g$								& dex		& float64		& Sect.~3.1 \\
feh\_sme  				& SME pseudo iron abundance [Fe/H]								& dex		& float64		& Sect.~3.1 \\
e\_feh\_sme  			& Uncertainty in SME pseudo iron abundance [Fe/H]						& dex		& float64		& Sect.~3.1 \\
vmic\_sme			& Microturbulence velocity $\xi$									& km/s		& float64		& Sect.~3.1 \\
e\_vmic\_sme			& Uncertainty of microturbulence velocity $\xi$							& km/s		& float64		& Sect.~3.1 \\
vsini\_sme			& Rotational and Macroturbulence Velocity $v_{\rm mac+rot}$				& km/s		& float64		& Sect.~3.1 \\
e\_vsini\_sme			& Uncertainty of $v_{\rm mac+rot}$									& km/s		& float64		& Sect.~3.1 \\
rv\_sme				& Radial velocity												& km/s		& float64		& Sect.~3.1 \\
e\_rv\_sme			& Uncertainty of radial velocity										& km/s		& float64		& Sect.~3.1 \\
chi\_sme				& $\chi^2$ of the spectroscopic stellar parameter fit						& 			& float64		& Sect.~3.1 \\
flag\_sme				& Spectroscopic quality flag										&			& float64		& Sect.~3.1 \\
alpha\_fe\_sme			& Combined $\upalpha$ element abundance [$\upalpha$/Fe]					& dex		& float64		& Sect.~3.5 \\
e\_alpha\_fe\_sme		& Uncertainty of combined $\upalpha$ element abundance [$\upalpha$/Fe]		& dex		& float64		& Sect.~3.5 \\  
x\_abund\_sme			& Abundance [X/Fe] of element X									& dex		& float64		& Sect.~3.5 \\  
e\_x\_abund\_sme		& Uncertainty of abundance [X/Fe] of element X						& dex		& float64		& Sect.~3.5 \\  
chi\_x\_abund\_sme		& $\chi^2$ of abundance fit for element X								& dex		& float64		& Sect.~3.5 \\  
flag\_x\_abund\_sme		& Spectroscopic quality flag for abundance of  X						& dex		& float64		& Sect.~3.5 \\  
comb\_x\_abund\_sme	& Number of spectra used to estimate x\_abund\_sme					& dex		& float64		& Sect.~3.5 \\  
age\_mean			& Mean age $\tau$												& Gyr		& float64		& ELLI, Sect.~3.3 \\
age\_std				& Uncertainty of age $\tau$										& Gyr		& float64		& ELLI, Sect.~3.3 \\
mass\_mean      		& Actual mass $m_\mathrm{act}$ (incl. mass loss)						& $\mathcal M_{\odot}$ 	& float64 		& ELLI, Sect.~3.3 \\
mass\_std	      			& Uncertainty of actual mass $m_\mathrm{act}$ (incl. mass loss)			& $\mathcal M_{\odot}$ 	& float64 		& ELLI, Sect.~3.3 \\
R\_kpc	      			& Galactocentric radius											& kpc		& float64		& Sect.~3.6 \\
Phi	      				& Galactocentric azimuth angle										& kpc		& float64		& Sect.~3.6 \\
z\_kpc	      			& Height above Galactocentric plane									& kpc		& float64		& Sect.~3.6 \\
vR\_kms	      			& Galactocentric radial velocity										& km/s		& float64		& Sect.~3.6 \\
vT\_kms	      			& Galactocentric tangential										& km/s		& float64		& Sect.~3.6 \\
vz\_kms	      			& Galactocentric vertical											& km/s		& float64		& Sect.~3.6 \\
x\_xyz\_kpc	      		& Galactocentric cartesian coordinate X								& kpc		& float64		& Sect.~3.6 \\
y\_xyz\_kpc	      		& Galactocentric cartesian coordinate X								& kpc		& float64		& Sect.~3.6 \\
z\_xyz\_kpc	      		& Galactocentric cartesian coordinate X								& kpc		& float64		& Sect.~3.6 \\
U\_LSR	      			& Galactocentric cartesian velocity U									& km/s		& float64		& Sect.~3.6 \\
V\_LSR	      			& Galactocentric cartesian velocity V									& km/s		& float64		& Sect.~3.6 \\
W\_LSR	      			& Galactocentric cartesian velocity W								& km/s		& float64		& Sect.~3.6 \\
jR\_kpckms\_mean	      	& Radial action													& kpc km/s	& float64		& Sect.~3.6 \\
jR\_kpckms\_std	      	& Uncertainty of radial action										& kpc km/s	& float64		& Sect.~3.6 \\
Lz\_kpckms\_mean	      	& Angular momentum / azimuthal action								& kpc km/s	& float64		& Sect.~3.6 \\
Lz\_kpckms\_std	      	& Uncertainty of angular momentum / azimuthal action					& kpc km/s	& float64		& Sect.~3.6 \\
jz\_kpckms\_mean	      	& Vertical action												& kpc km/s	& float64		& Sect.~3.6 \\
jz\_kpckms\_std	      	& Uncertainty of vertical action										& kpc km/s	& float64		& Sect.~3.6 \\
TD\_D	      			& Rel. kinematic probabilities the thick disk-to-thin disk 					& 			& float64		& Sect.~3.6. \\
TD\_H	      			& Rel. kinematic probabilities the thick disk-to-halo 						& 			& float64		& Sect.~3.6. \\
D\_H	      				& Rel. kinematic probabilities the thin disk-to-halo 						& 			& float64		& Sect.~3.6. \\
\end{longtable}
\tablefoot{AB+16 and SFD+98 are the abbreviations of \citet{Astraatmadja2016} and \citet{Schlegel1998} respectively.}
}

\end{appendix}

\end{document}